\documentclass{vgtc}                          

\ifpdf
  \pdfoutput=1\relax                   
  \usepackage{graphicx}                
  \DeclareGraphicsExtensions{.pdf,.png,.jpg,.jpeg} 
\else
  \usepackage{graphicx}                
  \DeclareGraphicsExtensions{.eps}     
\fi%

\usepackage{wrapfig}
\usepackage{dblfloatfix}
\usepackage{xspace}
\usepackage{multirow}
\usepackage{compactalgo}
\usepackage{subcaption}
\usepackage[font={sf,scriptsize}]{caption}
\usepackage{hyperref}

\graphicspath{{figures/}{pictures/}{images/}{./}} 

\usepackage[table]{xcolor}
\definecolor{c1}{rgb}{1,0.54,0.0}
\definecolor{c2}{rgb}{0.4,0,0.2}
\definecolor{c3}{rgb}{0.16, 0.5, 0.0}
\definecolor{c4}{rgb}{0.2 0.2, 1}
\definecolor{KSra}{RGB}{197, 92, 40}
\definecolor{KSdi}{RGB}{255, 190, 0}
\definecolor{JMP}{RGB}{112, 173, 71}
\definecolor{CRS}{RGB}{55, 86, 35}
\definecolor{KSte}{RGB}{0, 102, 202}

\newcommand{\mypar}[1]{\smallskip\noindent{\bfseries #1.}}
\newcommand{\etal}{\emph{et al.}\xspace}

\usepackage[utf8]{inputenc}
\usepackage{algorithm}
\usepackage[noend]{algorithmic}

\makeatletter
\newcommand\clearsubcaptionflag{%
  \@ifundefined{caption@chgflag}{}{%
    \caption@clrflag{subcaption}}}
\makeatother

\usepackage{microtype}                 
\PassOptionsToPackage{warn}{textcomp}  
\usepackage{textcomp}                  
\usepackage{mathptmx}                  
\usepackage{times}                     
\usepackage{cite}                      
\usepackage{tabu}                      
\usepackage{booktabs}                  

\onlineid{0}

\vgtccategory{Research}

\title{Stable Visual Summaries for Trajectory Collections}

\author{Jules Wulms\thanks{e-mail: jwulms@ac.tuwien.ac.at}\\ %
        \scriptsize TU Wien%
\and Juri Buchm\"uller\thanks{e-mail: juri.buchmueller@uni-konstanz.de}\\ %
     \scriptsize University of Konstanz %
\and Wouter Meulemans\thanks{e-mail: [w.meulemans$\,|\,$k.a.b.verbeek$\,|\,$b.speckmann]@tue.nl}\\ %
     \scriptsize TU Eindhoven %
\and Kevin Verbeek\footnotemark[3]\\ %
     \scriptsize TU Eindhoven %
\and Bettina Speckmann\footnotemark[3]\\ %
     \scriptsize TU Eindhoven %
}

\teaser{
  \centering
  \includegraphics[width=\linewidth]{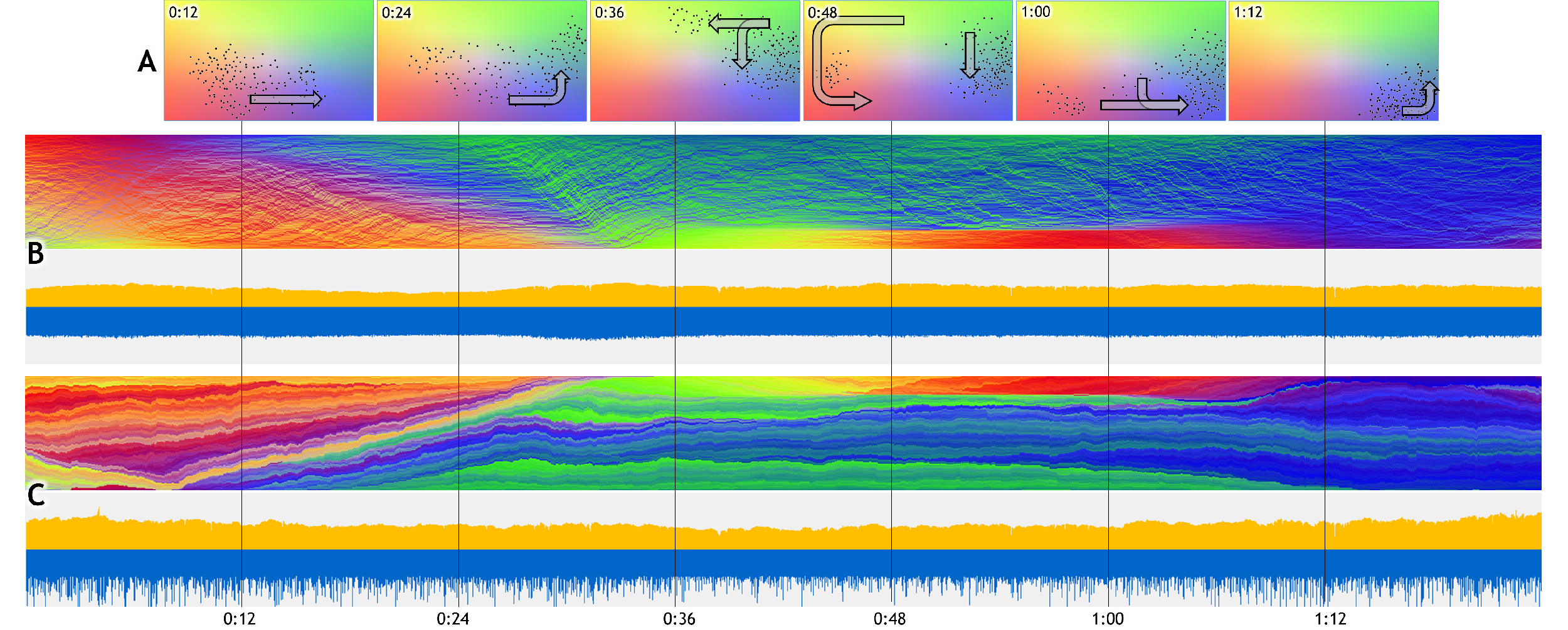}
  \caption{(A) Sample frames of movement; background indicates spatial coloring. 
  (B) Our Stable Principal Component method translates collections of moving entities at each time step into a vertical array of pixels; color indicates spatial location. Chart below indicates spatial quality (yellow) and stability (blue) per time step; high values indicate low quality. (C) Unstable ordering using Hilbert curve, note the artificial split between the green locations.}
	\label{fig:teaser}
}

\abstract{
The availability of devices that track moving objects has led to an explosive growth in trajectory data. When exploring the resulting large trajectory collections, visual summaries are a useful tool to identify time intervals of interest. A typical approach is to represent the spatial positions of the tracked objects at each time step via a one-dimensional ordering; visualizations of such orderings can then be placed in temporal order along a time line.
There are two main criteria to assess the quality of the resulting visual summary: \emph{spatial quality} -- how well does the ordering capture the structure of the data at each time step, and \emph{stability} -- how coherent are the orderings over consecutive time steps or temporal ranges?

In this paper we introduce a new Stable Principal Component (SPC) method to compute such orderings, which is explicitly parameterized for stability, allowing a trade-off between the spatial quality and stability. 
We conduct extensive computational experiments
that quantitatively compare the orderings produced by ours and other stable dimensionality-reduction methods to various state-of-the-art approaches using a set of well-established quality metrics that capture spatial quality and stability. We conclude that stable dimensionality reduction outperforms existing methods on stability, without sacrificing spatial quality or efficiency; in particular, our new SPC method does so at a fraction of the computational costs.
} 

\CCScatlist{
   \CCScatTwelve{Human-centered computing}{Visu\-al\-iza\-tion}{Visualization techniques}{};
   \CCScatTwelve{Mathematics of computing}{Probability and statistics}{Statistical paradigms}{Dimensionality reduction}
}

\begin{document}


\firstsection{Introduction}

\maketitle

\begin{figure*}[t!]
 \centering 
 \includegraphics[width=1\linewidth]{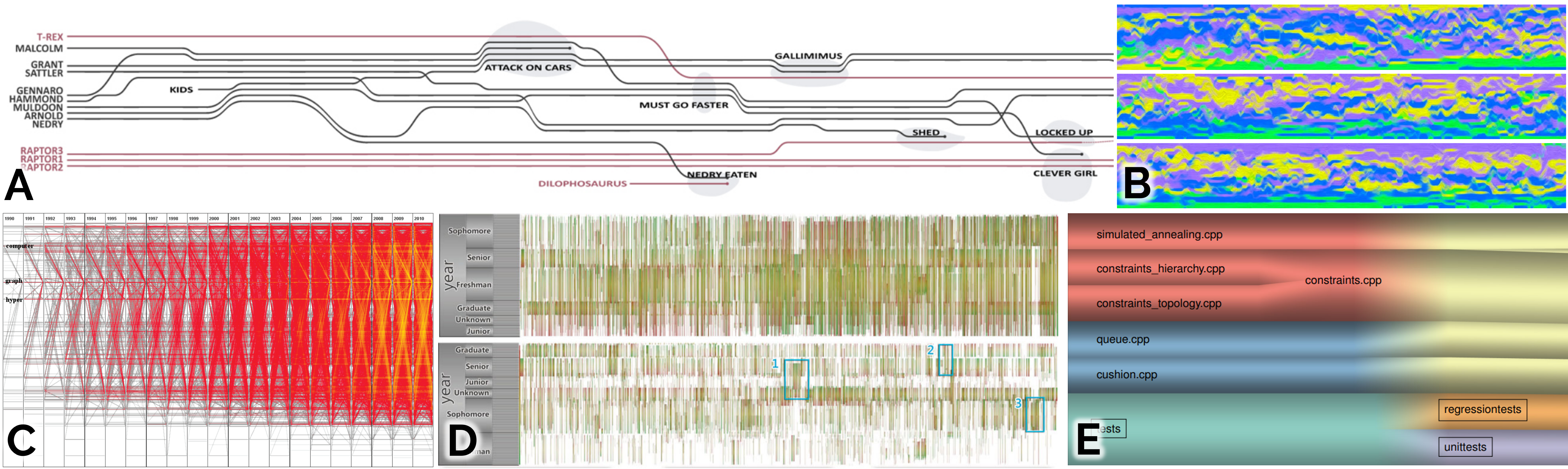}
 \caption{Examples of visual summaries in existing work: 
 (A) Dynamic StoryLine graph \cite{DBLP:journals/jgaa/DijkFFLMRSW17}, (B) Let It Flow for dynamic graphs \cite{DBLP:conf/apvis/CuiWLRMMG14}, (C)~Parallel Edge Splatting \cite{DBLP:journals/tvcg/BurchVBDW11}, (D) Extended Massive Sequence Views \cite{van2014dynamic} and (E) Temporal Treemaps \cite{DBLP:journals/tvcg/KoppW19}.}
 \label{fig:relworklabeled}
\end{figure*}

Over the past years the availability of devices that track moving objects –- GPS satellite systems, mobile phones, radio telemetry, surveillance cameras, RFID tags, and more –- has increased dramatically, leading to an explosive growth in trajectory data. Objects being tracked range from animals (for behavioral studies) and cars (for traffic prediction), to hurricanes, sports players (for video analysis of games), and suspected terrorists. 
When exploring the resulting large trajectory collections, visual summaries are a useful tool to identify time intervals for further consideration. A typical approach is to represent the spatial positions of the tracked objects at each time step via a one-dimensional ordering; visualizations of such orderings can then be placed in temporal order along a time line. 

MotionRugs by Buchm\"uller~\etal~\cite{BJC+19} (Fig.~\ref{fig:teaser}D) translate orderings of 2D moving objects into a compact grid-based visual summary; in a MotionRug each vertical strip of $n$ grid cells encodes $n$ objects at a specific time step. This representation is arguably as compact as possible and served as the initial motivation for our work. MotionRugs, and other previous work~\cite{BJC+19,guo2006spatial} which computes  orderings for moving entities, use either spatial subdivisions or clustering techniques. As a result, entities which are close in the ordering are also close in the input. However, the converse does not necessarily hold: elements which are separated in the  ordering can be close in the input. This causes unfortunate and visually salient artifacts (the so-called ``phantom splits'' in MotionRugs). Let us emphasize that these artifacts are not caused by the visual encoding in MotionRugs, but they are an inherent consequence of using spatial subdivision or clustering techniques to create orderings; any visualization which uses the resulting orderings will exhibit the same artifacts.

The algorithmic problem to be solved is in fact low-dimensional dimensionality reduction: how to adequately represent higher-dimension data in 1D? In this paper we hence propose to use actual dimensionality-reduction techniques to compute orderings of high \emph{spatial quality}: objects which are close in the input are also close in the ordering, avoiding the aforementioned artifacts. We also show that these techniques can produce coherent representations over whole temporal ranges, that is, they can be adapted to be \emph{stable}. 

\mypar{Formal problem statement}
Our input is a set $P = \{ p_1, \ldots, p_n\}$ of $n$ point objects. We sample their positions at $T$ consecutive time steps; each object $p_i$ is a sequence of $T$ locations. We use $p_{i}(t)$ to denote the location of $p_i$  at time $t$, $1 \leq t \leq T$, and, correspondingly, $P(t)$ to denote the complete point set at time $t$.
A visual summary $S$ of $P$ is a sequence of orderings of the points in $P$, one per time step. We denote the ordering at time $t$ by $S_t$, $S_t(p_i)$ denotes the rank of point object $p_i$ in the ordering at time $t$.
%
The quality of a visual summary $S$ is determined by two criteria: 
\begin{description}
\item [Spatial quality.] How well does $S_t$ capture the spatial structure of~$P(t)$? We characterize the spatial structure via local neighborhoods: we say that an ordering has high spatial quality if points that are spatially close in the input are also close in the ordering.
\item [Stability.] How consistent are the orderings over time? Here we can consider absolute changes between orderings or changes in local neighborhoods, as captured by nearest neighbors in the ordering. Both types of measures can be considered for consecutive time steps or over temporal ranges.
\end{description}
Clearly, a visual summary that uses the same ordering for all time steps is maximally stable. However, the spatial quality of this representation will typically be low. Conversely, optimizing spatial quality for each time step in isolation tends to result in unstable summaries which make it more difficult for the user to track objects. 

\mypar{Contributions}
In this paper we explore the use of dimensionality-reduction techniques to create orderings from trajectory collections. Our contributions are twofold.
\textbf{(1)} We introduce a new Stable Principal Component method [SPC] which is explicitly parametrized for stability, allowing a trade-off between spatial quality and stability. We chose to ``stabilize'' PCA since its principal component gives us an explicit representation of the shape of a trajectory collection at any point in time. We can interpolate between the first components of different time steps to achieve stability.
We also describe a stable Clustered Principal Component [CPC] method which is particularly well suited for data sets that exhibit clear clusters. For ease of explanation we describe our approaches in two-dimensions, however, they directly extend to three or higher dimensions. \textbf{(2)} We conduct extensive computational experiments, which allow us to conclude that stable dimensionality reduction outperforms existing methods on stability, without sacrificing spatial quality or efficiency. In particular, our new SPC methods does so at a fraction of the computational costs.

We discuss related work in Section~\ref{sec:related} and describe our Stable Principal Component method in Section~\ref{sec:stablePCA}.
Section~\ref{sec:setup} explains our experimental setup, including the ordering methods which we compare against (Section~\ref{sec:algorithms}), the quality metrics we use to capture spatial quality and stability (Section~\ref{sec:metrics}), and our real-world and synthetic data sets (Section~\ref{sec:data}).
In Section~\ref{sec:experiments} we report on the results of our experiments. 
In Section~\ref{sec:discussion} we close with a discussion of our results, as well as current limitations and directions for future work.

\section{Related work}\label{sec:related}

Visual summaries have been used for various different types of time-varying data.
For example, there are several methods that summarize time-varying graphs, such as Parallel Edge Splatting
~\cite{DBLP:journals/tvcg/BurchVBDW11} (Fig.~\ref{fig:relworklabeled}C) and Extended Massive Sequence Views
~\cite{van2014dynamic} (Fig.~\ref{fig:relworklabeled}D), which show the temporal evolution by drawing the graph at each time step in a narrow vertical strip. Similarly, Temporal Treemaps
~\cite{DBLP:journals/tvcg/KoppW19} (Fig.~\ref{fig:relworklabeled}E) encode hierarchies via (essentially) one-dimensional intervals and show the temporal evolution by placing these intervals consecutively along a line. Also, Storyline Visualizations~\cite{DBLP:journals/jgaa/DijkFFLMRSW17,DBLP:journals/tvcg/LiuWWLL13} (Fig.~\ref{fig:relworklabeled}A) use a compact representation at each time step (essentially a pixel per protagonist); these representations must be coherent between consecutive time steps and as such trace a trajectory for each actor.
An interesting variation of Storylines is presented by Muelder et al.~\cite{muelder}, who create so-called Behavioral Lines, which consist of a collection of features as 1D time series turned into comparable 2D lines, allowing users to identify similar and deviating behavior of the observed feature sets. J\"ackle et al.~\cite{temporalmds} introduce another technique which is based on one-dimensional orderings, namely window-based 1D MDS plots, which are arranged horizontally, grouping together similar events. The aforementioned techniques all rely on variations of similarity measures to determine the ordering of entities and thus, the visual outcome. Our approach takes order quality into account, as well as the stability of the visualization as a function of the changes in the entity orders.

Another set of techniques exploits one-dimensional mappings to produce dense representations of temporal data to generate insightful visual patterns. MotionRugs by Buchm\"uller~\etal~\cite{BJC+19} (Fig.~\ref{fig:teaser}C and D) computes orders from spatial locations for entities moving in 2D, whereas Cui~\etal~\cite{DBLP:conf/apvis/CuiWLRMMG14} (Fig.~\ref{fig:relworklabeled}B) use node degree to order dynamic graph data. Due to the packed representation of the ordering of the displayed data points, the visual outcomes of such techniques are specifically sensitive to ordering quality and, thus, could benefit directly from our approach.


In this paper we focus on computing orderings using dimen\-siona\-lity-reduction techniques. We perform experiments with PCA, Sammon mapping, and t-SNE. 
There are other dimensionality-reduction techniques, such as MDS~\cite{Kruskal1964}, Isomap~\cite{Tenenbaum2000}, and UMAP~\cite{umap2018}, but given the cost functions that they minimize, we believe that they give similar results (in fact, in the Euclidean plane, classical MDS is equivalent to PCA). Recently, Rauber~\etal~\cite{DBLP:conf/vissym/RauberFT16} also described Dynamic t-SNE: a more explicit way of making t-SNE stable over multiple time steps. Unfortunately, various issues prevented the inclusion of this method in our experiments; see Appendix~\ref{app:algorithms} and~\ref{app:summary-stats} for details.





\section{Stable Principal Component Analysis}\label{sec:stablePCA}

\begingroup
\setlength{\columnsep}{1.0em}
\begin{wrapfigure}{r}{0.2\linewidth}
  \includegraphics[width=\linewidth]{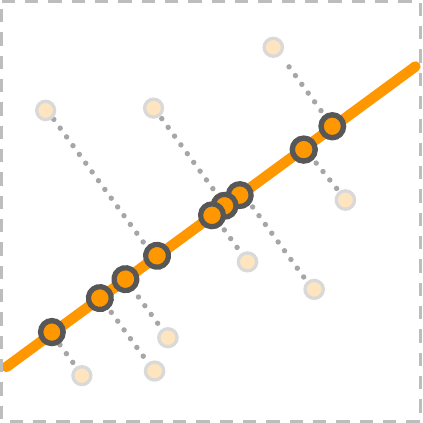}
\end{wrapfigure}
PCA was first introduced by Pearson \cite{pearson1901liii} and can be used for dimensionality reduction to 1D by projecting points onto the first principal component: a vector in the direction along which the point set has most variance. Projecting onto this vector maximally preserves spatial relations in the original point set. Based on this technique, we describe two algorithms that make such projections stable for moving entities.

\endgroup



Meulemans, Verbeek and Wulms~\cite{meulemans2019stability} study the trade-off between spatial quality and stability of orientation-based shape descriptors, including PCA, from a theoretical point of view. Their results show that the principal components of a set of moving points in 2D exhibit unstable behavior when the point set is not stretched, that is, the variance along the first and second principal component is similar. 
Our approach leverages this result by explicitly enforcing stability when the point set is not stretched. The intuition is as follows. If the variance along the first principal component is clearly higher than the variance along the second principal component, then the direction is very discriminative: the point set is clearly stretched in this direction and sorting the points along this vector tends to lead to high spatial quality. If this is not the case, then the point set is ``round'' and the spatial quality is roughly equivalent for other directions as well. Our goal is to smoothly interpolate the projection vector in those cases.

%

\begingroup
\setlength{\columnsep}{1.0em}
\begin{wrapfigure}{r}{0.2\linewidth}
  \includegraphics[width=\linewidth]{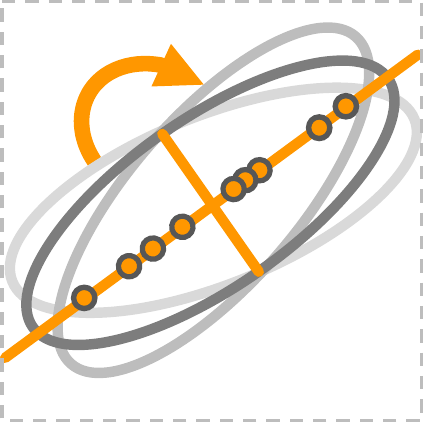}
\end{wrapfigure}
\mypar{[SPC$_\sigma$] Stable Principal Component} 
To create a stable version of PCA, we use the optimal direction (first principal component) as projection vector for any $t$ where $P(t)$ is stretched, as well as for the first and last time step. For all time steps in between (when the point set is not stretched) we linearly interpolate the orientation of the projection vector. We use a parameter $\sigma$ ($0 \leq \sigma \leq 1$) to control when we consider a point set as stretched.

\endgroup

Concretely, the Stable Principal Component algorithm is implemented as follows (see Algorithm~\ref{alg:stablepc} for an overview). To determine if a point set is stretched, we use the corresponding eigenvalues $v_1$ and $v_2$ of the first and second principal components, respectively. If $v_2 / v_1 \leq \sigma$, then the point set is stretched, and otherwise it is not. For the time steps $t$ where the point set is stretched (including $t = 1$ and $t = T$), we simply compute the principal component as projection vector $\textsc{pv}[t]$. Note that $-\textsc{pv}[t]$ is equally good as projection vector, but results in a mirrored representation. To avoid flipping, we therefore use the direction ($\textsc{pv}[t]$ or $-\textsc{pv}[t]$) that is most consistent with $\textsc{pv}[t-1]$ (computed using the dot product). For time steps $t$ where the point set is not stretched we also first compute the (consistent) first principal component. We use these vectors to keep track of the signed angle $\alpha$ describing how the orientation of the first principal component has changed since the last time $t'$ the point set was stretched (or $t' = 1$). Once we reach another time $t''$ where the point set is stretched (or $t'' = T$), we can linearly interpolate the orientation of the projection vector for all times $t$ with $t' < t < t''$. Although linear interpolation of orientations is not unique in general, we can use the accumulated signed angle $\alpha$ to uniquely interpolate the projection vector. Finally, we can project the point sets for all time steps onto the computed projection vectors $\textsc{pv}[t]$.

\begin{algorithm}[t]
  \caption{\textsc{StablePrincipalComponent}$(P,\sigma)$}\label{alg:stablepc}
  \begin{algorithmic}[1]
  \REQUIRE Point set $P$ over $T$ time steps, and $\sigma \in [0,1]$ 
  \ENSURE Visual summary $S$ for $P$
  
  \smallskip 
\STATE Set $\textsc{pv}[1]$ to the first principal component vector for $P(1)$
\STATE Set $t'$ to $1$ and $\alpha$ to $0$
\FOR{$t = 2$ to $T$}
    \STATE Set $\textsc{pv}[t]$ to the first principal component of $P(t)$ and compute eigenvalues $v_1,v_2$
    \STATE Add the signed angle between $\textsc{pv}[t]$ and $\textsc{pv}[t-1]$ to $\alpha$
    \IF{$v_2/v_1 \leq \sigma$ or $t = T$}
    \label{algline:criteria}
        \FOR{$t_s = t' + 1$ to $t - 1$}
            \STATE Set $\textsc{pv}[t_s]$ to $\textsc{pv}[t']$ rotated over $\alpha \cdot \frac{t_s - t'}{t - t'}$ \label{algline:interpolate}
        \ENDFOR
        \STATE Set $t'$ to $t$ and $\alpha$ to $0$ 
    \ENDIF
\ENDFOR
\FOR{$t = 1$ to $T$}
\STATE Define $S[t]$ by projecting $P(t)$ onto $\textsc{pv}[t]$
\ENDFOR
\RETURN $S$
\end{algorithmic}
\end{algorithm}

Since the eigenvalues and principal components of $n$ points in 2D can be computed in $O(n)$ time, it is easy to see that the entire algorithm runs in $O(nT)$ time.
The explicit trade-off between spatial quality and stability can be configured via parameter $\sigma$. If $\sigma$ is set to a value close to $1$, the focus of the algorithm is on spatial quality, and only when the point set is very ``round'', stability will be enforced; $\sigma=1$ eliminates interpolation and always uses the first principal component in every time step. 
However, if $\sigma$ is set closer to $0$, the focus will be on stability and even for moderately stretched point sets, linear interpolation can occur, thereby sacrificing spatial quality for stability; $\sigma = 0$ causes one interpolation, from the first principal component at $t = 0$ to the first principal component at $t = T$.

\begingroup
\setlength{\columnsep}{1.0em}
\begin{wrapfigure}{r}{0.2\linewidth}
  \includegraphics[width=\linewidth]{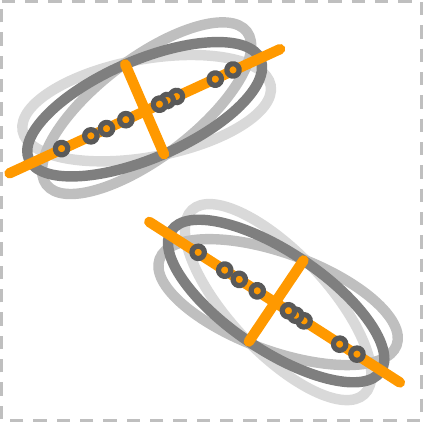}
\end{wrapfigure}
\mypar{[CPC$_\sigma$] Clustered Principal Component}
If a point set is strongly clustered, then we would expect an ordering of this point set with high quality to separate the different clusters. However, in the Stable Principal Component algorithm described above, two clusters may be interleaved if their projections happen to overlap. Therefore, we also propose the Clustered Principal Component algorithm, which is essentially a hybrid between SPC$_\sigma$ and a clustering algorithm (such hybrids have also been explored in~\cite{DBLP:journals/tvcg/WenskovitchCRHL18}). 

\endgroup


Intuitively, this algorithm performs SPC$_\sigma$ on the separate clusters. More specifically, for every frame we first perform Complete Linkage Clustering~\cite{gordon1987review} [CLC] on the point set, resulting in a hierarchical clustering. CLC is agglomerative and repeatedly merges the two clusters that are closest, where the distance between two clusters is determined by the farthest two points in different clusters. To obtain a partitioning of the points, we stop the process when the closest distance between clusters doubles with respect to the previous iteration. While this heuristic suffices to find salient clusters in our data sets, many other techniques exist to find a good partitioning in a hierarchical clustering~\cite{orford1976implementation}.  


Next, we perform SPC$_\sigma$ on the individual clusters, with two small adaptations, resulting in projection vectors $\textsc{pv}_C[t]$ for a cluster $C$. First, we end the linear interpolation of $\textsc{pv}_C[t]$ when the clustering changes and there is no longer a cluster with exactly the same points as $C$ (basically treating the time step as $t = T$). Second, it is no longer straightforward to determine the most consistent direction ($\textsc{pv}_C[t]$ or $-\textsc{pv}_C[t]$) for a cluster when the clustering changes. Here we use the projection vector used by the majority of the points in the cluster at time $t - 1$ to determine the most consistent direction.  


To find the global ordering at time $t$, we use the first principal component of the whole set $P$ to project the cluster centers. The orderings of points within a cluster are then placed around the projection of its cluster center. Although this approach may still result in overlap between two clusters in the projection, we can easily separate the clusters in the eventual ordering: first, we order the clusters according to their cluster centers, and then we order the points within a cluster according to their internal ordering. 

\section{Experimental Setup}\label{sec:setup}

We aim to quantitatively evaluate methods for computing 1D orderings, based on the resulting spatial quality and stability, to understand the trade-offs that are likely to exist between these methods, as well as to understand how our own parametrized methods make a trade-off between spatial quality and stability. Below, we briefly describe the methods we compare, the measures we use to compare them, and the data that is used in this evaluation.



\subsection{Algorithms}\label{sec:algorithms}

\begin{figure}[b]
 \centering 
 \includegraphics[width=1\linewidth]{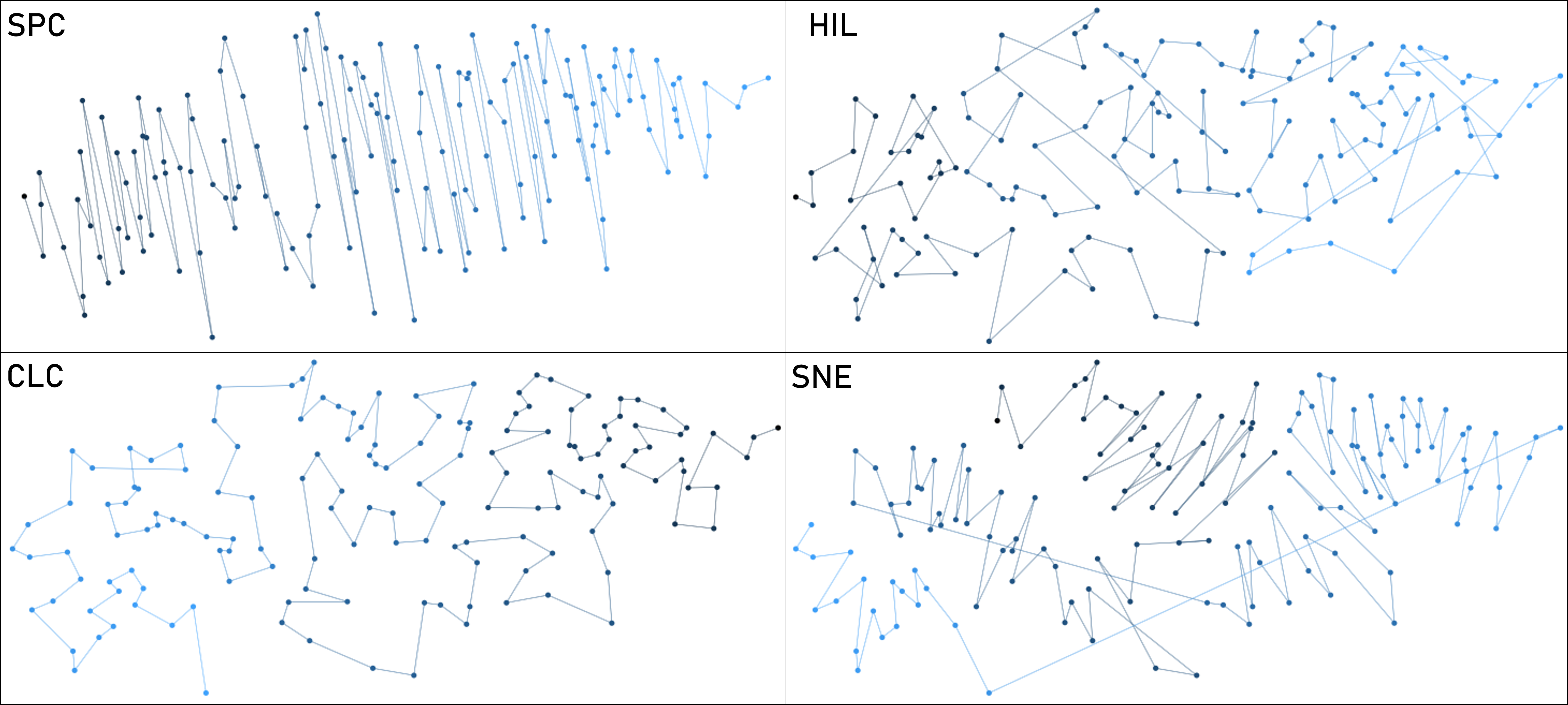}
 \caption{
 Visualization of orderings generated for one data frame using dimensionality reduction (SPC and SNE), space-filling curves (HIL) and clustering (CLC). Note the very different ways in which space is transformed into an ordering.}
 \label{fig:orderings}
\end{figure}

Previous work by Guo and Gahegan~\cite{guo2006spatial} and Buchm{\"u}ller~\etal~\cite{BJC+19} which compute 1D orderings for moving entities in 2D use either spatial subdivisions or clustering techniques. For our experiments, we chose the algorithms that performed best in their experiments. 
We also include a baseline algorithm [FXD] that is solely focused on stability.
Fig.~\ref{fig:orderings} shows examples of the orderings generated by some of the algorithms, including dimensionality-reduction techniques, for one time step of our test data. We give a short overview here; details can be found in Appendix~\ref{app:algorithms}.

\mypar{[FXD] Fixed order} 
Outputs the same arbitrary linear order for every time step; each horizontal line represents one moving entity.

\mypar{Spatial subdivisions}
Several well-known 1D ordering approaches used for spatial indexing, such as tree data structures and space-filling curves, are based on iterating through some spatial subdivision. 
Many variations exist; see \cite{DBLP:journals/debu/LuO93} for an overview.
Here we focus on four established, representative techniques from this area: \textbf{[HIL]} Hilbert curve~\cite{hilbert1891ueber}, \textbf{[ZOR]} Z-order curve, \textbf{[PQR]} Point Quadtree~\cite{DBLP:journals/acta/FinkelB74,DBLP:books/daglib/0032640} and \textbf{[RTR]} R-tree~\cite{DBLP:conf/sigmod/Guttman84}.

\mypar{Clustering}
Another approach
is to first compute a hierarchical clustering on the point set, and then order the points in such a way that clusters stay together. These algorithms are defined by how the points are clustered, and how the linear order is computed from the clustering. We use the aforementioned \textbf{[CLC]} Complete Linkage Clustering~\cite{gordon1987review} and \textbf{[SNN]} Shared Nearest Neighbors~\cite{Jarvis1973} to cluster points and derive an ordering from the cluster hierarchy as follows. 

The hierarchical clustering is represented by a tree with the individual points stored in the leaves. We aim to order the leaves of such a tree without changing the cluster structure, that is, by only changing the order of the children of any internal node.
We follow the algorithm by Bar-Joseph~\etal~\cite{Bar-Joseph2003} to efficiently compute the optimal order that minimizes the length of the path formed by visiting the input points in that order. 

\mypar{Dimensionality reduction}
We also consider dimensionality reduction techniques to compute 1D ordering. In our experiments we specifically consider \textbf{[SAM]} Sammon mapping~\cite{Sammon1969} and \textbf{[SNE]} t-SNE~\cite{tsne2008}, next to the PCA-based techniques described in Section~\ref{sec:stablePCA}.

Both Sammon mapping and t-SNE use gradient descent to minimize a cost function. Normally, this gradient descent is started with a random initial solution, but this may result in poor stability over time. To improve the stability of both algorithms, we initialize them with the solution of the previous time step, resulting in two stable versions, \textbf{[SAMp]} and \textbf{[SNEp]}. The rationale is that, if the local minimum found in the previous time step still exists, but has slightly shifted, then this approach will likely find this local minimum again rather than any other local minimum.


\subsection{Metrics}\label{sec:metrics}

In this section we discuss the quality metrics we use to capture spatial quality and stability. We choose measures that focus on the preservation of local neighborhoods. For applications where other types of measures are preferred, we refer to the survey in \cite{espadoto2019towards}.

\subsubsection{Spatial Quality}


Spatial quality measures the correspondence between $P(t)$ and the linear order $S_t$. We capture this by considering the local neighborhood of a point, as characterized by its nearest neighbors.
One way to measure changes in local neighborhoods is using an evaluation of dimensionality reduction via persistent homology as introduced by Rieck and Leitte \cite{DBLP:journals/cgf/RieckL15}. However, we choose not to use this type of measure. While this approach is more recent than the measure we are using, it does not compare to older results, it is more complex, and most importantly it does not directly relate input to output, but measures through an intermediate topological structure. Hence, we use the \emph{Keys Similarity} measures as described by Guo and Gahegan \cite{guo2006spatial} to directly measure the changes in nearest neighbors. 

To simplify notation, we omit dependencies on time step $t$, as the metrics consider each time step in isolation.
Thus, $P$ denotes a point set in the plane, and $S$ denotes a linear order.
Let $n(i,j) \in P$ denote the $j^{th}$ nearest neighbor of $p_i$ in $P$, for each $j$ with $1 \leq j \leq k$ for some constant $k$. 
We use $r(i,j)$ to denote the neighbor rank in $S$ between $p_i$ and $n(i,j)$. However, the difference in rank $|S(n(i,j)) - S(p_i)|$ is not unique. There are two neighbors at rank difference 1, two at rank difference 2, until we reach one end of a linear order. To avoid arbitrariness, we do not break ties but rather consider each pair with the same rank difference to have the same value for $r(i,j)$. Thus, there are two nodes with $r(i,j) = 1$ (rank difference 1), two nodes with $r(i,j) = 3$ (rank difference 2), etc.

Generally, Keys Similarity at time $t$ is then defined as 
\[ KS(P, S) = \frac{\sum_{p_i \in P} \sum^k_{j=1} w(i,j) \cdot r(i,j)}{\sum_{p_i \in P} \sum^k_{j=1} w(i,j)}, \]
where $w(i,j)$ denotes the weight or importance of maintaining the $j^{th}$ nearest neighbor of $p_i$ at time $t$ -- note that these weights need not be the same at every time step.
We use two variants of Keys Similarity, see Appendix~\ref{app:metrics} for the exact formulas.

\mypar{[KSra] Rank-weighted Keys Similarity}
We define $w(i,j) = 1/j$ inversely proportional to the rank; hence maintaining the closest neighbors is more important than maintaining the distant neighbors.



\mypar{[KSdi] Distance-weighted Keys Similarity}
We define $w(i,j) = 1/\| p_i - n(i,j) \|$ inversely proportional to the Euclidean distance, such that maintaining close neighbors is more important than maintaining distant neighbors. In contrast to KSra, this variant does not treat neighbors at (nearly) identical distances differently.


\mypar{Other facets}
Our metrics focus on combinatorial aspects of the position of the point objects. 
Spatial structure in general knows many other facets, such as distances and directions between points, as well as density. For projections into a single dimension, distances and density can factor into spatial quality. However, a linear order inherently does not lend itself to represent such concepts. 


\subsubsection{Stability}

Stability or temporal coherence measures the similarity between consecutive orders in $S$. In our evaluation, we use the following three measures for stability. The first two are based on absolute changes in the order and match the measures used by Buchm\"uller~\etal~\cite{BJC+19} to evaluate MotionRugs. The latter uses neighborhoods, based on the concepts by Guo and Gahegan \cite{guo2006spatial}. 

We aim to compare the similarity between two linear orders, $S_t$ and $S_{t+1}$ for each $t$ with $1 \leq t < T$. We could easily use the same metrics to compare nonconsecutive orders, but this provides little insight for such inherently sequential data. To consider the stability over a temporal range $[t,t']$, we use standard summary statistics (e.g., average, minimum, or maximum) over all consecutive pairs.

\mypar{[JMP] Jump distance}
We quantify the jump distance for a single point object $p_i$ as the difference between its ranks in the two orders, that is, $|S_t(p_i) - S_{t+1}(p_i)|$. The jump distance between two orders is then the sum over all jump distances for each point object.
\[ \mathit{JMP}_t(P, S) = \sum_{p_i \in P} |S_t(p_i) - S_{t+1}(p_i)| \]
The value for $\mathit{JMP}_t(P,S)$ lies between $0$ (perfectly stable) and $\frac{n(n-1)}{2}$ (complete inversion of the order).

\mypar{[CRS] Crossings}
Whereas JMP penalizes any change in the order, many points moving up together may not constitute much change. Instead we may count the number of inversions or crossings in the order, that is, the pairs $p_i$, $p_j$ for which $S_t(p_i) < S_t(p_j)$ and $S_t(p_i) > S_t(p_j)$.
The metric $\mathit{CRS}_t(P,S)$ lies between $0$ (perfectly stable) and $\frac{n(n-1)}{2}$ (complete inversion of the order).

Buchm\"uller et al. \cite{BJC+19} also use Kendall's $\tau$ coefficient to evaluate stability. We choose to omit this, as it is equivalent to $1-2\cdot \mathit{CRS}_t(P,S)/(n(n-1)/2)$. That is, Kendall's $\tau$ is the same as CRS up to normalization to the range $[-1,1]$.

\mypar{[KSte] Temporal Keys Similarity}
We may also take the same approach as for spatial similarity and consider the similarity of local neighborhoods in both orders. As distances are not inherently meaningful in the combinatorial order and simply correspond to ranking differences, we use only the rank-weighted version of Keys Similarity. Also for this metric $\mathit{KSte}_t(P,S)$, we do not break ties in either order, but rather give them the same rank.

\subsection{Data}\label{sec:data}

For comparability, we use the same data as MotionRugs~\cite{BJC+19} along with two synthetic data sets, one generated using Netlogo~\cite{netlogo} and another generated with the well known Reynolds model~\cite{reynolds}. Note that we use MotionRugs solely to visualize the output of our algorithms. The algorithms as well as the quantitative evaluation are independent of the visualization that uses the resulting ordering.

The first data set tracks 151 fish of the Notemigonus crysoleucas species (Golden shiner). Golden shiner fish live in large groups called ``shoals'', moving in coordination at almost any given time. 
%
The 151 fish were tracked optically while moving through a 2.1m by 1.2m shallow water tank, thus avoiding movement in the third dimension. The tank did not feature any obstacles or hindrances besides the side walls. Different movement patterns can be observed in the data, which allows us to test quality in different situations. Among these patterns are uniform group movements, partial and complete changes of direction, circular movement patterns, splitting off in separate clusters, and changes in group density, speed, and acceleration. This data set is quite large, hence we use two excerpts of 2000 frames of movement, which were recorded at a rate of 25 frames per second, each resulting in 80 seconds of available collective movement data. For each frame, the spatial coordinates of each fish are recorded in a Cartesian coordinate system. Fig.~\ref{fig:fishdatalarge} in the appendix is a visual summary of the full data set.

\mypar{Fish 1} In the first excerpt, the fish first move around the boundary of the tank and finally enter a so-called milling formation, moving in a circular shape. The fish always form a single cluster.

\mypar{Fish 2} In contrast, the second excerpt shows the fish splitting in separate clusters, as can be seen in Fig.~\ref{fig:teaser}A.

\smallskip
\noindent
In addition to analyzing Fish 1 in full, we also zoom in on a small part of the movement of the fish, which
triggers so-called ``phantom splits''~\cite{BJC+19} for certain ordering methods, most notably HIL, PQR, and SNEp (see Fig.~\ref{fig:strategycomparison}). The shoal of fish appears to split, but this is purely an artifact of the method and not reflecting the data.

\mypar{Netlogo} The Netlogo data set is generated using the Flocking model~\cite{netlogoflocking} from the openly available Models Library within the Netlogo application. Minimal adaptations were made to the model to ensure the boundaries of the canvas do not wrap around, and the trajectories of the moving entities could be extracted easily.

\mypar{Reynolds} The final data set, which we use to demonstrate clustered movement, is generated by an adaptation from the well known Reynolds model~\cite{reynolds}, where between the movers of the three visible clusters only repulsion forces apply, but no attraction, keeping the clusters separate. The generator code by Piljek~\cite{piljek} is public. 

Since the results are similar across all data sets, we mainly focus on Fish 1 and Reynolds, while highlighting notable difference in the results of the other data sets. A full analysis for the remaining data sets is given in Appendices~\ref{app:fish2}--\ref{sec:reynolds}.

\section{Experimental Results}\label{sec:experiments}

\addtocounter{figure}{2}

\begin{figure*}[b]
    \centering
    \includegraphics[width=\linewidth]{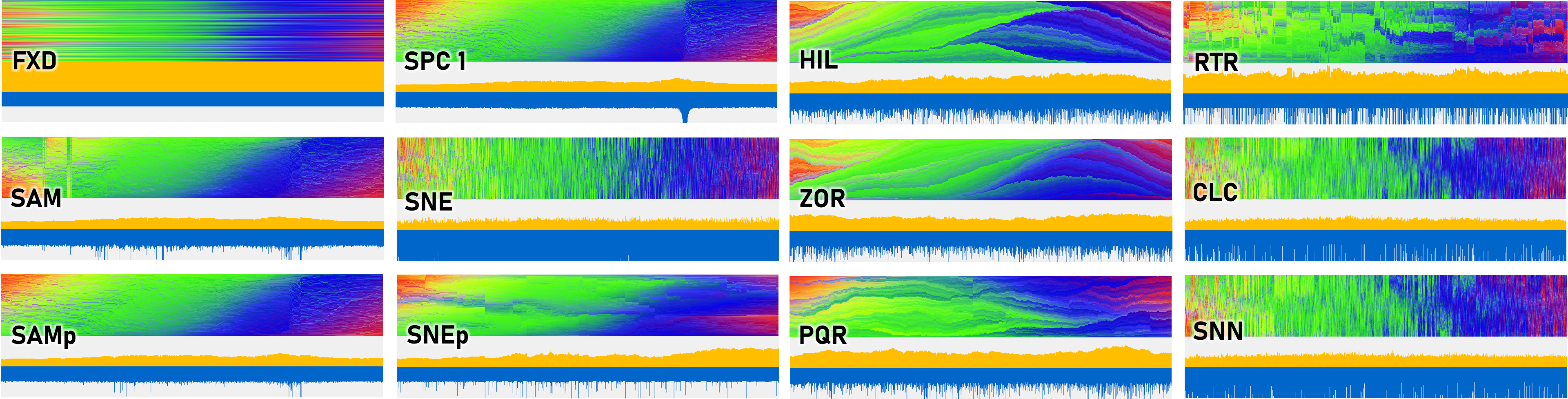}
    \caption{Visual summaries for Fish 1, focused on so called phantom splits for SNEp, HIL, ZOR, PQR and RTR. Below each we show KSdi (yellow) and KSte (blue), capped at 37.5 and 6.25, respectively. More gradual color changes in the MotionRug relate to better quality.
    }
    \label{fig:strategycomparison}
\end{figure*}


Here we report on the results of the quantitative experiments described in Section~\ref{sec:setup}. 
Tables~\ref{tab:experiments}--\ref{tab:experiments4} in Appendix~\ref{app:summary-stats} provide summary statistics over all time steps and for each metric, for all data sets.
We first determine the most effective measures for spatial quality and stability. We then explore how the parameter for SPC effects a trade-off between spatial quality and stability. Finally we briefly discuss the computational efficiency.

\addtocounter{figure}{-3}

\begin{figure}[t]
    \centering
    \includegraphics[width=\linewidth]{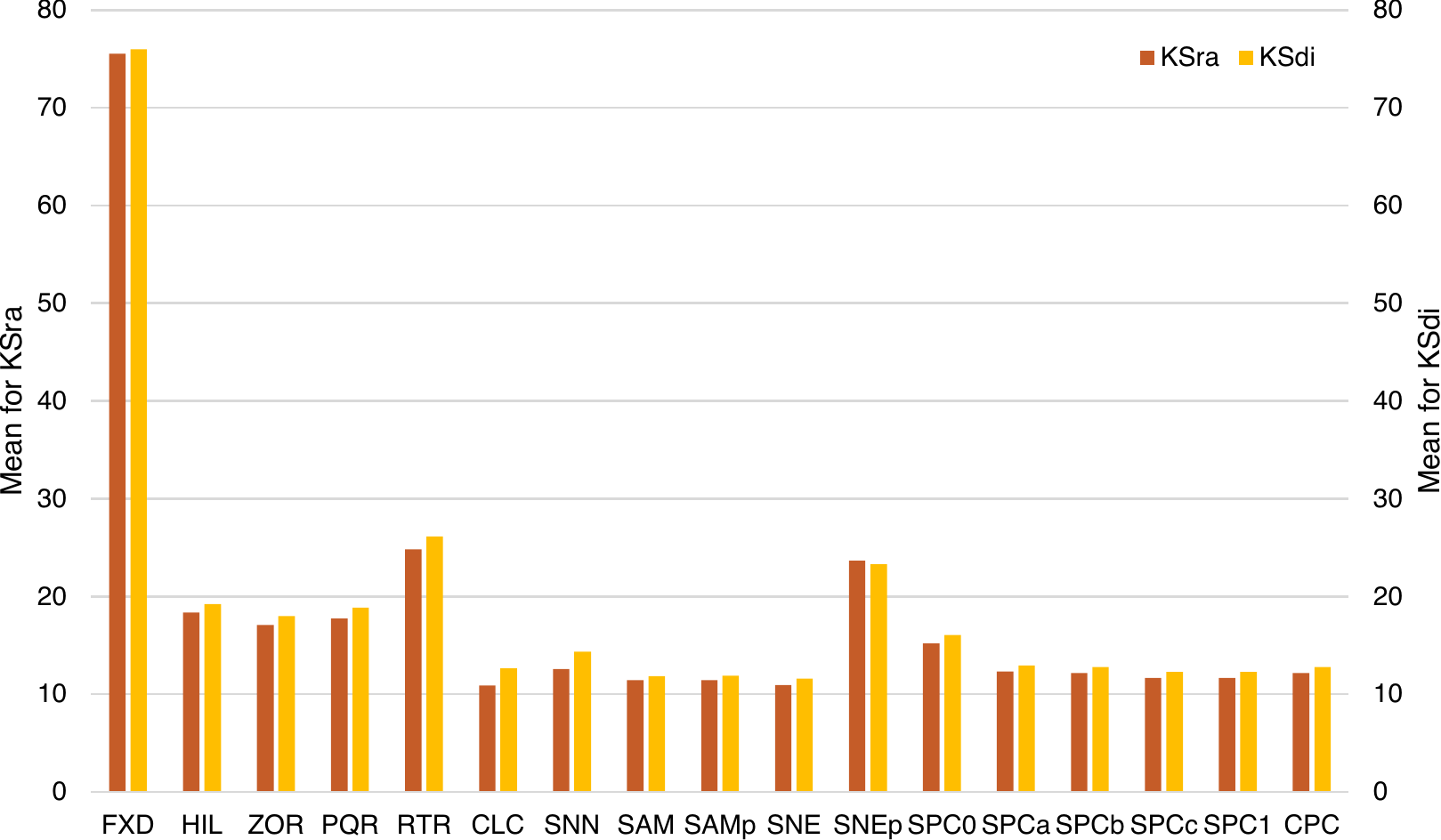}
    \caption{Spatial-quality metrics: mean KSra (left) and KSdi (right) for all algorithms over all frames of Fish 1. Lower numbers indicate better overall spatial quality of the ordering.}
    \label{fig:spatialquality-chart}
\end{figure}

Our experiments explore how these methods and quality criteria relate. To illustrate the resulting orderings, we use MotionRugs using a 2D RGB colormap as introduced in \cite{buchmuller2020spatial}. As can be seen in Fig.~\ref{fig:teaser}, in the orderings (B) and (C) the data points are colored according to their 2D position in (A). In visual summaries of high spatial quality, data points that are close in 2D should be close in 1D, hence similar colors should end up close to each other. Furthermore, in stable summaries the neighborhoods do not change much in the orderings, and hence the colors should smoothly change over time. Thus, this graphical representation of the orderings allows us to also visually assess spatial quality and stability.

\subsection{Quality Results -- Fish 1}
Fig.~\ref{fig:strategycomparison}~and~\ref{fig:spcoverview} show visual summaries for all algorithms for Fish 1. The MotionRugs are accompanied by a visualization of the mean KSdi and KSte values for each frame,
cut off slightly above the mean values of most algorithms. This ensures that the differences between the average behavior of the algorithms becomes visible at a glance.
Below, we first discuss spatial quality and stability statistics separately, along with a discussion on phantom splits. We follow up with an exploration into the effects of the parameter value
on the outcome of SPC and finally consider the trade-off between spatial quality and stability for all methods.

\mypar{Spatial quality}
Fig.~\ref{fig:spatialquality-chart} compares the spatial-quality measures KSra and KSdi, as measured on all algorithms for Fish 1. For both measures lower values indicate higher spatial quality. Overall, we see that the KSra measurements are slightly lower for all algorithms, except SNEp where KSdi has a minimal edge over KSra. 
As expected FXD achieves the worst spatial quality. 
Furthermore, SNEp and the algorithms using spatial subdivisions are outperformed by the clustering algorithms, and other dimensionality-reduction techniques. 
Comparing the spatial quality of SPC and CPC to the algorithms that perform best on spatial quality, we see that SPC and CPC both achieve comparable spatial quality. 
The choices for parameter $\sigma$ of SPC on Fish 1 are 0, 1, and variables $a=0.35, b=0.53, c=0.78$, while for CPC we chose $\sigma = 0.53$. The choice for the intermediate values $a,b$ and $c$ is different for the various data sets and will be justified in the parameter exploration. For CPC we choose a parameter value that according to the parameter experiment performs well on both spatial quality and stability, again different for every data set.
Due to the strong correlation of both spatial quality measures, we focus on only KSdi in the remainder.

\begin{figure}[t]
    \includegraphics[width=\linewidth]{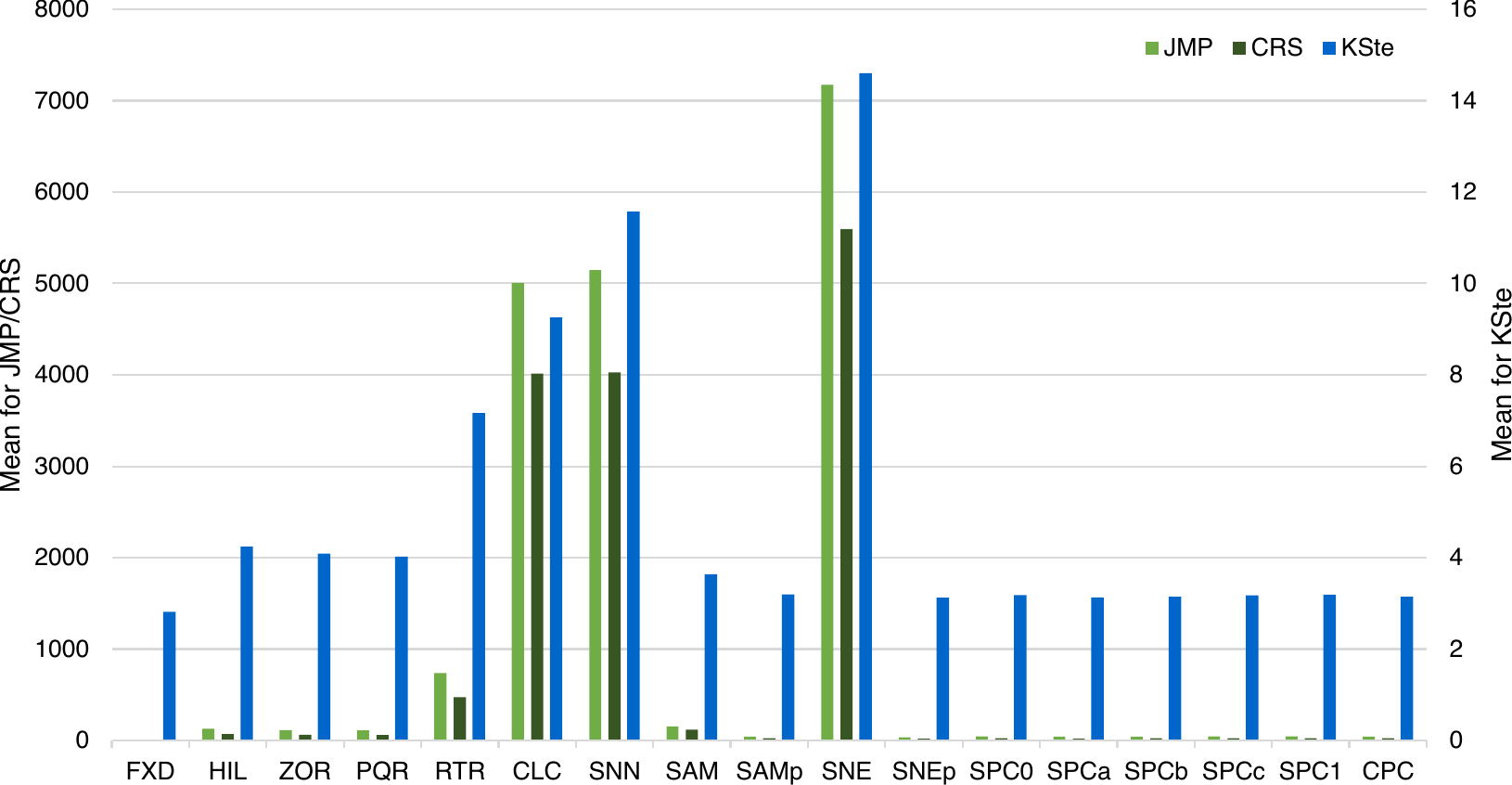}
    \caption{Stability metrics: mean JMP, CRS (left axis), and KSte (right axis) for all methods over all frames of Fish 1. Lower numbers indicate better overall ordering stability.}
    \label{fig:stability-chart}
\end{figure}

\mypar{Stability}
Fig.~\ref{fig:stability-chart} compares the stability measures: JMP, CRS and KSte. While JMP and CRS measure absolute changes between orders, KSte captures changes in local neighborhoods. For each measure lower values indicate higher stability. We see that CRS results in lower values than JMP, which is expected: two entities can jump to different positions in the next frame without crossing, but they cannot cross each other without jumping. We do see some differences between the two data sets, as opposed to the results for spatial quality. 
For FXD the result is again obvious: all measures are at their minimum. 
While JMP and CRS are generally low, CLC, SNN and SNE show very high numbers. Those three algorithms also perform worst according to the KSte metric. 
Another outlier that performs poorly on KSte is RTR, which also performs comparatively poorly on JMP and CRS. Of the remaining algorithms, the spatial subdivisions perform worst on KSte. The SAM, SAMp and SNEp algorithms, the SPC variants and CPC show similar and very low mean values of KSte.
Again, we observe a strong correlation between the metrics, and thus consider only KSte in the remainder.

\addtocounter{figure}{1}

\begin{figure*}[t]
    \centering
    \includegraphics[width=.95\linewidth]{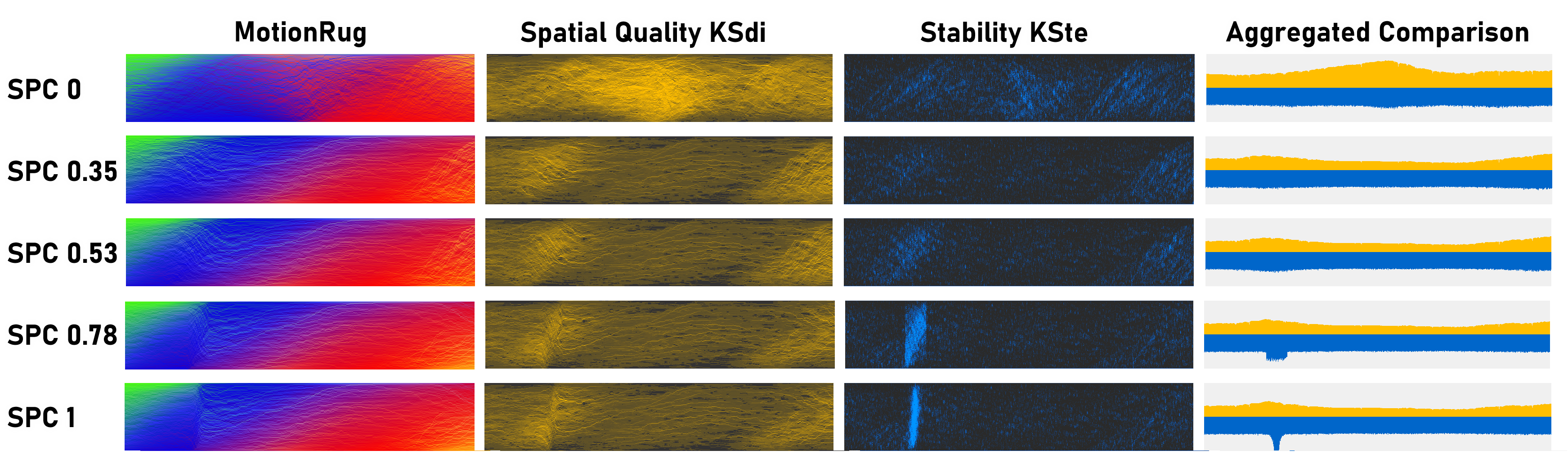}
    \caption{Visual summaries for Fish 1, focused on instabilities to show how $\sigma$ affects interpolation in SPC. The summaries in the second and third columns show how each data point adds to the KSdi (yellow) and KSte (blue) measures. Brighter colors indicate worse spatial quality and stability. The measures are aggregated per frame in bar charts on the right.}
    \label{fig:spcoverview}
\end{figure*}

\mypar{Phantom splits}
In Fig.~\ref{fig:strategycomparison} each frame of movement data is represented by a column of pixels, where each fish corresponds to a pixel. The pixels are colored according to the position of the fish in 2D, as shown in Fig.~\ref{fig:teaser}.
The ordering method clearly defines the resulting visual patterns. 
An anomaly can be identified in the subdivision methods (HIL, ZOR, PQR, RTR) and SNEp, the so called phantom splits \cite{BJC+19}. These visual summaries suggest that the shoal of fish somehow splits, but this is not the case. Such patterns are hence undesirable, as they convey false information. Other algorithms do not seem to be prone to these kind of visual artifacts or generally produce visual results too fuzzy for such patterns to appear.
Some algorithms, such as CLC and SNE, result in cluttered visuals despite having good spatial quality. This clutter is caused by instability: the summaries fail to convey patterns over time despite individual frames being objectively good.

\begin{figure}[b]
\centering
\begin{minipage}{0.45\linewidth}
    \centering
    \includegraphics[width=\linewidth]{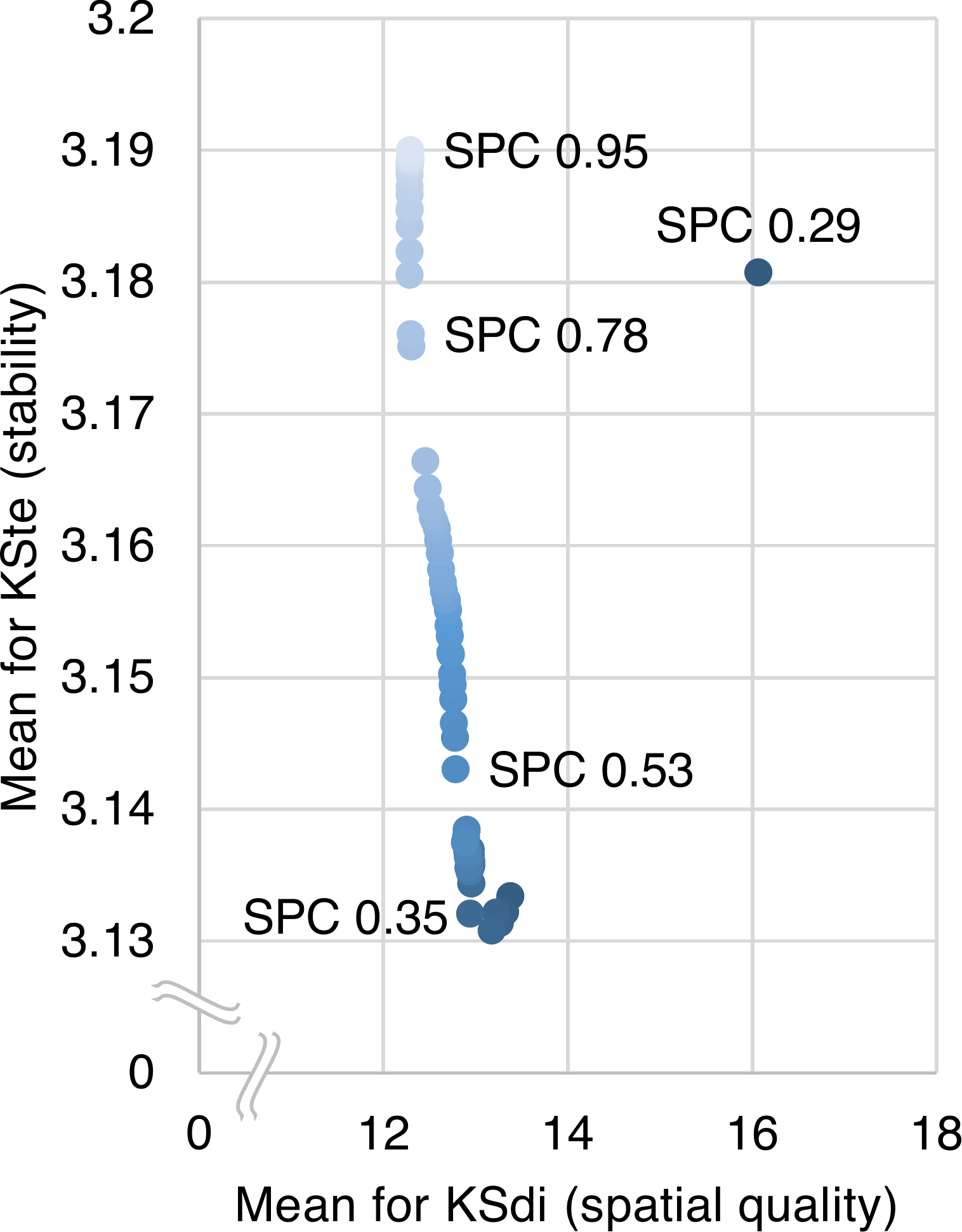}
\end{minipage}
\hfill
\begin{minipage}{0.45\linewidth}
    \centering
    \includegraphics[width=\linewidth]{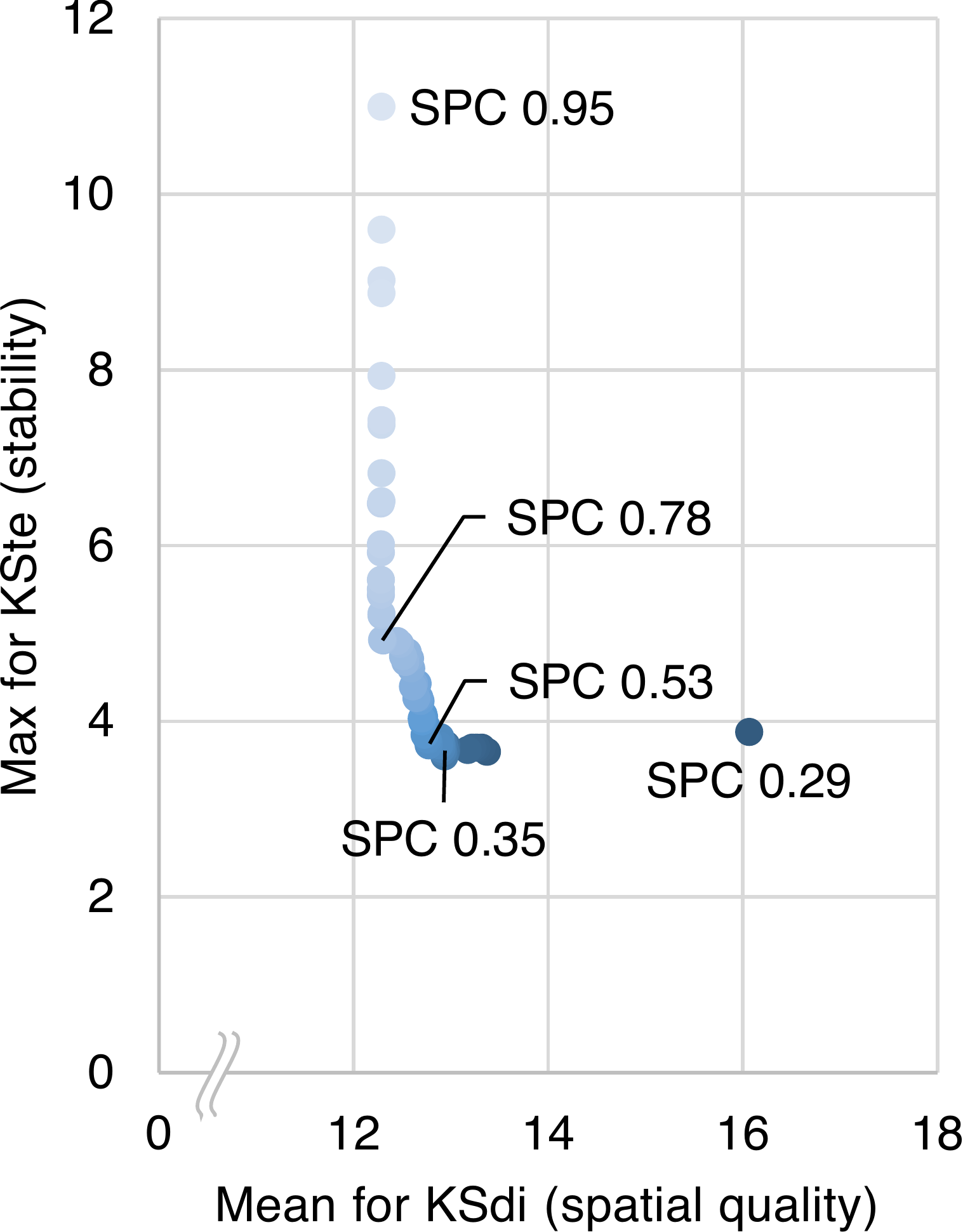}
\end{minipage}
\caption{Comparing the mean and mean (left) as well as max and mean (right) for KSte (stability) and for KSdi (spatial quality), for uniformly distributed parameter settings of SPC on Fish~1.}
\label{fig:parameter}
\end{figure}

\mypar{SPC parameter}
We now investigate the parameter $\sigma$ of SPC and its effect on the results.
We run SPC for $101$ different values for $\sigma$ from $0$ to $1$ with increments of $0.01$. 
As discussed before, we use KSdi to measure the spatial quality of the visual summaries, and specifically we use the mean over all frames. For stability we use the mean as well as the max KSte to quantify stability. As we saw before, mean KSte captures cohesion over time, while max KSte should be low to prevent visual artifacts from disrupting temporal patterns.
The results for Fish 1 are shown in Fig.~\ref{fig:spcoverview} and~\ref{fig:parameter}. 
Note that the highest plotted value of $\sigma$ is $0.95$, while the lowest is $0.29$. Values above and below these extremes are identical to results with $0.95$ and $0.29$ respectively. 
The $\sigma$ values indicated by labels in the figures are chosen as representatives, and used in our other experiments.

\begin{figure*}[t]
    \begin{minipage}[b]{\columnwidth}
    \centering
    \includegraphics[width=\textwidth]{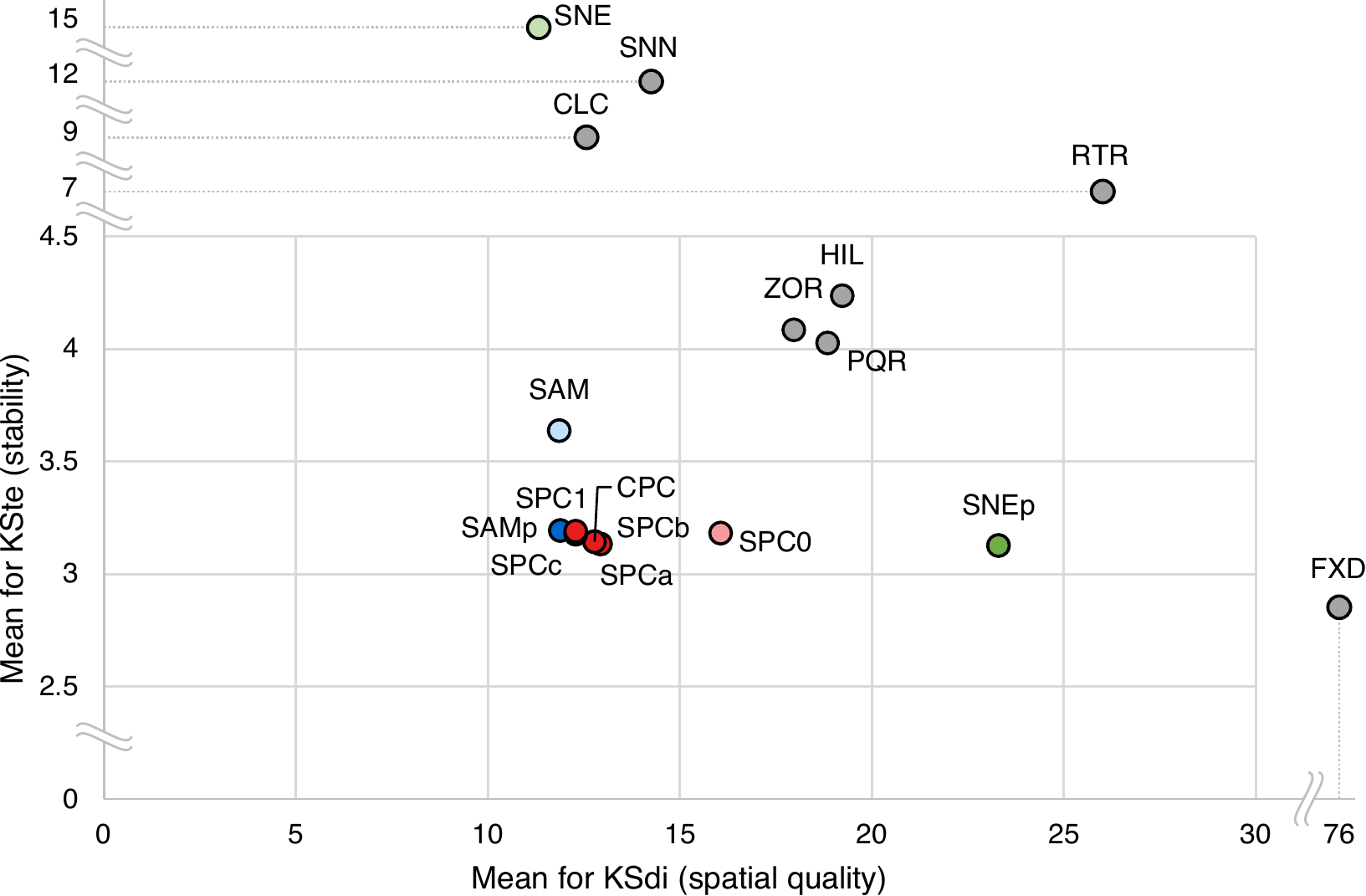}
    \caption{Comparing the mean for KSte (stability) and for KSdi (spatial quality) for all algorithms on Fish 1.}
    \label{fig:spatialqualityVSstability-chart}
    \end{minipage}
    \hfill
    \begin{minipage}[b]{\columnwidth}
    \centering
    \includegraphics[width=\textwidth]{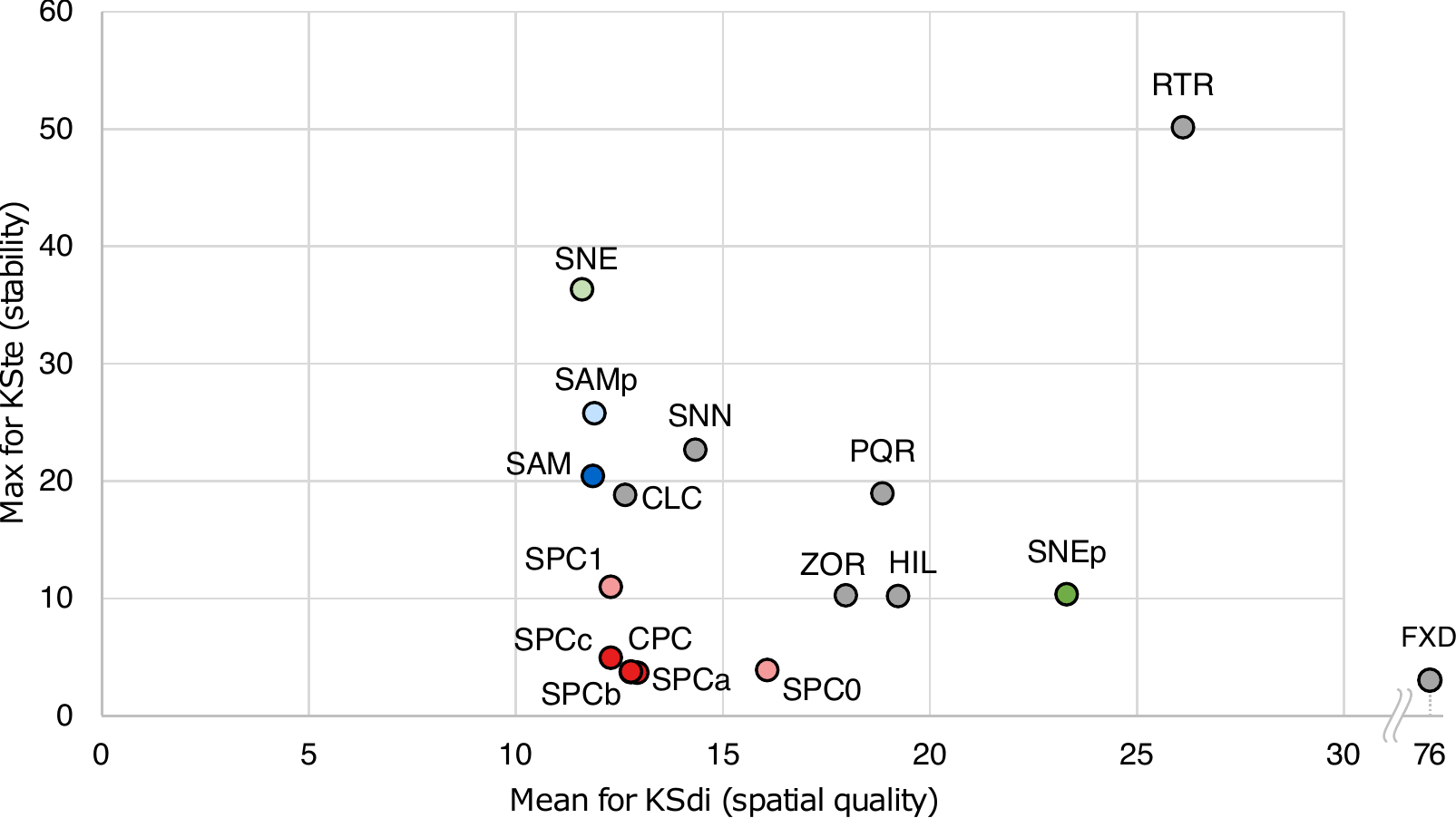}
    \caption{Comparing between the max for KSte (stability) and the mean for KSdi (spatial quality) for all algorithms on Fish 1.}
    \label{fig:spatialqualityVSmaxstability-chart}
    \end{minipage}
\end{figure*}

Overall, we see an inverse relation between stability and spatial quality. Values of $\sigma$ closer to $1$ result in better spatial quality, while values closer to $0$ sacrifice some spatial quality for more stability. 
This is to be expected, as SPC$_{1}$ always projects the fish to the first principal component; this will likely lead to the best spatial quality that can be achieved for any parameter value.

As $\sigma$ is decreased, SPC increasingly uses interpolated lines for projection instead.
This interpolation smooths changes in angle of the line, but the projection reflects spatial relations less accurately as a result. 
When $\sigma$ drops below $0.30$, the interpolation happens purely between the first and last frame of the data set. 
Contrary to expectation, this negatively affects both spatial quality and stability: the first principal component rotates both clockwise and counterclockwise at varying speeds, not matching the uniform interpolation over such a long time period; as a result, the interpolated lines do not correspond at all to the first principal components, neither in angles nor in rotation direction. 
This mismatch in angles leads to poor spatial quality per frame, while the mismatch in rotation direction also decreases stability.


Finally, we explicitly show the effects of changing $\sigma$ on the resulting visual summaries using Fig.~\ref{fig:spcoverview}. In this figure, we visualize spatial quality in yellow and stability in blue. On the left we show how much every point contributes to the measures, with brighter colors indicating worse spatial quality/stability, while darker colors show placements in the ordering of high spatial quality or stability. On the right the aggregated values over all points in the data are visualized in a histogram. Fig.~\ref{fig:spcoverview} specifically shows an instability that occurs in Fish 1, using the same intermediate values for $\sigma$ as before. When the first principal component is used without introducing stability ($\sigma = 1$), we see a short burst of instability, along with slightly elevated measurements in spatial quality. As $\sigma$ decreases and more stability is introduced by interpolating the direction of the first principal component over larger time frames, we see that the instability is distributed over more frames and decreases. The spatial quality is not negatively impacted by this, until $\sigma$ becomes too low: when we interpolate over too many frames at once, spatial quality will drastically deteriorate, as seen for $\sigma = 0$.

\mypar{Trade-offs}
Our main goal is to investigate the trade-off between spatial quality and stability. Fig.~\ref{fig:spatialqualityVSstability-chart} shows a scatterplot on the means of KSdi and KSte of all algorithms. Since lower values indicate better quality for both, methods in the bottom-left corner perform well on both aspects. In both figures SPC variants and CPC are colored in shades of red, SAM variants in blue and SNE variants in green. The ``best'' variants have fully opaque colors, while unstable variants or those of worse spatial quality have a lighter shade.

The results for Fish 1 clearly show that methods based on spatial subdivisions (ZOR, HIL) and space-filling curves (PQR, RTR), albeit fast to compute, perform poorly on spatial quality and stability. 
The clustering methods (CLC and SNN) as well as SNE, on the other hand, perform well on spatial quality, but exhibit very poor stability.

The fixed order (FXD) and SNEp are on the other extreme, having good stability, but very poor spatial quality. Furthermore, the strong influence of initialization for t-SNE stands out. When initialized with random coordinates (SNE), the spatial quality is very good, but the stability is extremely poor. On the other hand, initializing t-SNE with the embedding of the previous time step (SNEp) greatly improves stability, but spatial quality suffers greatly.

That leaves SAM, SAMp, SPC variants, and CPC, which perform well on both aspects. We note that SAM and SAMp perform very similarly on KSdi (difference of $0.03$), but SAMp performs significantly better in terms of stability. 
SPC variants also strike a good balance between spatial quality and stability. All SPC variants have slightly worse spatial quality than SAM variants, but improve stability. However, recall that SPC is significantly faster to compute than the Sammon mapping algorithms SAM and SAMp. Finally, CPC performs similarly to SPC, which is expected since the fish stay grouped in a single cluster, hence CPC  and SPC have very similar outcomes on this data set.

It is also important for stability to be consistently low, to avoid visual artifacts and ensure that visual patterns in the summary point to actual movement patterns. As such, bursts of high instability are undesirable. We hence also consider the maximum value of KSte; see Fig.~\ref{fig:spatialqualityVSmaxstability-chart} for a scatterplot.
The overall composition remains similar, but differences in stability are highlighted. Note that SAM, SAMp and SNEp are deteriorating with respect to other methods; we can also see clear bursts of instability in Fig.~\ref{fig:strategycomparison} for these methods.

Interestingly, SAMp performs worse here on stability than SAM, unlike for other data sets. This shows that, although initializing the gradient descent with the solution of the previous time step generally improves stability, there is no guarantee that it will always do so; there may be outliers, as is the case here.


Fig.~\ref{fig:spatialqualityVSmaxstability-chart} also highlights stability differences between SPC variants. SPC$_1$ always uses the first principal component, which can behave erratically for round point sets, decreasing stability. The other variants of SPC overcome this problem by interpolating over these bursts of instability. Indeed, SPC is largely unaffected for lower parameter values, having the smallest standard deviation overall (see Table~\ref{tab:experiments} in Appendix~\ref{app:summary-stats}.

\subsection{Quality results -- Reynolds}
In this section we present the results of our experimental evaluation for the Reynolds data set. For spatial quality and stability we only sketch our results. A more elaborate analysis, including figures, can be found in Appendix~\ref{sec:reynolds}.

\mypar{Spatial quality \& stability}
Overall, the relative performance of most algorithms on both spatial quality and stability is similar for Reynolds and Fish 1. However, on spatial quality measures clustering techniques and CPC perform better relative to the other algorithms, making CPC one of the best algorithms to achieve spatial quality. Since there are multiple cluster in this data, which do not interact with each other, it is not surprising that these techniques perform so well. In terms of stability, lower values are also measured for the absolute changes for clustering techniques, and CPC performs relatively better than on other data sets. Especially when considering the max values for KSte we see that the clustering techniques and SNE perform better than on the other data sets. The clustering technique CLC even beats some spatial subdivision techniques (HIL and ZOR) on this measure. All of these results are to be expected: since the clusters do not interact and contain the same points in every frame, the clustering techniques also perform very well on stability. Finally, SNE is less of an outlier for this data set, as SNN also performs worse on the mean KSte.

\begin{figure}[b]
\centering
\begin{minipage}{0.45\linewidth}
    \centering
    \includegraphics[width=\linewidth]{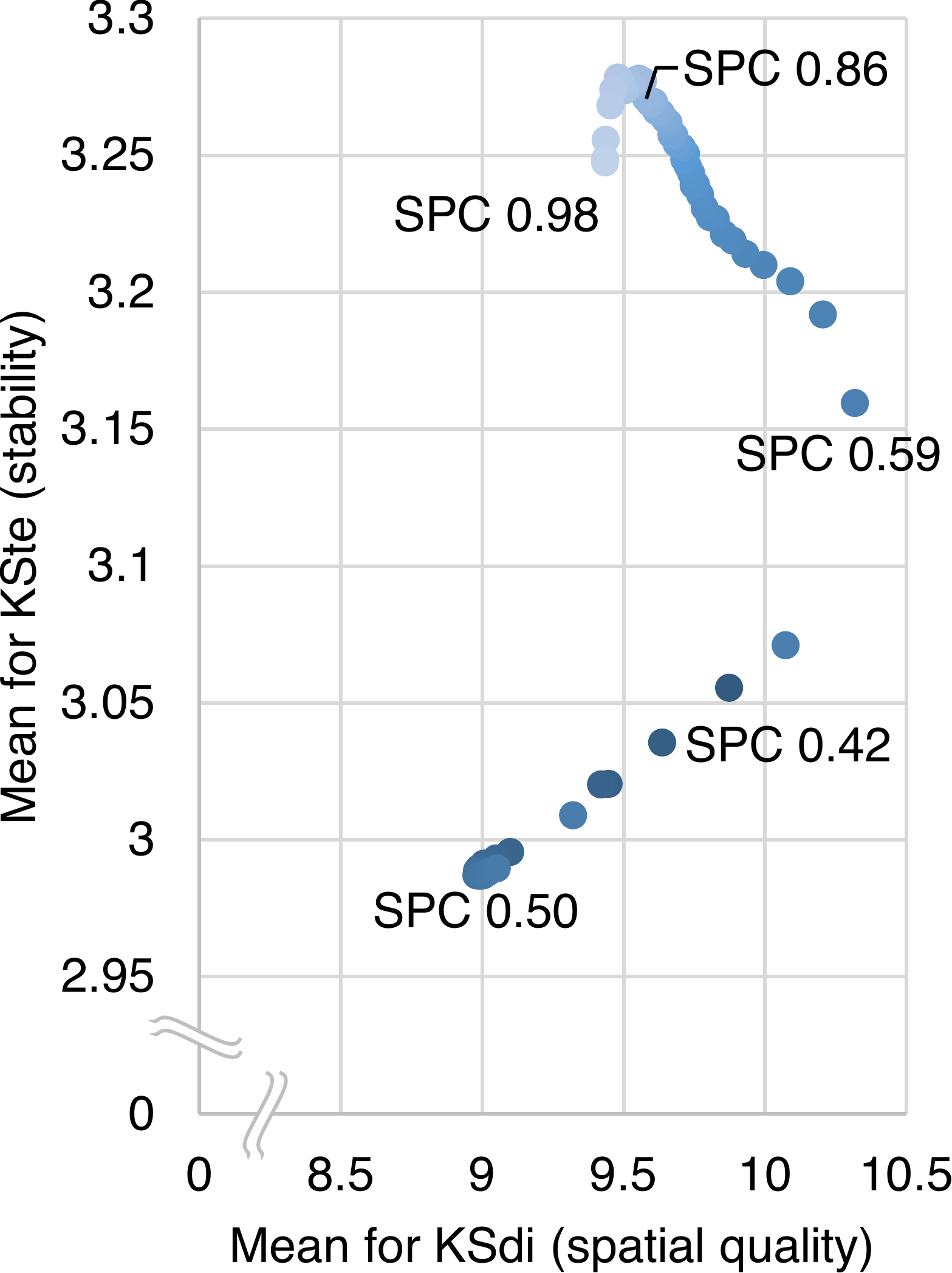}
\end{minipage}
\hfill
\begin{minipage}{0.45\linewidth}
    \centering
    \includegraphics[width=\linewidth]{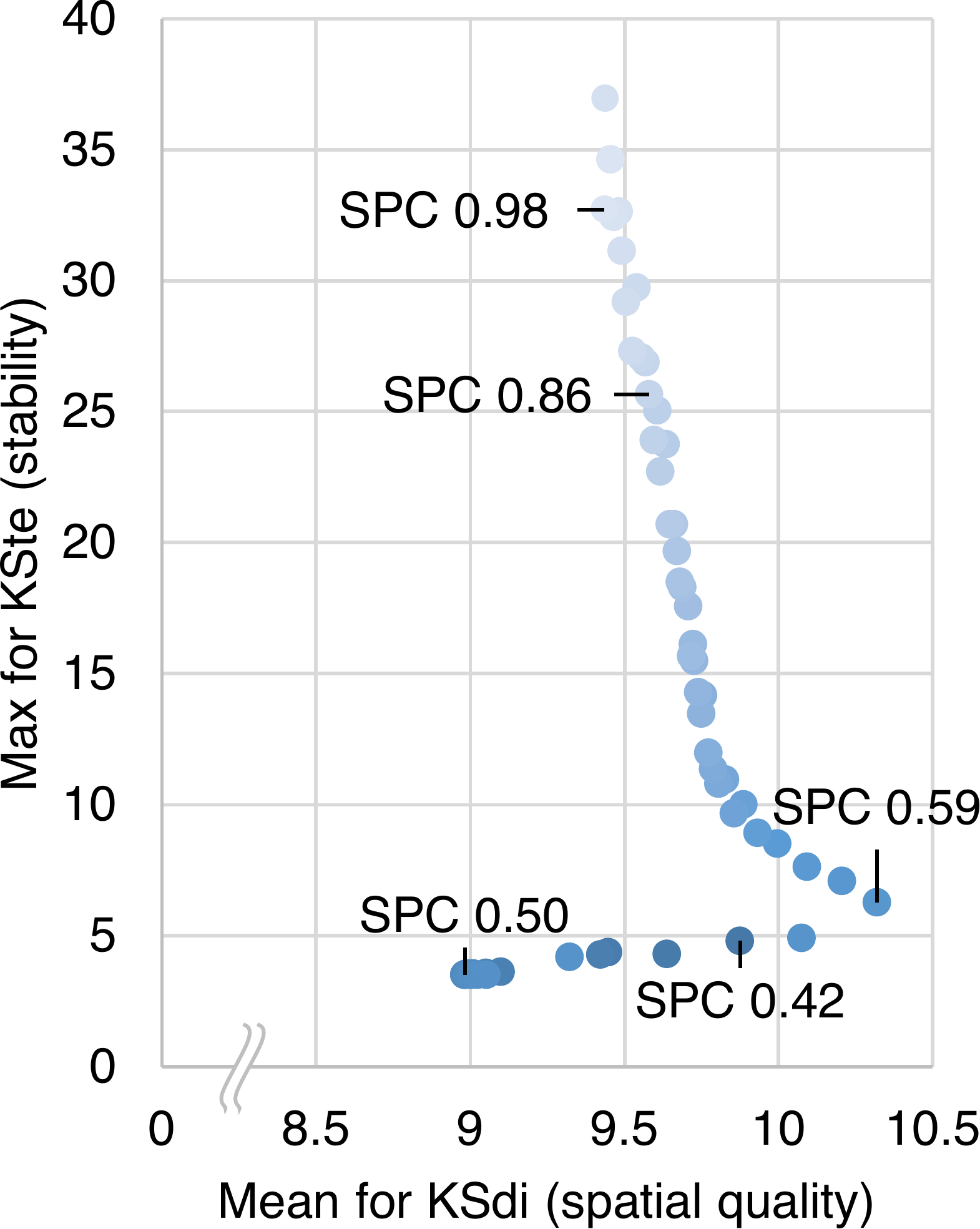}
\end{minipage}
\caption{Comparing the mean and mean (left) as well as max and mean (right) for KSte (stability) and for KSdi (spatial quality), for uniformly distributed parameter settings of SPC on Reynolds.}
\label{fig:cluster-parameter}
\end{figure}

\mypar{Parameter experiment}
The parameter experiment gave some surprising results for Reynolds. Fig.~\ref{fig:cluster-parameter} shows charts containing the results. The cut-off values are $0.98$ and $0.42$ for this data set, meaning that every value above $0.98$ and below $0.42$ uses the same projection vectors as the visual summaries using the cut-off values. The parameter values that are indicated by labels in the figures are the values we used in our other experiments. As intermediate values we choose $a=0.50$, $b=0.59$, and $c=0.86$. 

First we consider the parameter values between $0.86$ and $0.59$. These values show the inverse relation between spatial quality and stability: as the parameter value decreases, the stability increases while the spatial quality deteriorates. This is the expected behaviour, which we already saw for Fish 1.

Between values $0.59$ and $0.50$ we see that lowering $\sigma$ improves both the spatial quality as well as the stability. The first principal component does not seem to be the vector that results in the best spatial quality here, hence interpolating more can give better spatial quality, while improving stability. Lowering $\sigma$ further results in worse spatial quality and stability, as we saw before.

Finally between $0.98$ and $0.86$ we see better stability when increasing $\sigma$. While this is counter-intuitive in general, it can be explained for this data set. As $\sigma$ increases we interpolate over less frames and interpolate over configurations where the point set is very rounded. This has a positive effect on the stability in this data set, since it prevents 2 clusters from overlapping a lot: When interpolating, we get a lot of frames where the projected points of two clusters interleave, while the points move in opposite directions. This causes many changes in the neighborhood of all the points in those two clusters. If we interpolate less, this behaviour is less prominent and contained in a few frames, resulting in better stability.

\begin{figure}[t]
\centering
\begin{minipage}{0.45\linewidth}
    \centering
    \includegraphics[width=\linewidth]{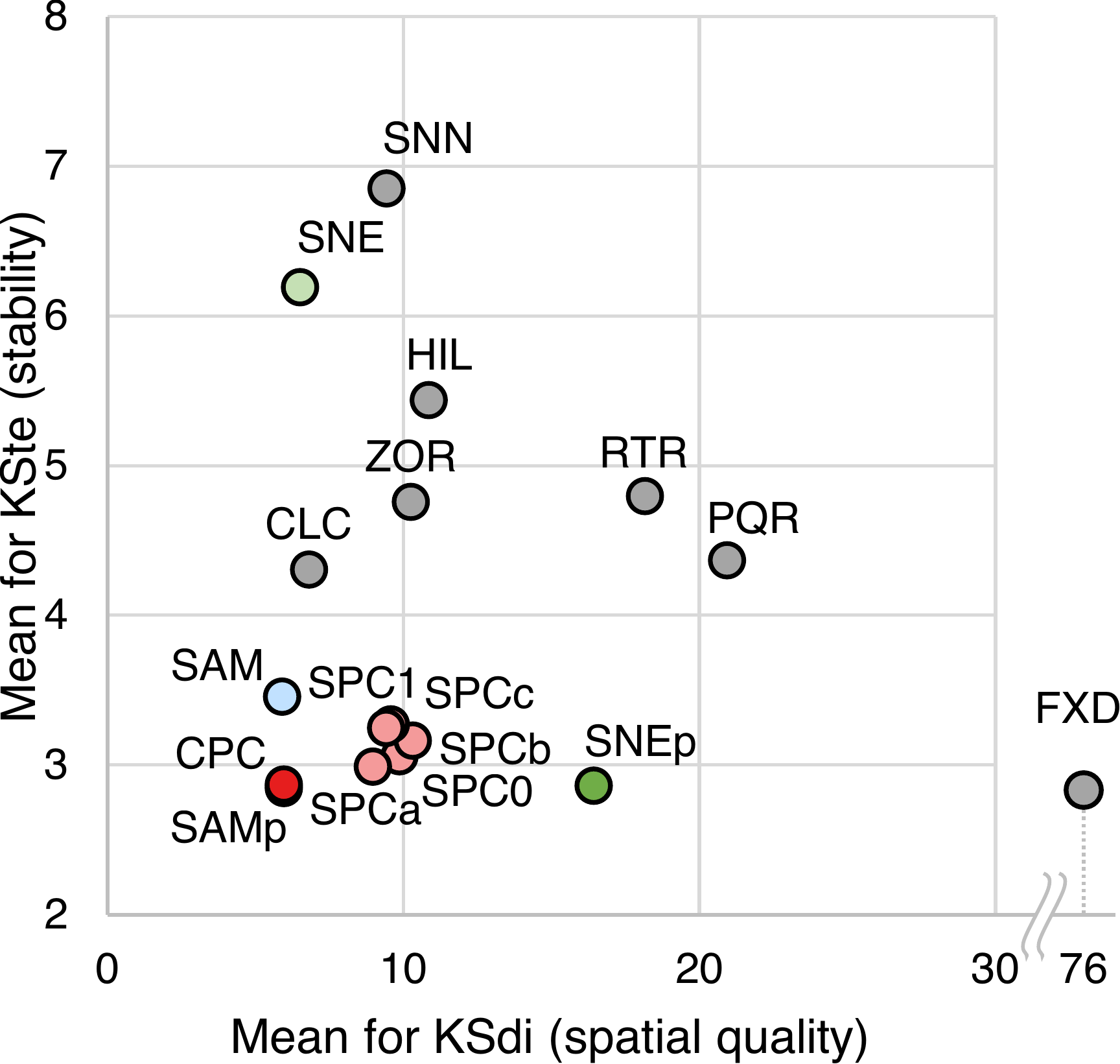}
\end{minipage}
\hfill
\begin{minipage}{0.45\linewidth}
    \centering
    \includegraphics[width=\linewidth]{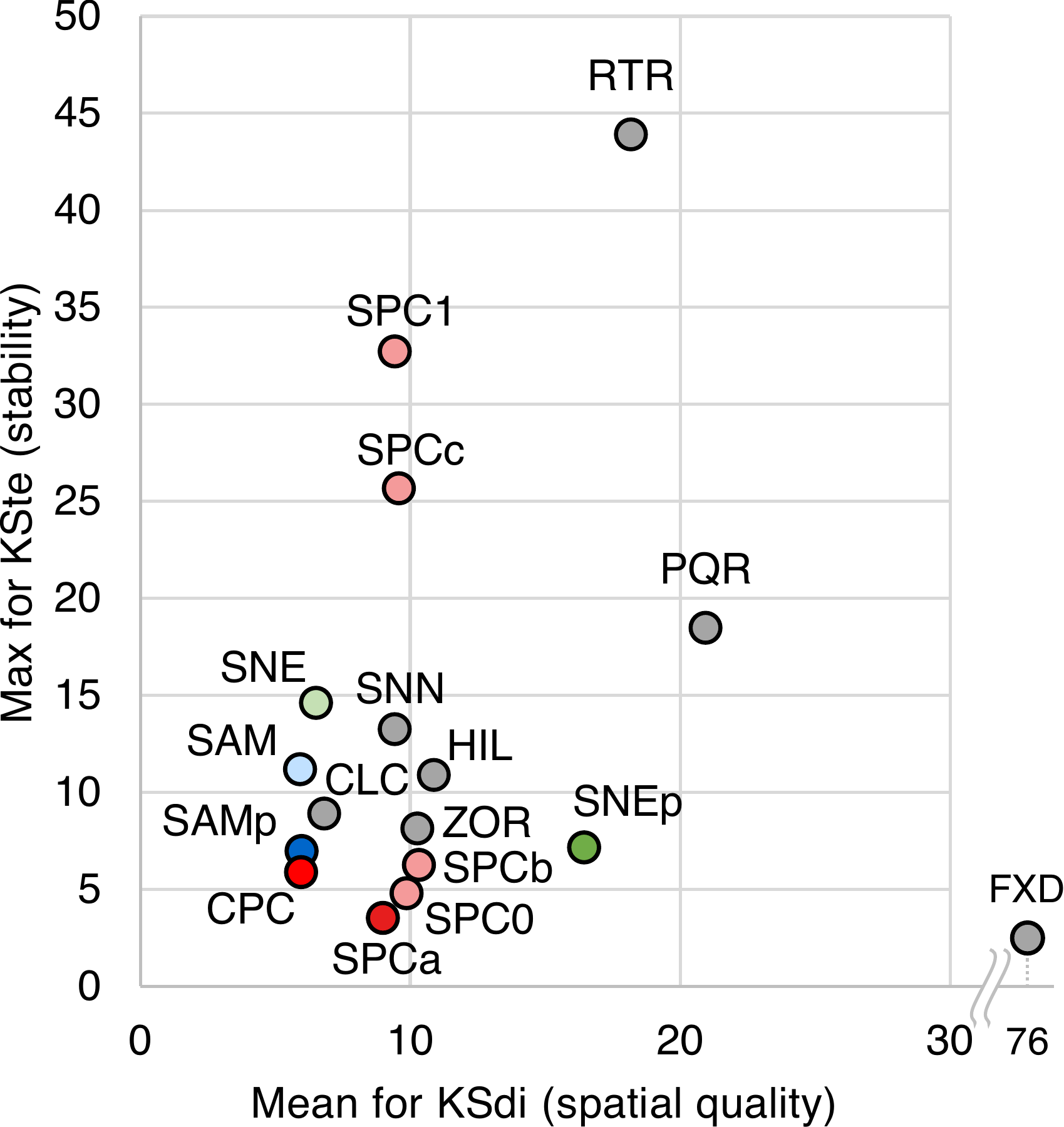}
\end{minipage}
\caption{Comparing the mean for KSdi (spatial quality) and the mean (left) and max (right) for KSte (stability), for all algorithms on Reynolds.}
\label{fig:cluster-VS-charts}
\end{figure}

\mypar{Trade-offs}
The trade-offs between spatial quality and stability for the clustered data set can be observed in Fig.~\ref{fig:cluster-VS-charts}. As we already observed when considering stability in isolation, clustering techniques and SNE perform really well on this data set, especially when considering maximum values for KSte. These techniques end up in the bottom left corner, making them viable techniques for data sets that are clustered. However, they are still outperformed by SPC variants for low $\sigma$ values, SAMp, SNEp and CPC, when it comes to stability. For $\sigma = 0.50$ SPC performs particularly well, even better than SAMp and CPC on maximum KSte. However, SAMp and CPC also have very good spatial quality, making them the best techniques for this data set.

\subsection{Quality results -- Fish 2 \& Netlogo}
For the remaining data sets, the values for spatial quality and stability are very similar to the previous data sets. We therefore give a short overview of the parameter experiment and trade-offs.

\mypar{Parameter experiment} For both data sets there is mostly an inverse relation between stability and spatial quality. However, for the Netlogo data set this relation is absent for $0.45 \leq \sigma \leq 0.59$.
Increasing $\sigma$ in this interval leads to both worse stability and worse spatial quality, like we saw in the Reynolds data set.
During instabilities we observe that at certain frames where SPC projects to an interpolated line, the spatial quality is better than when we project to the first principal component. 
This explains how more interpolation (for decreasing $\sigma$) can improve spatial quality.
For the Netlogo data set we further observe that both spatial quality and stability change erratically for $\sigma < 0.40$, caused by large intervals of interpolation.

\mypar{Trade-offs} For both data sets we see similar results as before, except for CPC, which seems to perform worse. In general, whenever the outcome of the clustering in CPC changes, this significantly changes the ordering and results in a burst of instability. Fish 1 and Reynolds are not affected by this, as the clustering does not change in those data sets.

\subsection{Running Time}

We implemented and executed all algorithms in Java 11 on a workstation with two Intel Xeon E5-2687W CPUs at 3.10GhZ, 16 Cores, 128GB Ram and an NVidia Quadro M600 GPU, running Windows 10.
We measure running times only for computing the orderings, excluding reading input, color mapping and rendering.
The running times range from a few milliseconds for the Z-Order curve (ZOR) to just over 8 hours for t-SNE (SNE). 
General observations include comparably good performance for the subdivision methods (ZOR, HIL, PQR, RTR), with values under one second. 
Only SPC variants are on par with this speed. 


\subsection{Conclusion}
Overall, stable dimensionality reduction methods such as SAMp, SNEp, and SPC for parameter values lower than $1$ perform very well in terms of average and worst-case stability, while only marginally sacrificing spatial quality in the case of SAMp and SPC. SPC does so at a fraction of the computational cost necessary for more complex dimensionality reduction techniques. Considering all the above, we conclude that stable dimensionality reduction methods are the best for computing visual summaries of time-varying data.

\section{Discussion \& Future Work}
\label{sec:discussion}

Here we discuss our results and future work in the context of data properties, algorithm performance, and visual summaries. 

\mypar{Movement characteristics}
Our results show that algorithm efficacy is influenced by the characteristics of the moving entities. SPC works particularly well for a a single, roughly convex cluster (Fish 1) or with only a few such clusters (Fish 2), even though the method emphasizes cluster order and uses a suboptimal axis within each (Fish 2). We can consider many clusters with only a few entities to be effectively the same as a single cluster, as the order within a cluster has little to no influence on the spatial quality.

With multiple, reasonably sized clusters (Reynolds), separating the different clusters in the linear order can be desirable.
By their nature, clustering-based methods will perform better in this regard.
But our experiments show that such methods nonetheless struggle to find a good, stable order within the clusters.
Our hybrid CPC method combines the advantages of clustering with SPC.
However, when the cluster composition changes, stability is now harder to achieve as CPC does not interpolate between the different cluster compositions. 
We leave to future work how such a hybrid method can be turned into a ``clairvoyant'' algorithm \cite{meulemans2019stability} that already aligns the SPC axes of clusters before a change in clusters is actually occurring.

For a complexly shaped cluster, we face yet another issue. Cluster detection might not find an adequate structure. Neither does a single, straight projection axis necessarily capture proximity or neighborhood structure well and is hence likely to give unsatisfactory results as well.
Perhaps methods from topological persistence can play an important role in identifying the structures of these clusters.
We leave the development and evaluation of algorithms for more complex data as future work.
Our results show the potential here, for adapting existing methods to explicitly consider stability.

\mypar{Beyond spatial data}
Our stable methods can be used in any situation with time-varying data in at least two (numeric) dimensions, to determine the ordering and construct visual summaries from these. 
In other words, our techniques may thus be useful for providing an overview also for nonspatial data. However, we expect it to be primarily useful when proximity (or more generally, neighborhoods) of items are meaningful and of interest.
Investigating precise conditions under which this approach is effective is left to future work.



\mypar{Enriching summaries}
We can augment a visual summary with indicators of its spatial quality and stability.
In this paper we used juxtaposed bar charts (Fig.~\ref{fig:teaser} and~\ref{fig:strategycomparison}) or color (Fig.~\ref{fig:spcoverview}). 
Various other encodings could also be considered, e.g. reducing the saturation of the colors or underlining the summary with two lines where the pixel colors indicate the spatial and temporal quality. 
Such augmentations carry information about uncertainty introduced by the dimensionality reduction.
How to best visually convey such information, and how this affects interpretation are left to future work.

\begin{figure*}[t]
    \centering
    \includegraphics[width=\linewidth]{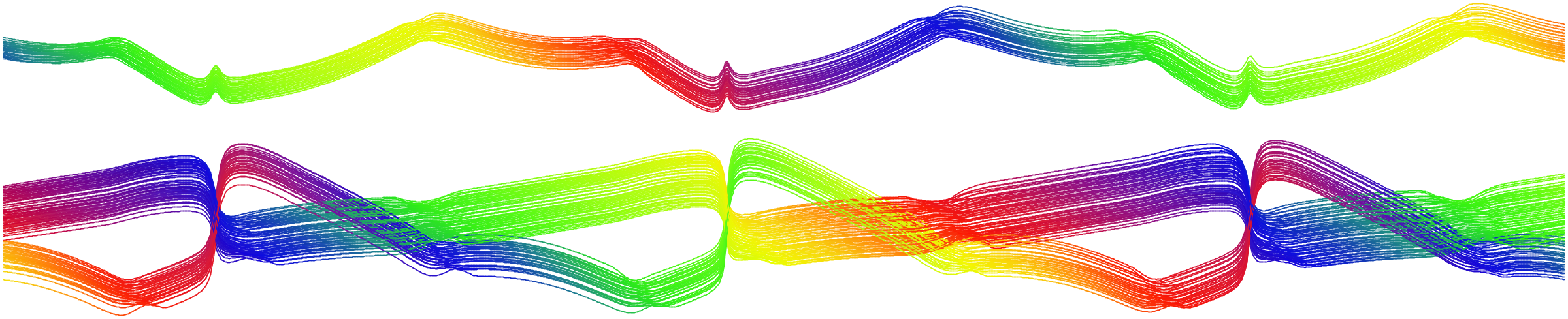}
    
    \vspace{\baselineskip}
    \includegraphics[width=\linewidth]{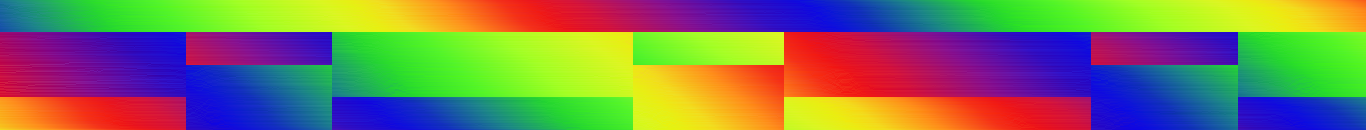}
    
    \caption{A MotionLines and a MotionRug of the clustered data set, using 1D representations produced by our CPC algorithm ($\sigma = 0.5$).}
    \label{fig:motionlines}
\end{figure*}

If the graphical space and the number of elements allow for it, we can also use not just the order but the actual resulting 1D representation of dimensionality reduction. Connecting the different positions for each object with a single line then gives us what we think of as ``MotionLines'', which combine ideas from MotionRugs \cite{BJC+19} and Story Lines \cite{DBLP:journals/jgaa/DijkFFLMRSW17}. As shown in Fig.~\ref{fig:motionlines}, the additional space can be used to communicate more information, such as relative distances or cluster structure. This hints at a possible trade-off between the compactness and expressiveness of visual summaries, which can be further investigated in future work.


\mypar{Overview-first}
Visual summaries are primarily an overview-first tool.
They give an analyst a rough idea of what happens during the motion of the entities, as a first entry point to find time spans or sets of entities to further investigate.
It is thus important to understand how movement patterns relate to patterns visible in the summary and vice versa.
To ensure that \emph{collective} movement of subgroups leads to observable patterns in a visual summary, we need the attribute used for coloring to be similar for spatially close entities.
Without a relation between spatial proximity and attribute value, the colors may jump and it becomes difficult to follow entities or subgroups.

Generally, an efficacious visual summary is coherent: its observable visual patterns are meaningful in the actual movement and vice versa. Though we leave the visual saliency and meaning of patterns to future work, our results point towards algorithms that have good measured performance in the spatial and temporal domain.

\medskip

\mypar{Acknowledgments}
The authors wish to thank Prof. Dr. Iain Couzin and Dr. Alex Jordan of the Max-Planck-Institute for Ornithology, Radolfzell, Germany, and Dominik J\"ackle for helpful insights and discussions, and for providing the fish data set.

W. Meulemans and J. Wulms were partially supported by the Netherlands eScience Center (NLeSC); grant no.~027.015.G02. J. Wulms was partially supported by the Austrian Science Fund (FWF), grant~P~31119. B. Speckmann and K. Verbeek were partially supported by the Dutch Research Council (NWO); project no.~639.023.208 and no.~639.021.541, respectively.
\bigskip

\bibliographystyle{abbrv-doi}
\bibliography{references}

\newpage
\appendix
\section{Computing 1D Orderings}
\label{app:algorithms}

\begin{figure}[h]
	\centering 
	\includegraphics[width=1\linewidth]{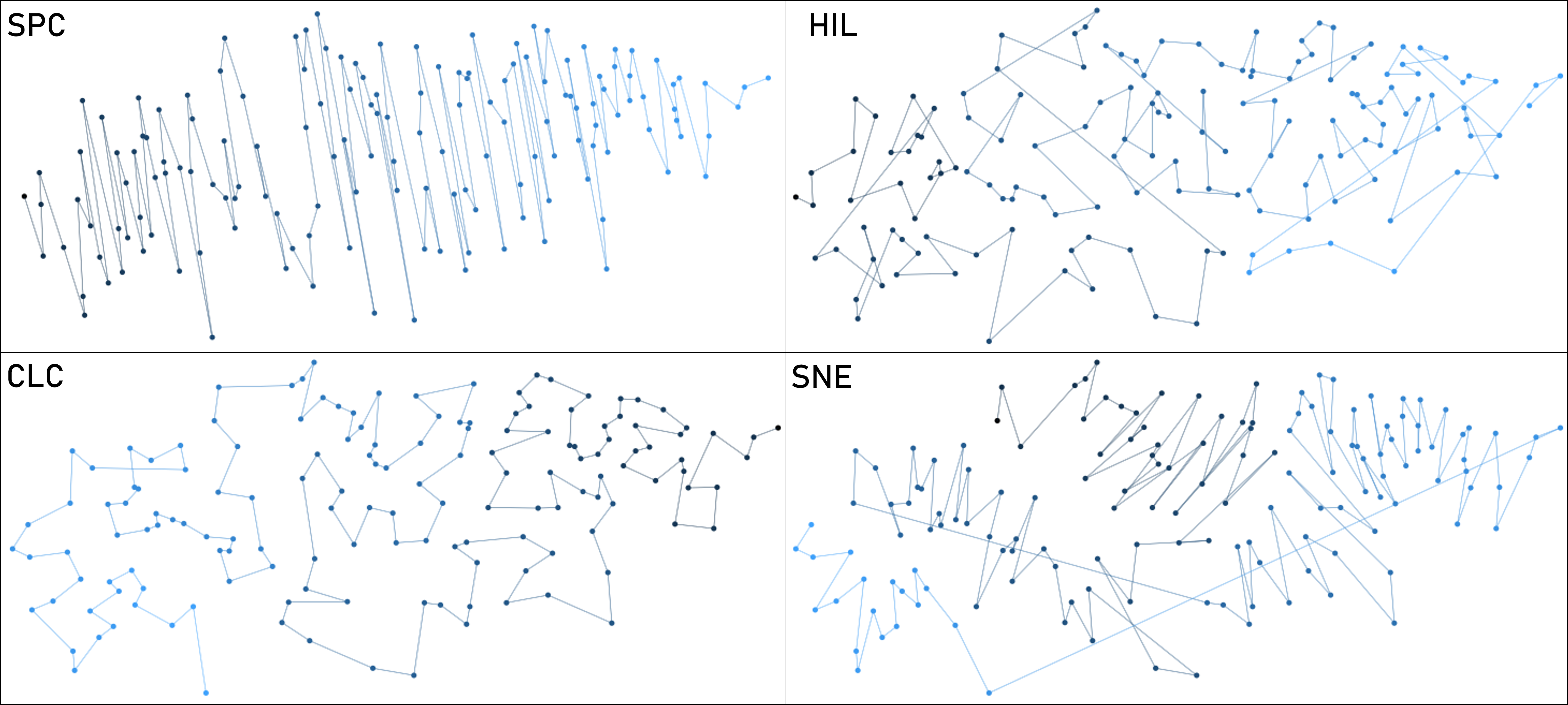}
	\caption{
		Orderings for one time step generated using linear (SPC) and nonlinear (SNE) dimensionality reduction, space-filling curves (HIL) and clustering (CLC).}
	\label{fig:orderings2}
\end{figure}

In addition to dimensionality reduction, various other techniques can be found in literature that are used to compute linear orders for a set of points.
We compare dimensionality-reduction techniques and our new adaptations not only to each other, but also to several of such existing algorithms. 
Most of these algorithms are not designed to be stable and typically consider different time steps in isolation. 
To select suitable algorithms for our comparison, we choose the algorithms that performed best in the experiments by Guo and Gahegan~\cite{guo2006spatial} and Buchm{\"u}ller~\etal~\cite{BJC+19}. 
We may classify these remaining algorithms based on how they compute a linear order: (1) via spatial subdivisions; (2) via clustering. 
Finally, we also include a baseline algorithm that is solely focused on stability.
Figure~\ref{fig:orderings2} shows an example of the orderings generated by a selection of the algorithms, including the dimensionality-reduction techniques, for one time step of our test data for reference.

\bigskip

\mypar{[FXD] Fixed Order} 
This algorithm outputs the same arbitrary linear order for every time step and hence serves as reference baseline for our experiments. With FXD, each horizontal line always represents the same moving entity.

\subsection{Spatial Subdivisions}

Several well-known linearization approaches, which are primarily used for spatial-indexing applications, are based on the principle of iterating through some spatial subdivision. 
These approaches encompass tree data structures and space-filling curves. 
We focus on four established, representative techniques from this area, though many variations exist; see \cite{DBLP:journals/debu/LuO93} for an overview.

\newpage

\begingroup
\setlength{\columnsep}{1.0em}
\begin{wrapfigure}{r}{0.2\linewidth}
	\includegraphics[width=\linewidth]{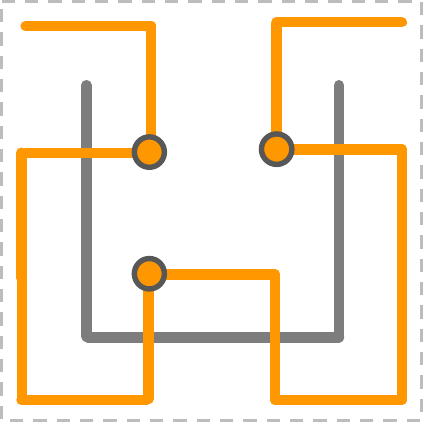}
	
	\vspace{5pt}
	
	\includegraphics[width=\linewidth]{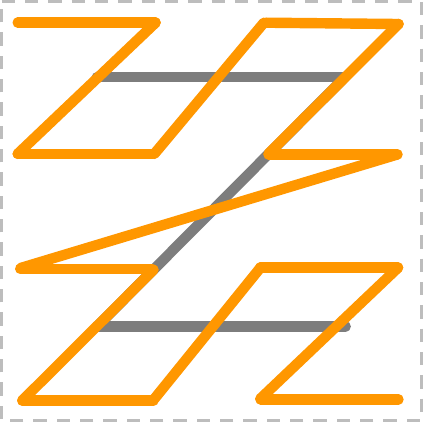}
\end{wrapfigure}
\mypar{[HIL] Hilbert curve and [ZOR] Z-order curve}
The Hilbert curve~\cite{hilbert1891ueber} is a continuous space-filling curve. It can be applied to cover a spatial region in arbitrary precision by repeating the construction pattern recursively. A set of points in space can then be linearized by sampling the curve and noting the order in which the points are encoded on the curve. Another representative of space-filling curves is the Z-order curve, which differs from the Hilbert curve in its geometrical construction pattern resembling a Z shape, where the space is partitioned in four quadrants in the order NW, NE, SW, SE (see figure on the right). Both approaches differ in neighborhood retention and construction complexity, as Lu and Ooi describe~\cite{DBLP:journals/debu/LuO93}. In their comparison, Guo and Gahegan~\cite{guo2006spatial} found that Hilbert curves avoid long jumps better than the Z-Order curve, which in turn outperforms the Hilbert curve in the average of the compared metrics. Since both produce visually different outcomes, we include both strategies in the comparison.

\endgroup

\begingroup
\setlength{\columnsep}{1.0em}
\begin{wrapfigure}{r}{0.2\linewidth}
	\centering
	\includegraphics[width=\linewidth]{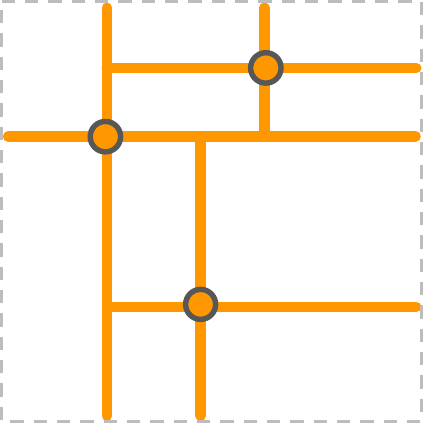}
\end{wrapfigure}
\mypar{[PQR] Point Quadtree}
Quadtrees~\cite{DBLP:journals/acta/FinkelB74} partition space recursively in four parts, until each part contains only a single point. Consequently, sparse areas cause fewer splits than dense areas. Standard quadtrees divide the space in equal parts, while Point Quadtrees split at an input point and thus potentially unevenly in terms of area. 
To derive the 1D ordering, a depth-first tree-iteration strategy is used; given the neighborhood structure in the tree, this is more suitable than a breadth-first strategy. See \cite{DBLP:books/daglib/0032640} for details on tree-iteration strategies.
The standard quadtree essentially reflects a Z-Order curve linearization if the same quadrant iteration is applied. 
Hence, we use the point quadtree variant which produces different orderings due to the intermittent partition.

\endgroup

\begingroup
\setlength{\columnsep}{1.0em}
\begin{wrapfigure}{r}{0.2\linewidth}
	\includegraphics[width=\linewidth]{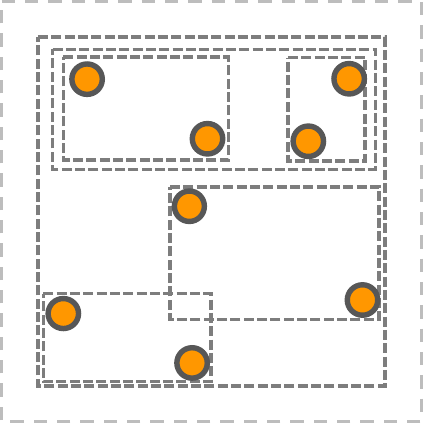}
\end{wrapfigure}
\mypar{[RTR] R-tree}
In R-Trees~\cite{DBLP:conf/sigmod/Guttman84} objects are stored recursively in minimum bounding rectangles (MBR). Each MBR can hold at most a predefined number of objects, thus ensuring a minimum fill. In comparison to quadtrees, more complex balancing is necessary, recomputing the MBRs, when the object limit is reached. Note that MBRs can overlap. Again, a depth-first iteration strategy is used to order points in an R-Tree.

\endgroup

\subsection{Clustering}
Another method to compute a linear order from a point set is to first compute a hierarchical clustering on the point set, and then order the points in such a way that clusters stay together. Algorithms of this type are defined by two aspects: (1) how the points are clustered, and (2) how the linear order is computed from the clustering. In the algorithms we consider below, we always use the following method to compute the linear order from the clustering. 

The hierarchical clustering is represented by a tree with the individual points stored in the leaves. We aim to order to leaves of such a tree without changing the cluster structure: we can change only the order of the children of any internal node.
We follow the algorithm by Bar-Joseph~\etal~\cite{Bar-Joseph2003} to compute the order that minimizes the length of the path formed by visiting the input points in that order. 
The algorithm uses dynamic programming to efficiently find the optimal order for every subtree placing two specific points at the first and last position in the order. 

\begingroup
\setlength{\columnsep}{1.0em}
\begin{wrapfigure}{r}{0.2\linewidth}
	\includegraphics[width=\linewidth]{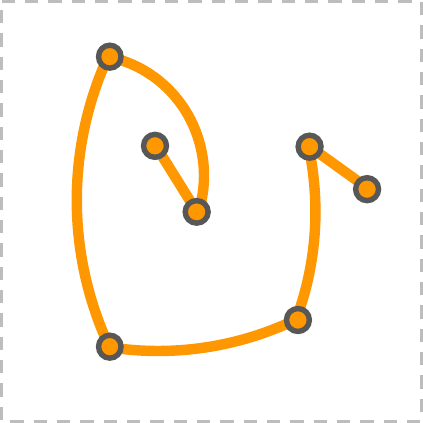}
\end{wrapfigure}
\mypar{[CLC] Complete Linkage Clustering}
Initially, every point is considered as a separate cluster to be hierarchically merged in a bottom-up fashion~\cite{gordon1987review}. We do so by repeatedly merging the closest two clusters, until we obtain a single cluster. Distance between clusters is measured as the distance between their farthest points.

\endgroup

\begingroup
\setlength{\columnsep}{1.0em}
\begin{wrapfigure}[6]{r}{0.2\linewidth}
	\includegraphics[width=\linewidth]{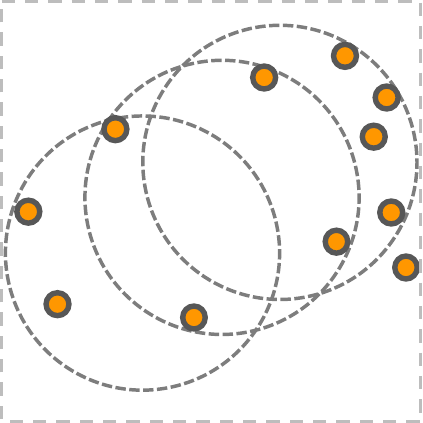}
\end{wrapfigure}
\mypar{[SNN] Shared Nearest Neighbors}
This clustering algorithm~\cite{Jarvis1973} works the same as CLC, but it uses a different metric than Euclidean distance to measure the dis(similarity) between two points. For two points $p$ and $q$, we first count the number of points $x$ that are in the set of $k$ nearest neighbors for both $p$ and $q$. We then define the \emph{shared nearest neighbor} (SNN) distance between $p$ and $q$ as $1/(x+1)$. The SNN clustering is computed using SNN distance instead of Euclidean distance. In case of ties in SNN distance, we use Euclidean distance to break ties. In our experiments we use $k = 10$.

\endgroup

\subsection{Gradient-descent methods for dimensionality reduction}\label{subsec:dimreduc}

Many dimensionality-reduction techniques define a cost function to describe the spatial quality of the resulting representation, and aim to minimize that function. For example, Sammon mapping uses a function that measures how well distances are preserved, while t-SNE uses a function that measures how well local neighborhoods are preserved. Since finding the global minimum of such a cost function is typically hard, they often use local search heuristics (usually gradient descent) to find a good solution. In our experiments we consider two such dimensionality-reduction techniques: Sammon mapping and t-SNE. There are other dimensionality-reduction techniques, such as MDS~\cite{Kruskal1964} and Isomap~\cite{Tenenbaum2000}, but based on their cost functions we believe that they give similar results (in fact, in the Euclidean plane, classical MDS is equivalent to PCA). We first recall Sammon mapping and t-SNE for a static point set, before explaining our adaptations for improved stability.

\begingroup
\setlength{\columnsep}{1.0em}
\begin{wrapfigure}{r}{0.2\linewidth}
	\includegraphics[width=\linewidth]{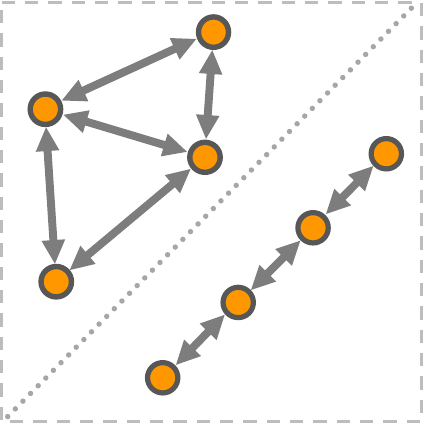}
\end{wrapfigure}
\mypar{[SAM] Sammon Mapping}
Sammon mapping \cite{Sammon1969} aims to preserve distances. Let $d_{ij}$ denote the Euclidean distance between points $p_i$ and $p_j$, denote the resulting (1D) coordinates by $x_i$, and let $\delta_{ij} = |x_i - x_j|$. Sammon mapping computes coordinates $x_i$, attempting to minimize this cost function:
\[
C = \frac{1}{\sum_{1 \leq i < j \leq n} d_{ij}} \sum_{1 \leq i < j \leq n} \frac{(d_{ij} - \delta_{ij})^2}{d_{ij}}
\]
The cost $C$ is then minimized using a gradient descent starting from a random initial solution.

\endgroup

\begingroup
\setlength{\columnsep}{1.0em}
\begin{wrapfigure}{r}{0.2\linewidth}
	\includegraphics[width=\linewidth]{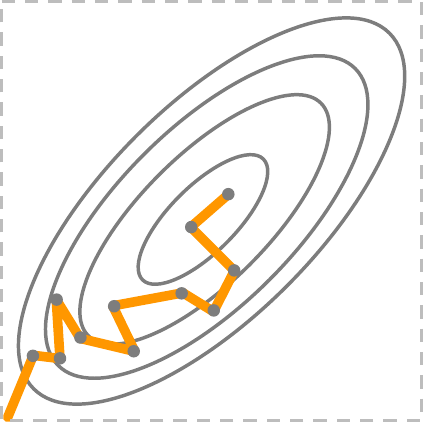}
\end{wrapfigure}
\mypar{[SNE] t-Distributed Stochastic Neighbor Embedding}
The goal of t-SNE~\cite{tsne2008} is to preserve local neighborhoods in the dimensionality reduction. Again, let $d_{ij}$ denote the Euclidean distance between points $p_i$ and $p_j$. Similarities between points are captured by a probability distribution:
\[
\mathcal{P}_{j | i} = \frac{\exp\left(-\frac{d_{ij}^2}{2 \sigma_i^2}\right)}{\sum_{k \neq i} \exp\left(-\frac{d_{ik}^2}{2 \sigma_i^2}\right)}
\]
The values $\sigma_i$ are chosen depending on the predefined \emph{perplexity} $\kappa$ (see~\cite{tsne2008} for details); in our experiments we use $\kappa = 40$.
We further define $\mathcal{P}_{ij} = \frac{1}{2n}(\mathcal{P}_{j | i} + \mathcal{P}_{i | j})$ and we set $\mathcal{P}_{ij} = 0$ if $i = j$. Denote the resulting (1D) coordinates by $x_i$, and define $\delta_{ij}$ as
\[
\delta_{ij} = \frac{(1 + |x_i - x_j|^2)^{-1}}{\sum_{k \neq l} (1 + |x_k - x_l|^2)^{-1}}
\]
The cost function is defined by the Kullback-Leibler divergence as:
\[
C = \sum_{i \neq j} \mathcal{P}_{ij} \log\frac{\mathcal{P}_{ij}}{\delta_{ij}}
\]
Finally, the cost $C$ is again minimized using a (momentum-based) gradient descent\footnote{We tried using the existing implementation at \url{https://github.com/lejon/T-SNE-Java} to compute the t-SNE mapping. This implementation uses approximations to speed up the computation, which lead to artifacts in our results. We therefore implemented the default version of t-SNE ourselves. See Appendix~\ref{app:summary-stats} for more details.} starting from a random initial solution.

\endgroup

\mypar{Stability improvements}
Both Sammon mapping (SAM) and t-SNE (SNE) start the gradient descent with a random initial solution, which may result in low stability over time. To improve the stability of both algorithms, we initialize them with the solution of the previous time step, resulting in two stable versions, \textbf{[SAMp]} and \textbf{[SNEp]}. The rationale is that, if the local minimum found in the previous time step still exists, but has slightly shifted, then this approach will likely find this local minimum again rather than a random other local minimum. 

Recently, Rauber~\etal~\cite{DBLP:conf/vissym/RauberFT16} described Dynamic t-SNE: a more explicit way of making t-SNE stable over multiple time steps. Their approach performs a global optimization over all time steps simultaneously, using a separate copy of each point for each time step. They enforce temporal coherence by adding a term to the optimization function depending on the distance between two copies of the same point at consecutive time steps. 
For two reasons we were not able to include this algorithm in our experiments. First, it is very slow. The paper reports a running time of about 6 minutes per time step. Although a single time step of our data consists of only hundreds of points, we consider thousands of time steps, making their algorithm prohibitively slow for our experiments. Second, the implementation of Dynamic t-SNE rarely gives meaningful output when run on our data\footnote{The implementation often gives NaN as output. The authors~\cite{Telea2019} have verified that this is a known problem with the implementation.}, which changes much faster than the data experimented on in~\cite{DBLP:conf/vissym/RauberFT16}. We further believe that Dynamic t-SNE would converge slowly on a time-varying data set with many time steps: it would take at least $T$ gradient descent iterations for frames that are $T$ time steps apart to affect each other. Since t-SNE is already known to converge quite slowly, the combination may simply require too many iterations to obtain a reasonable solution. 
Thus, Dynamic t-SNE exacerbates the usual downsides of t-SNE, namely black-box parameter tuning and slow convergence.

\section{Metrics}
\label{app:metrics}
\mypar{[KSra] Rank-weighted Keys Similarity}
We define $w(i,j) = 1/j$ inversely proportional to the rank, such that maintaining the closest neighbors is considered more important than the more distant neighbors. With $H_k$ denoting the $k^{th}$ harmonic number, we obtain:
\[ KSra(P,S) = \frac{\sum_{p_i \in P} \sum^k_{j=1} r(i,j) / j}{\sum_{p_i \in P} \sum^k_{j=1} 1/j} = \frac{\sum_{p_i \in P} \sum^k_{j=1} r(i,j) / j}{n \cdot H_k} \]


\mypar{[KSdi] Distance-weighted Keys Similarity}
We define $w(i,j) = 1/\| p_i - n(i,j) \|$ inversely proportional to the Euclidean distance, such that maintaining close neighbors is considered more important than distant neighbors. In contrast to KSra, this variant does not treat neighbors at (nearly) identical distances differently.

\[ KSdi(P,S) = \frac{\sum_{p_i \in P} \sum^k_{j=1} r(i,j) / \| p_i - n(i,j) \|}{\sum_{p_i \in P} \sum^k_{j=1} 1/\| p_i - n(i,j) \|} \]

\section{Experimental evaluation Fish 2}\label{app:fish2}
\begin{figure*}
	\begin{tabu} to \linewidth {XX}
		\subcaptionbox{The two quality metrics: mean for KSra (left) and KSdi (right column) for all algorithms over all data set frames of Fish 2.\label{fig:merge-spatialquality-chart}}[\linewidth]{\includegraphics[width=\linewidth]{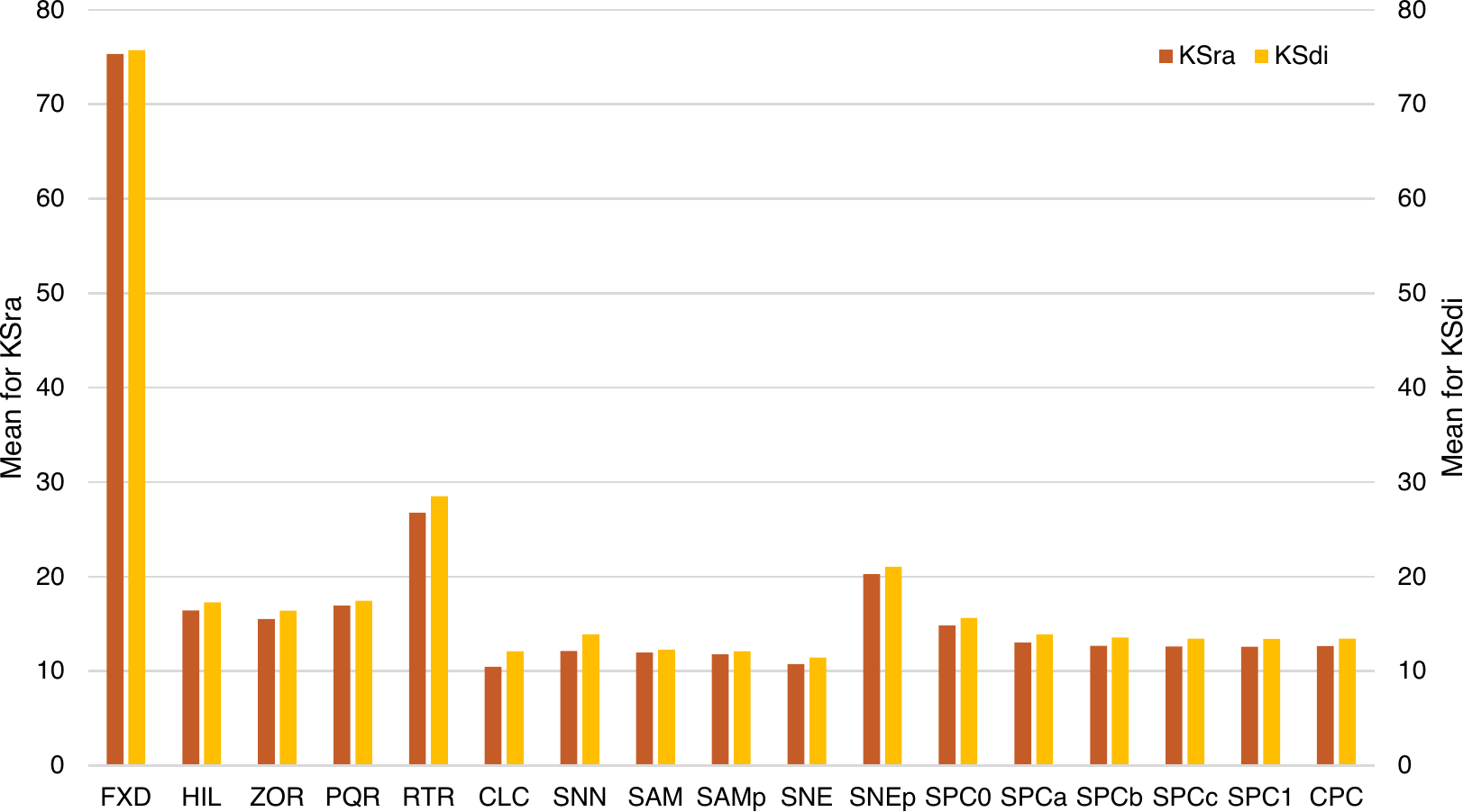}} &
		\subcaptionbox{The three stability metrics: mean for JMP, CRS (left), and KSte (right) for all algorithms over all data set frames of Fish 2.\label{fig:merge-stability-chart}}[\linewidth]{\includegraphics[width=\linewidth]{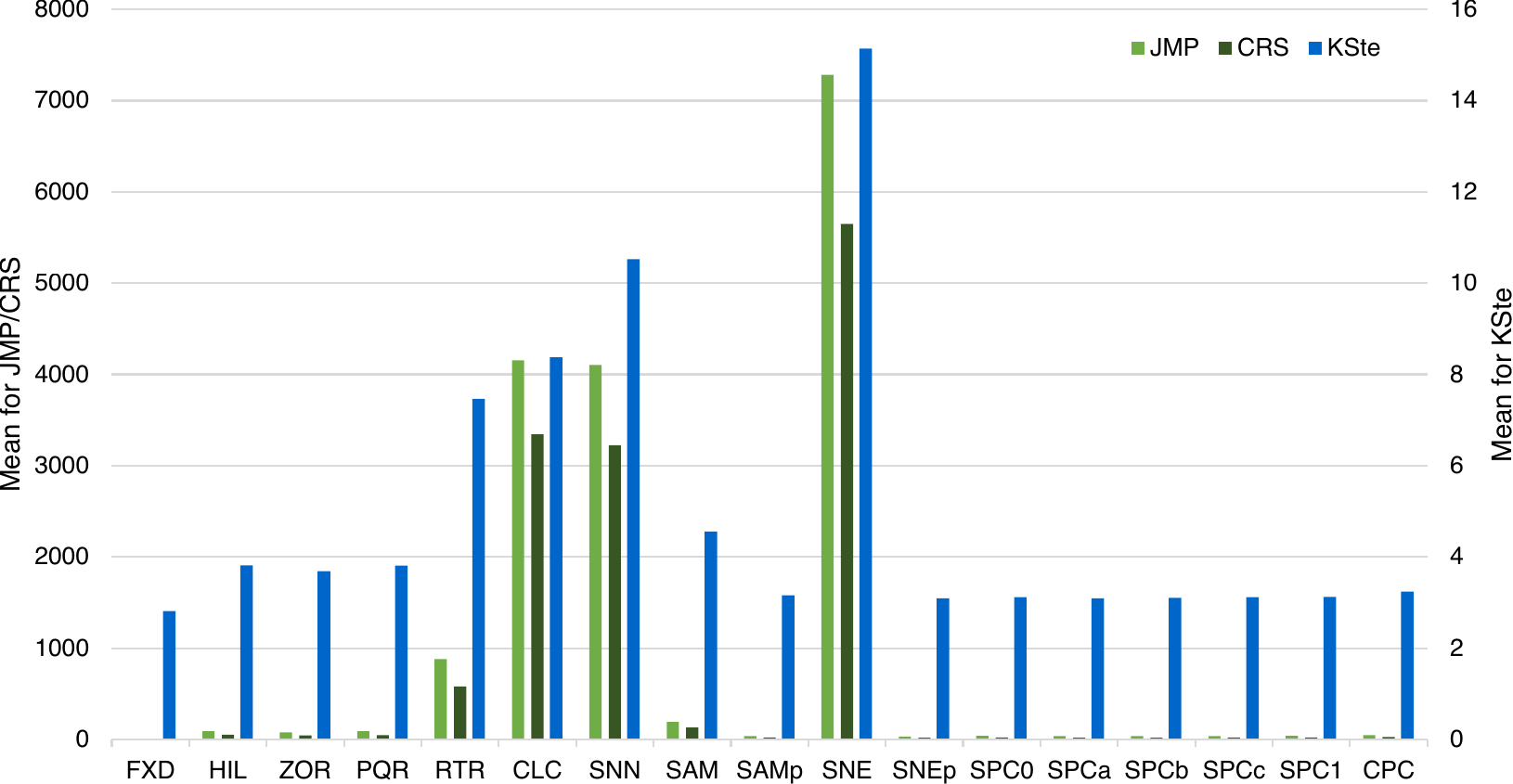}}\\
		\\
		\\
		\subcaptionbox{A comparison between the mean for KSte and for KSdi, for uniformly distributed $\sigma$ of SPC$_{\sigma}$ on Fish 2. \label{fig:merge-parameter-mean}}[\linewidth]{\includegraphics[width=\linewidth]{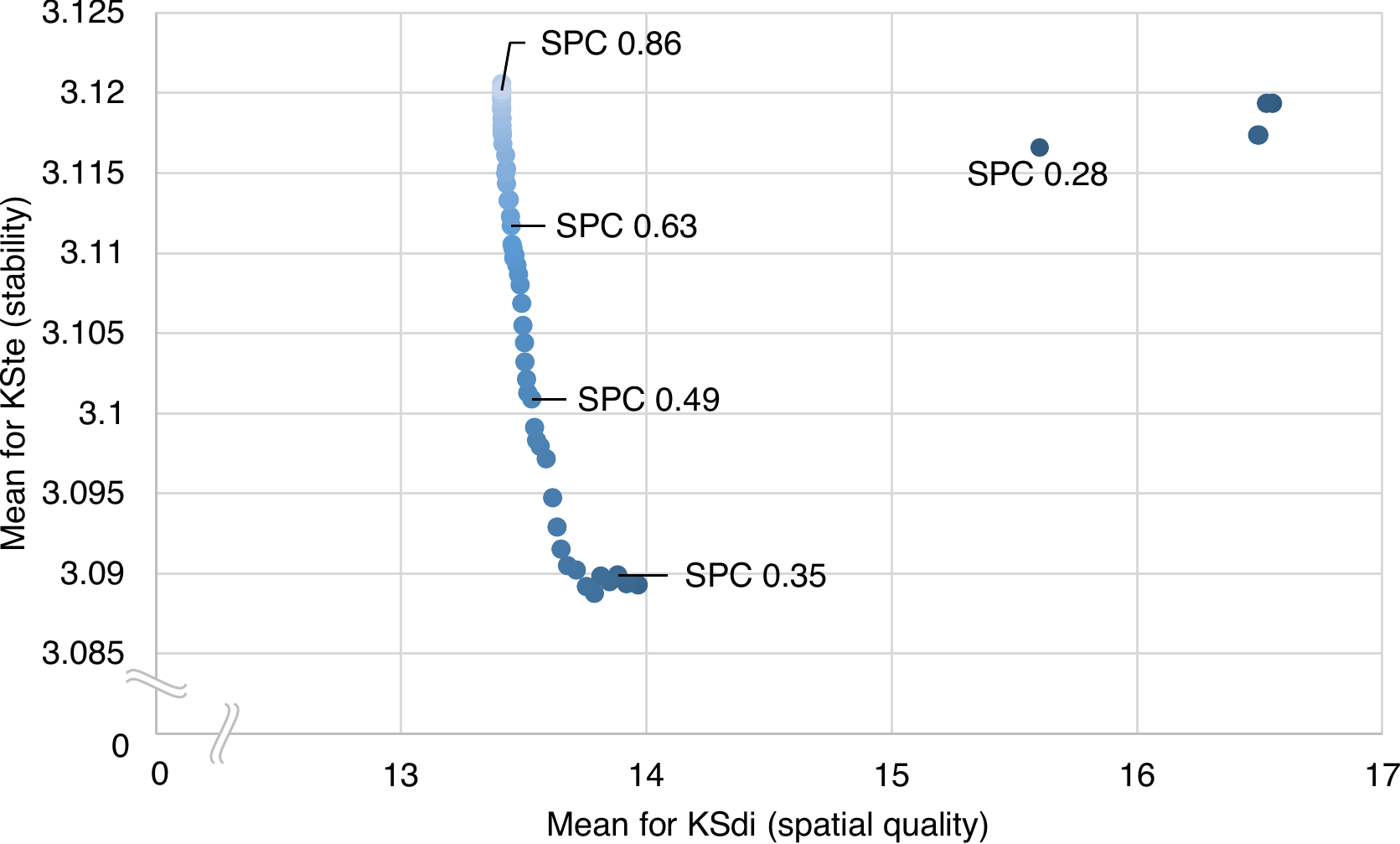}} &
		\subcaptionbox{A comparison between the mean for KSte and for KSdi for all algorithms on Fish 2. \label{fig:merge-spatialqualityVSstability-chart}}[\linewidth]{\includegraphics[width=\linewidth]{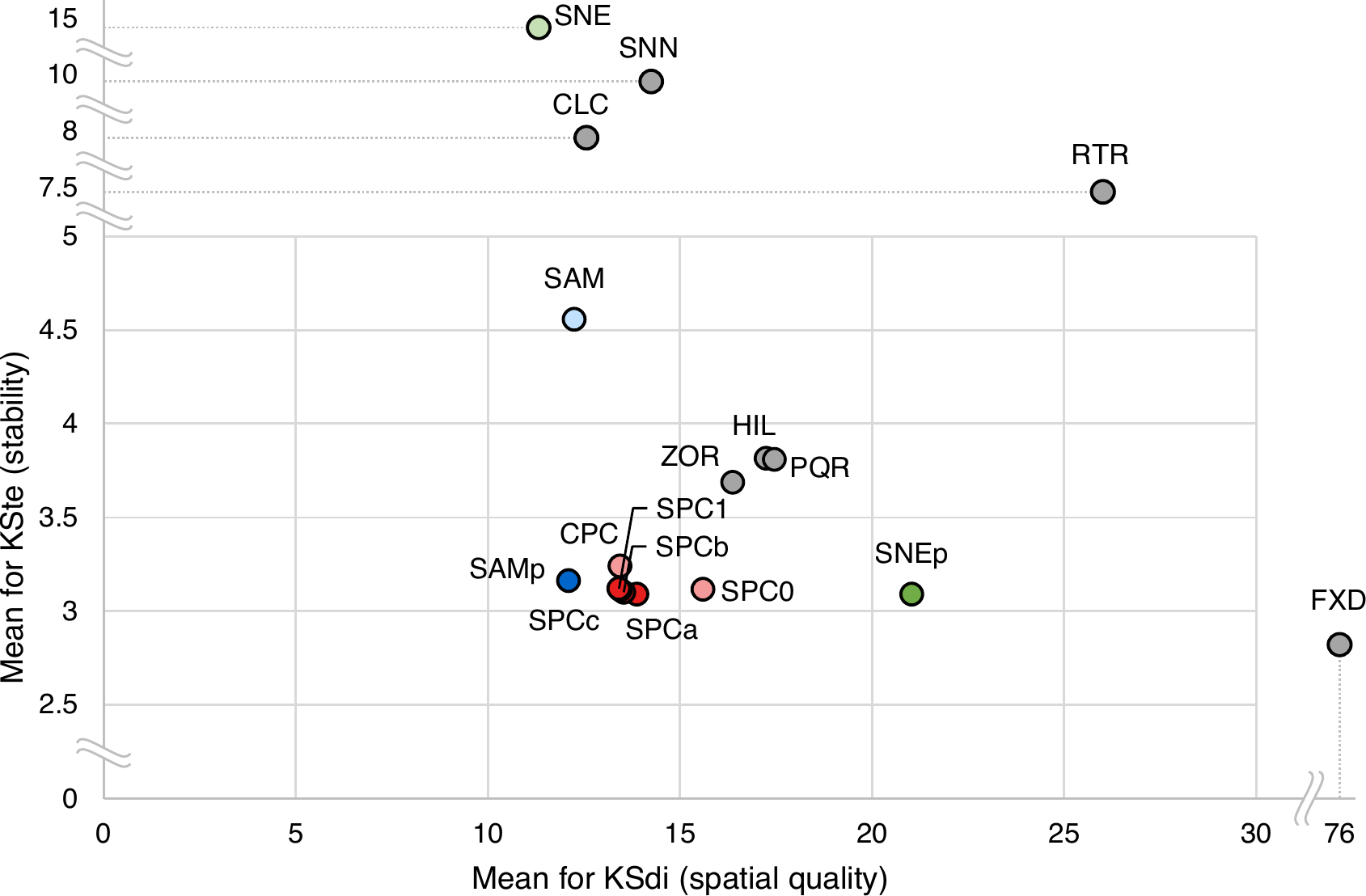}}\\
		\\
		\\
		\subcaptionbox{A comparison between the max for KSte and the mean for KSdi, for uniformly distributed $\sigma$ of SPC$_{\sigma}$ on Fish 2. \label{fig:merge-parameter-max}}[\linewidth]{\includegraphics[width=\linewidth]{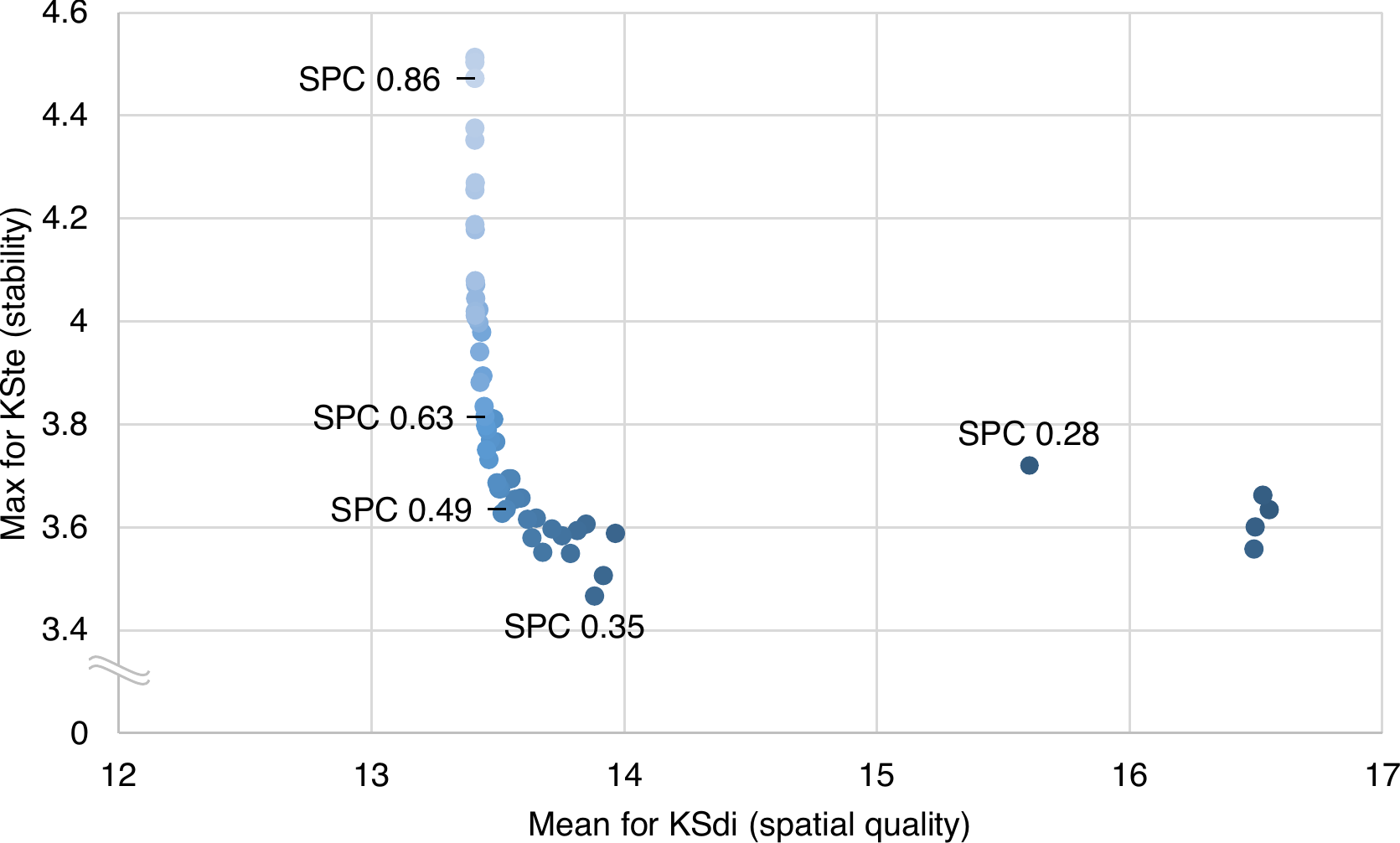}} &
		\subcaptionbox{A comparison between the max for KSte and the mean for KSdi for all algorithms on Fish 2. \label{fig:merge-spatialqualityVSmaxstability-chart}}[\linewidth]{\includegraphics[width=\linewidth]{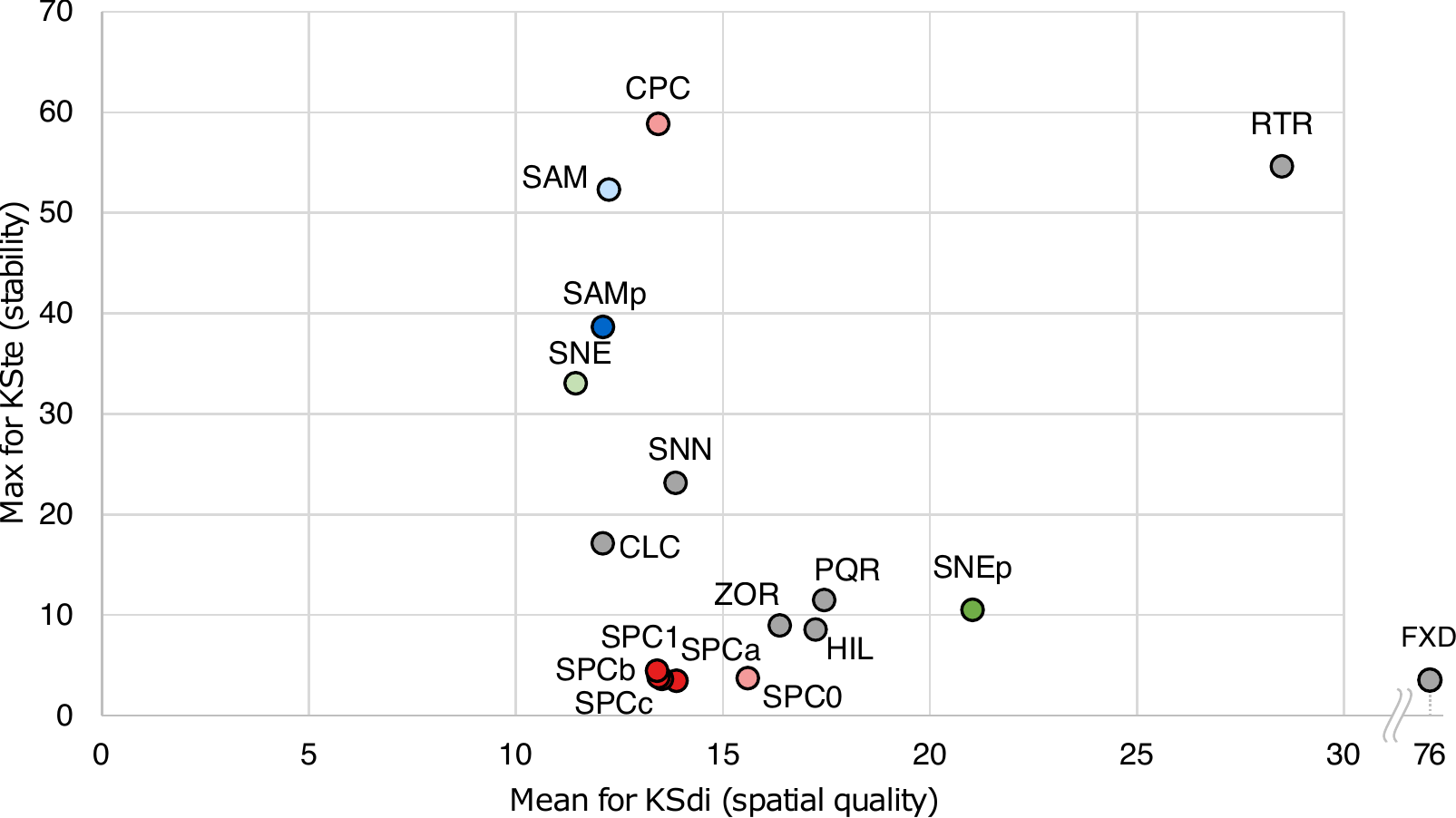}}\\
	\end{tabu}
	\caption{Quantitative evaluation of Fish 2.}
	\label{fig:mergeevaluation}
\end{figure*}

This section gives an in-depth explanation of the results found in the experimental evaluation for Fish 2. In this data set, the fish start moving as a single cluster, after which they split into two separate clusters, and eventually merge into a single cluster again. We investigate the spatial quality and stability in isolation, followed by an exploration of the parameter for SPC. We conclude this section with discussion of the trade-off between spatial quality and stability. All results can be found in Figure~\ref{fig:mergeevaluation}.

\mypar{Spatial quality}
In Figure~\ref{fig:merge-spatialquality-chart} we can see the spatial quality measurements for Fish 2, which are nearly identical to the results for Fish 2. Even though the movement patterns of the fish are quite different in the two excerpts, this does not seem to affect the overall spatial quality of the resulting orderings. One small but notable difference in the results is that SNEp no longer has a higher KSra than KSdi measurements for Fish 2.

\mypar{Stability}
A similar overview of the stability for this data set can be found in Figure~\ref{fig:merge-stability-chart}. Again we see very similar results in comparison to Fish 1. Differences can be seen for SAM, and the spatial subdivision and clustering techniques. These have both higher absolute values for all stability measures, as well as higher values relative to the other techniques.

\mypar{Parameter experiment}
The results of the parameter experiment can be found in Figures~\ref{fig:merge-parameter-mean} and \ref{fig:merge-parameter-max}. We choose the intermediate values of the parameter $\sigma$ to be $a=0.35, b=0.49$ and $c=0.65$ for Fish 2. The cut-off values are $0.30$ and $0.86$, meaning all values below and above the respective cut-offs result in the same summaries. While the plotted measurements for all values of $\sigma$ look very similar to those of Fish 1, there are some small differences. We choose the intermediate values to be in similar positions in the plot, so that they are nicely spread over all 101 measurements. However, the values are different than for Fish 1, which indicates that we need different parameter values to achieve similar spatial quality and stability. Overall we see that the maximum values for KSte are also a lot higher for Fish 1. This is probably due to the fact that in Fish 1 the fish become very rounded, hence forcing the first principal component changes quickly. The difference in intermediate values is therefore not unexpected: Even though the stability measurements of Fish 2 are very similar to the measurements for Fish 1, individual instabilities require different parameter settings to result in stable orderings.

\mypar{Trade-offs}
The trade-off between spatial quality and stability measures is shown in Figures~\ref{fig:merge-spatialqualityVSstability-chart} and~ ~\ref{fig:merge-spatialqualityVSmaxstability-chart}. As we already saw in the overall stability, SAM performs a lot worse on Fish 2, compared to Fish 1. Furthermore, CPC performs a lot worse on maximum stability. This is due to the fact that the fish in Fish 2 split into different clusters. The changes in the clustering of CPC will cause big changes in the ordering, leading to instability at the few frames where the clustering changes. Finally we can see that SNE performs better, relative to its performance on Fish 1, especially for maximum values of KSte.

\begin{figure*}[b]
	\scriptsize{
		\hfill
		\rotatebox{-90}{
			\begin{minipage}{664.4pt}
				\includegraphics[width=664.4pt]{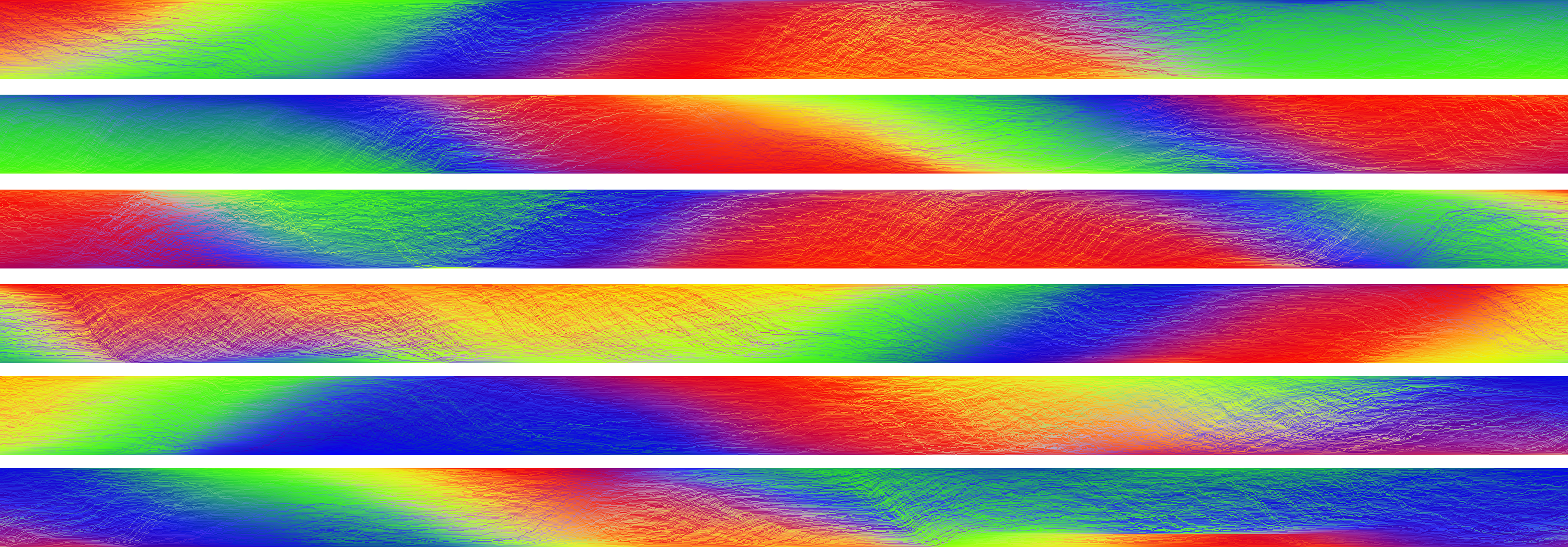}
				\caption{Visual summary of the whole fish data set, cut into 6 pieces to fit on a single page. Fish 1 is taken at the start of the first line, while Fish 2 can be recognized near the end of the last line.}
				\label{fig:fishdatalarge}
			\end{minipage}
		}
		\hfill
	}
\end{figure*}

\mypar{Visual summary}
In our experiments we use two excerpts from a large fish data set. To put the two excerpts in perspective, we show a visual summary of the full data set in Figure~\ref{fig:fishdatalarge}. The visual summary uses 1D orderings and is generated using SPC with $\sigma = 0.5$. Fish 1 is located at the start of the data set, while Fish 2 is taken close to the end of the data set.

\section{Experimental evaluation Netlogo}\label{app:netlogo}
\begin{figure*}
	\begin{tabu} to \linewidth {XX}
		\clearsubcaptionflag
		\subcaptionbox{The two quality metrics: mean for KSra (left) and KSdi (right column) for all algorithms over all Netlogo data set frames.\label{fig:spatialquality-chart2}}[\linewidth]{\includegraphics[width=\linewidth]{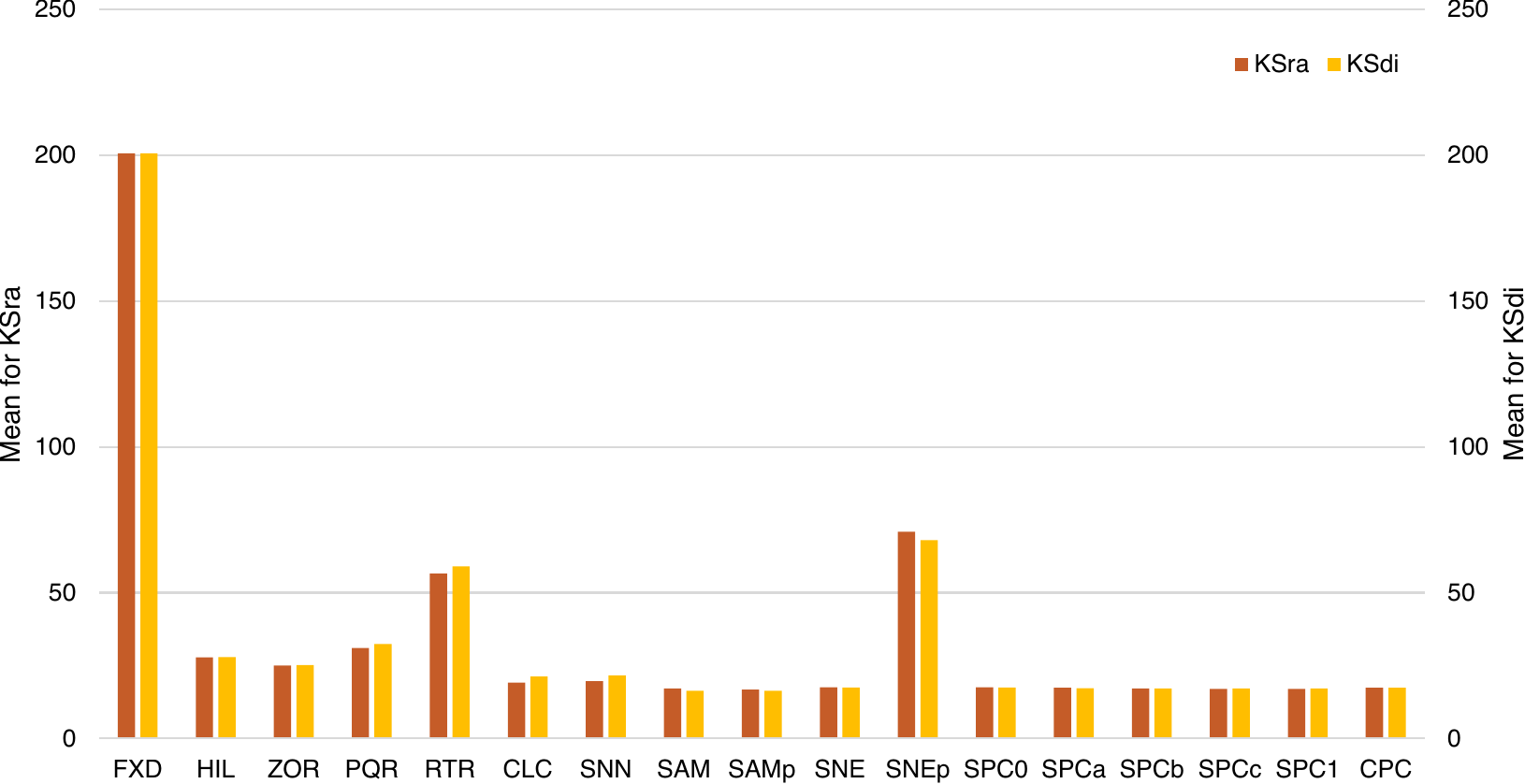}} &
		\subcaptionbox{The three stability metrics: mean for JMP, CRS (left), and KSte (right) for all algorithms over all Netlogo data set frames.\label{fig:stability-chart2}}[\linewidth]{\includegraphics[width=\linewidth]{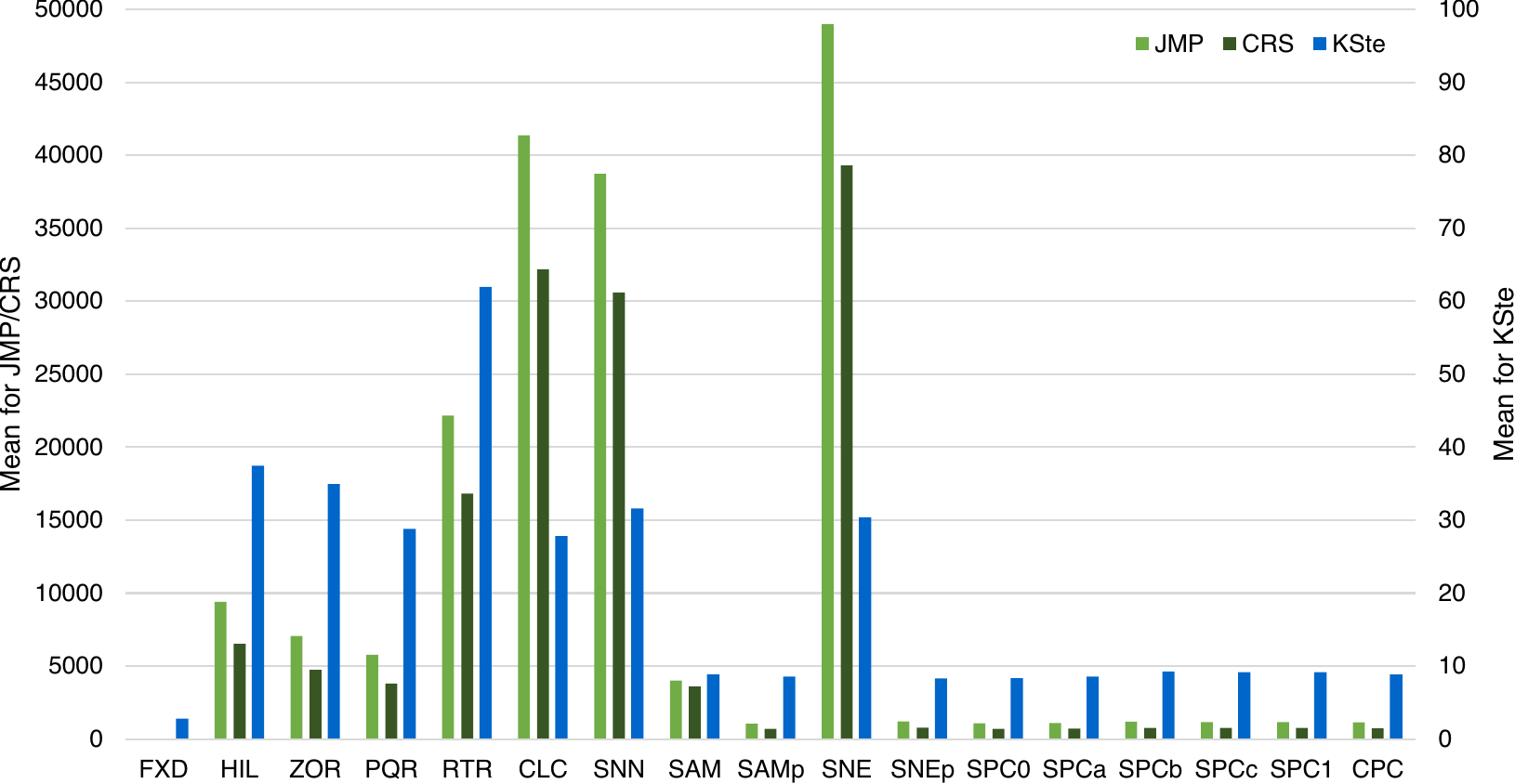}}\\
		\\
		\\
		\subcaptionbox{A comparison between the mean for KSte and for KSdi, for uniformly distributed $\sigma$ of SPC$_{\sigma}$ on the Netlogo data set. \label{fig:parameter-mean2}}[\linewidth]{\includegraphics[width=\linewidth]{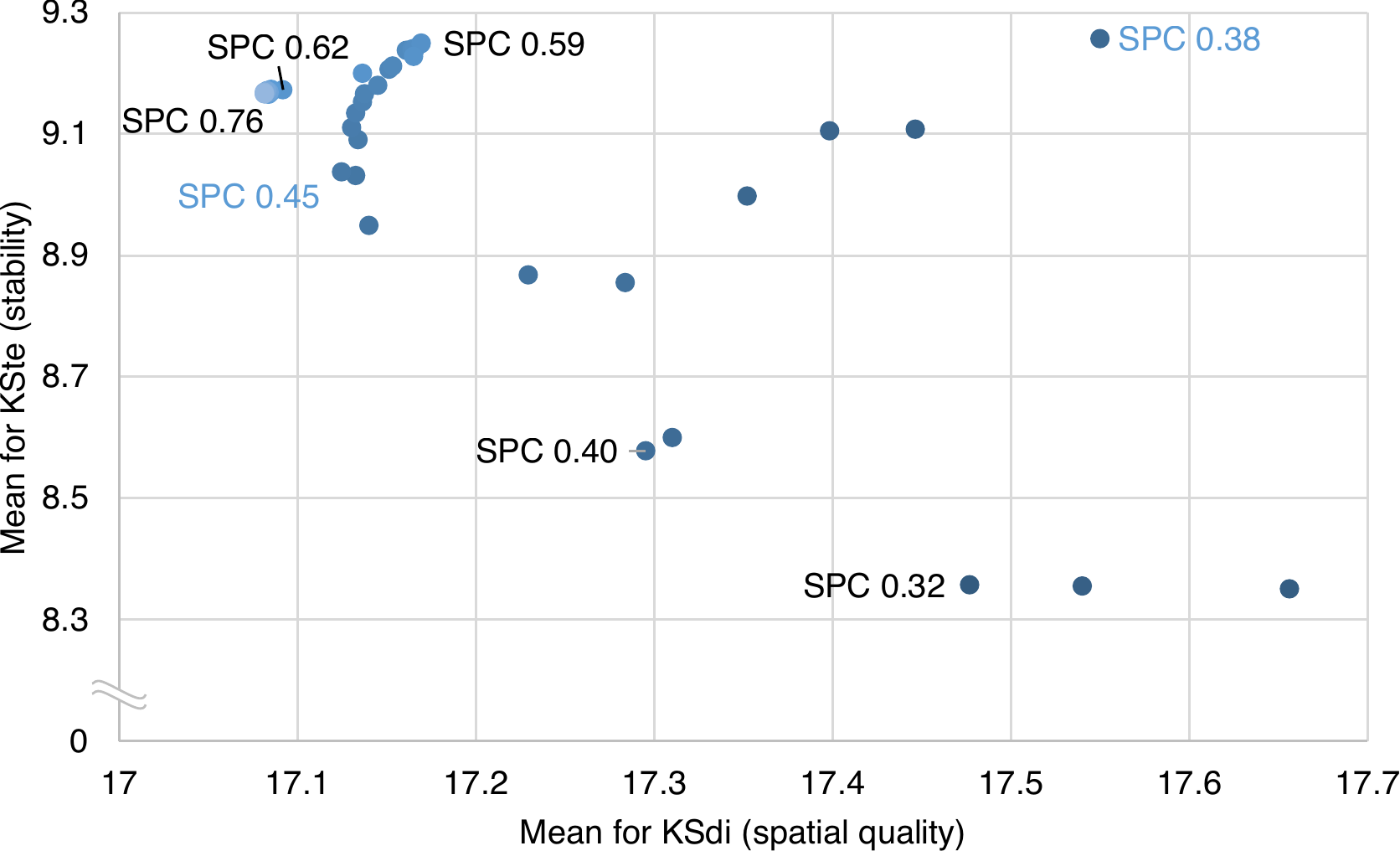}} &
		\subcaptionbox{A comparison between the mean for KSte and for KSdi for all algorithms on the Netlogo data set. \label{fig:spatialqualityVSstability-chart2}}[\linewidth]{\includegraphics[width=\linewidth]{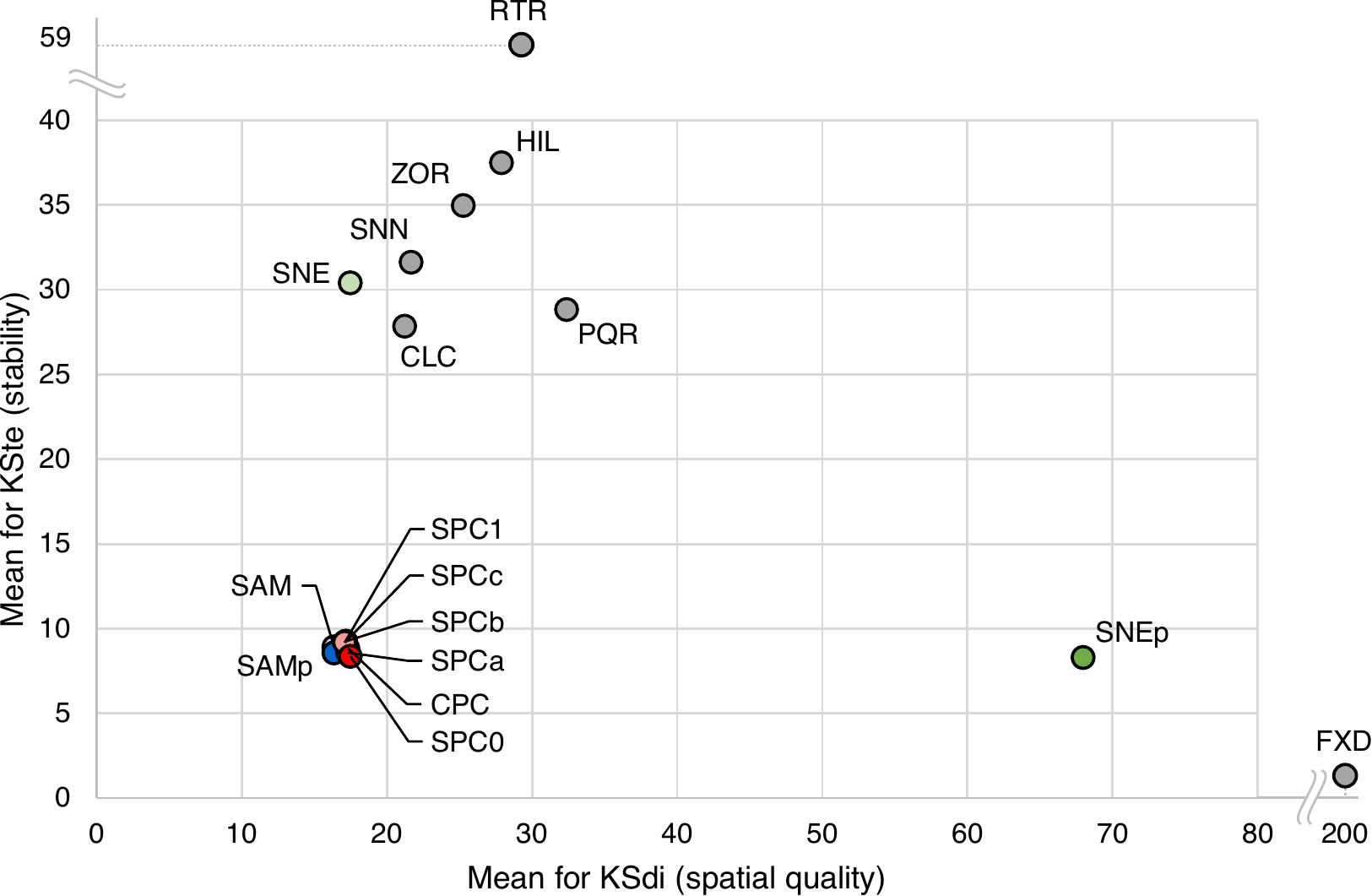}}\\
		\\
		\\
		\subcaptionbox{A comparison between the max for KSte and the mean for KSdi, for uniformly distributed $\sigma$ of SPC$_{\sigma}$ on the Netlogo data set. \label{fig:parameter-max2}}[\linewidth]{\includegraphics[width=\linewidth]{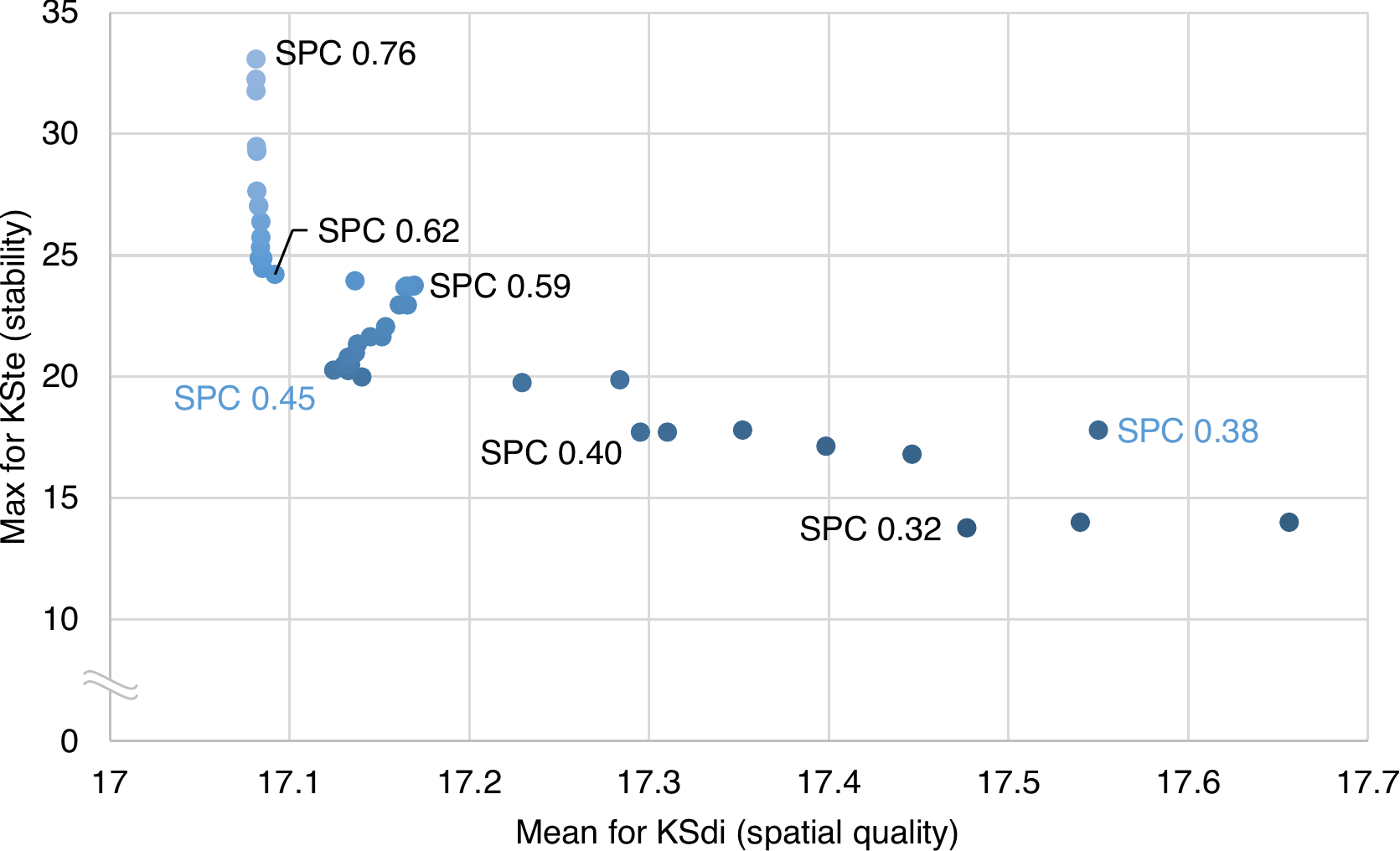}} &
		\subcaptionbox{A comparison between the max for KSte and the mean for KSdi for all algorithms on the Netlogo data set. \label{fig:spatialqualityVSmaxstability-chart2}}[\linewidth]{\includegraphics[width=\linewidth]{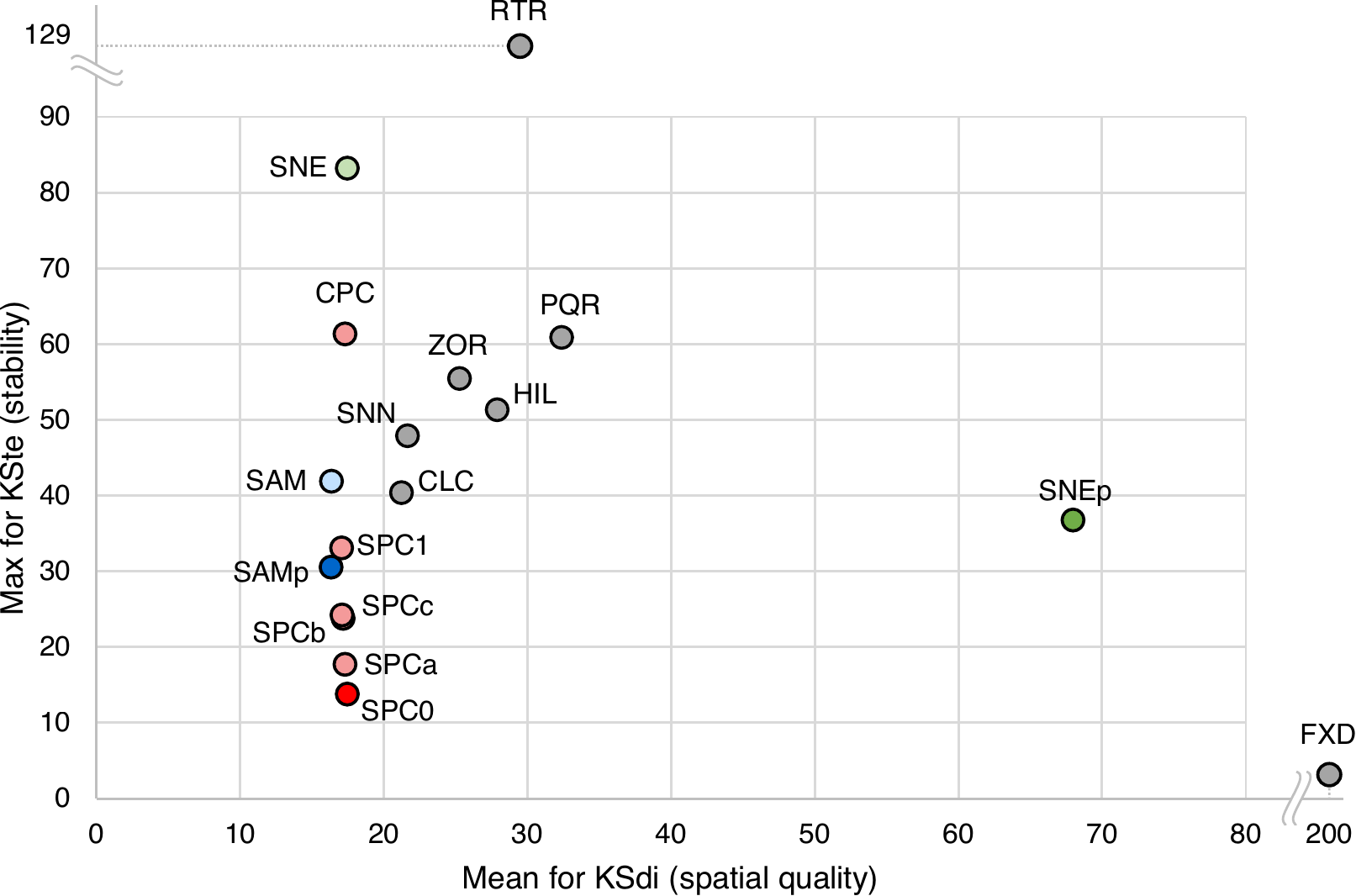}}\\
	\end{tabu}
	\caption{Quantitative evaluation of the Netlogo data set.}
	\label{fig:netlogoevaluation}
\end{figure*}

We examine the statistics for the Netlogo data set used in our experiments in more detail in this section. Again, we first consider the spatial quality and stability separately, followed by the parameter exploration for SPC. Finally we discuss the trade-off between spatial quality and stability as observed on Netlogo. Charts of all statistics can be found in Figure~\ref{fig:netlogoevaluation}.

\mypar{Spatial quality}
As can be seen in Figure~\ref{fig:spatialquality-chart2}, while the absolute values for Netlogo are higher than for Fish 1 and Fish 2, the relative values are very similar. The fixed order gives very bad results on spatial quality, followed by SNEp and RTR. The spatial subdivision techniques all perform similarly, and are slightly better than the previously mentioned techniques. Of the remaining algorithms, the clustering techniques (CLC and SNN) perform slightly worse than all remaining dimensionality reduction techniques (SAM, SAMp, SNE, SPC and CPC). For Netlogo we have chosen different parameter values for SPC, specifically $a=0.40, b=0.59$ and $c=0.62$.


\mypar{Stability}
In Figure~\ref{fig:stability-chart2} we plot the stability statistics for Netlogo. While the chart looks quite different from the stability chart for Fish 1 and Fish 2, this is mostly due to the fact that JMP and CRS count the absolute number of changes in the orders, whereas KSte is normalized. Since Netlogo behaves less stable than Fish 1 and Fish 2, all metrics show higher values. However, Netlogo also contains more moving entities, which increases the absolute number of changes even further.

Comparing the statistics of JMP and CRS, we see very similar performances of all algorithms, with SNE being the least stable, while SAMp is the most stable. On Fish 1 and Fish 2 SNE was the least stable method overall, and while it still has the highest number of absolute changes on Netlogo, it performs a lot better according to KSte, meaning neighborhoods are perserved relatively well over time. Now we see that RTR performs worst on KSte, followed by the space-filling curves (HIL and ZOR) and the clustering algorithms (CLC and SNN) together with PQR. The SPC variants perform relatively worse than on Fish 1 and Fish 2, with parameter values close to but lower than 1 being optimal for stability. Finally SAM, SAMp, SNEp, and CLC are the most stable according to KSte.


\mypar{Parameter experiment}
As already explained in the main text, the results of the parameter experiment are slightly different for Netlogo. In Figures~\ref{fig:parameter-mean2}~and~\ref{fig:parameter-max2} the results are plotted. For Netlogo, the cut-off values are $0.76$ and $0.32$, so everything above $0.76$ uses exactly the first principal component per frame, and similarly for values below $0.32$ we always interpolate between the first and last frame. The parameter values that are indicated by labels in the figures are the values we used in our other experiments ($a=0.40$, $b=0.59$, and $c=0.62$). Note that there are two blue labels, which represent other values of interest that we will use in the analysis below.

We will consider the results between the values indicated by black labels in the figures. Starting from the lowest parameter value $0.32$ we see that increasing $\sigma$ has chaotic effects on both spatial quality and stability up to $0.40$ where this fickle behaviour ends. On closer inspection, the values between $0.32$ and $0.40$ constantly pick up more frames where the entities are stretched enough to use the actual first principal component. Since the intervals between which interpolation happens, constantly change, the results do not steadily change, but are quite erratic. Parameter value $0.38$ shows the worst combination, having both bad spatial quality and stability (max and mean).


From $0.40$ to $0.45$ we see a steady decrease in stability and increase in spatial quality, as expected when increasing $\sigma$. Further increasing the parameter to $0.59$ has negative effects on both the spatial quality and stability. On closer inspection, this increase in spatial quality can be attributed to properties of Netlogo. During instabilities we can observe that at certain frames where SPC projects to an interpolated line, the spatial quality is better than when we project to the first principal component. While we expect projections to the first principal component to have high spatial quality, it is not always the case, as we see here. In Netlogo this occurs when the cluster of points changes direction and shortly does not form a convex shape. It is therefore not unreasonable that interpolating less (and using first principal component more) when increasing $\sigma$ from $0.46$ to $0.56$ can negatively effect spatial quality.

At $0.61$ SPC splits the interpolation over the two consecutive instabilities that were seen as one big instability. This split improves both spatial quality and stability, up to $0.62$ where there are a couple of non-interpolation frames between the instabilities. Increasing the parameter further leads to certain instabilities not being interpolated over any longer, which negatively affects the maximum KSte values observed for those runs of SPC.

\mypar{Trade-offs}
For Netlogo the trade-offs between spatial quality and stability can be found in Figures~\ref{fig:spatialqualityVSstability-chart2} and \ref{fig:spatialqualityVSmaxstability-chart2}. When considering the mean values for KSte and KSdi, we see a similar spread as before: SAM and SPC variants along with CPC are in the bottom left corner, SNEp and FXD have worse spatial quality but good stability, while the remaining techniques (spatial subdivision, clustering and SNE) have relatively good spatial quality but bad stability. However, SNE is more stable than we have seen for Fish 1 and Fish 2, outperforming all spatial subdivision and clustering techniques, except for PQR and CLC. For the maximum values of KSte, we again see CPC perform worse than on Fish 1. Netlogo mostly consists of a single cluster, but the occasional outlier can trigger the clustering in CPC to find different clusters, resulting spikes of instability.

\section{Experimental evaluation Reynolds}
\label{sec:reynolds}
\begin{figure*}
	\begin{tabu} to \linewidth {XX}
		\clearsubcaptionflag
		\subcaptionbox{The two quality metrics: mean for KSra (left) and KSdi (right column) for all algorithms over all Reynolds frames.\label{fig:clusters-spatialquality-chart}}[\linewidth]{\includegraphics[width=\linewidth]{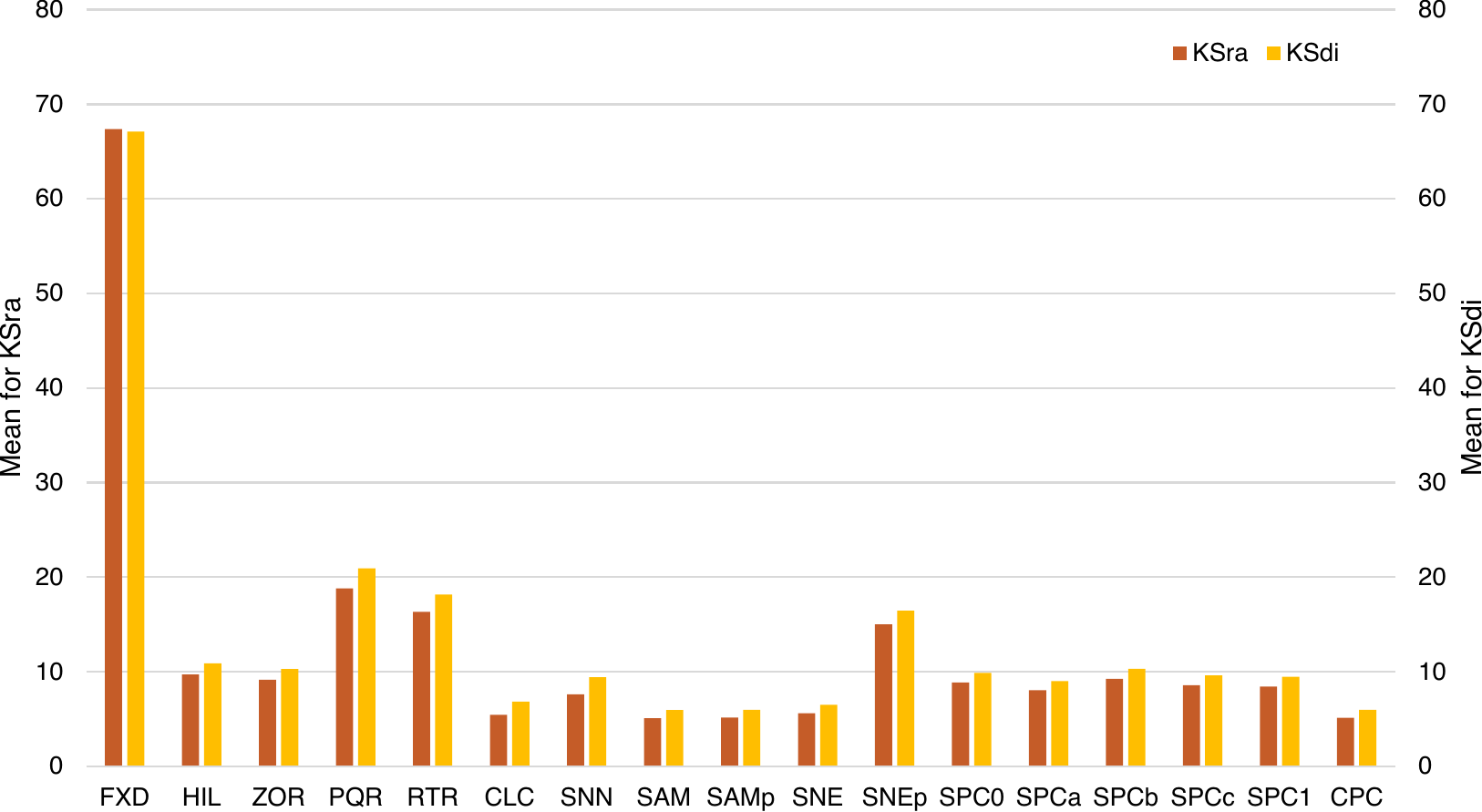}} &
		\subcaptionbox{The three stability metrics: mean for JMP, CRS (left), and KSte (right) for all algorithms over all Reynolds frames.\label{fig:clusters-stability-chart}}[\linewidth]{\includegraphics[width=\linewidth]{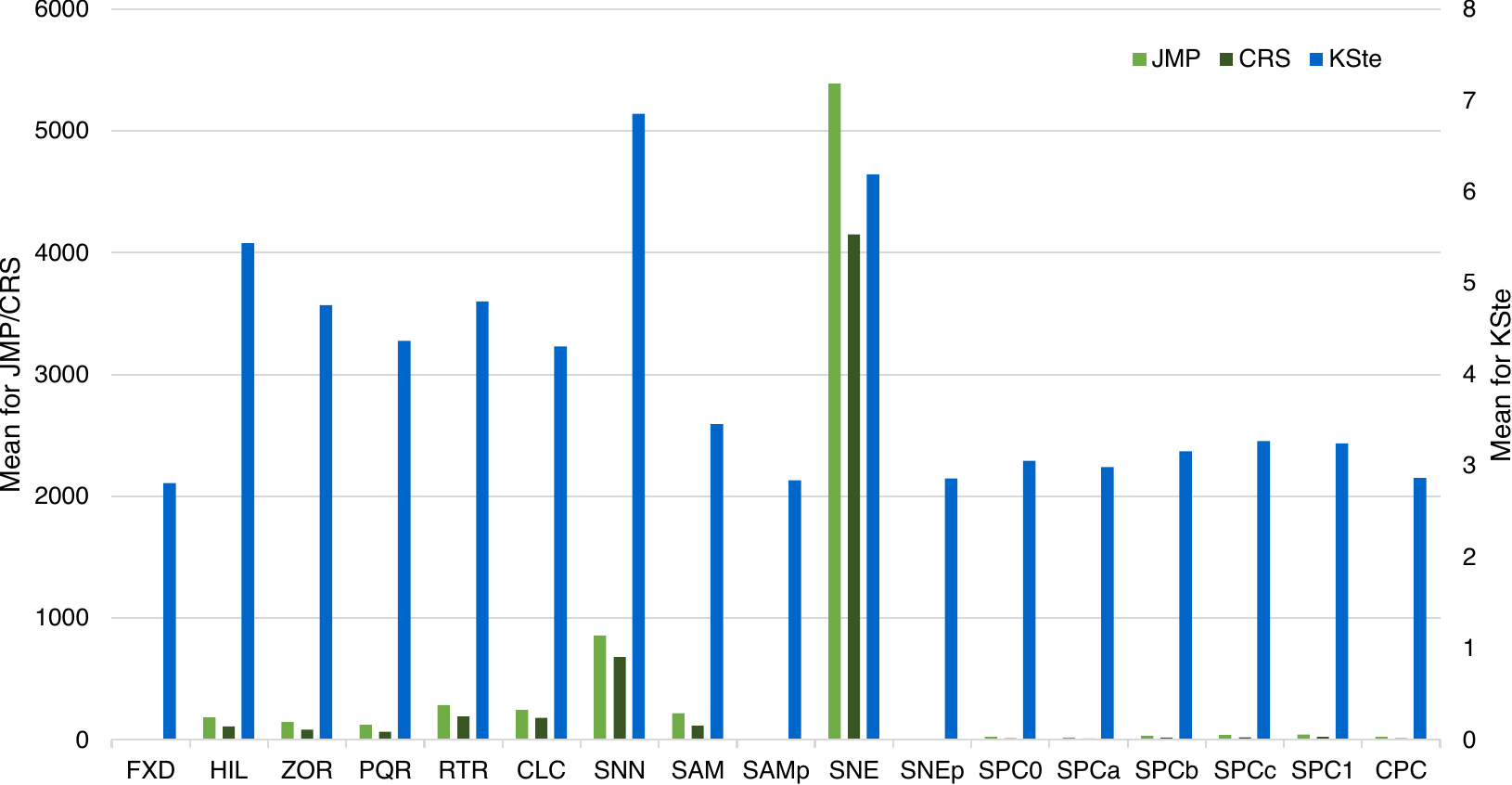}}\\
		\\
		\\
		\subcaptionbox{A comparison between the mean for KSte and for KSdi, for uniformly distributed $\sigma$ of SPC$_{\sigma}$ on Reynolds. \label{fig:clusters-parameter-mean}}[\linewidth]{\includegraphics[width=\linewidth]{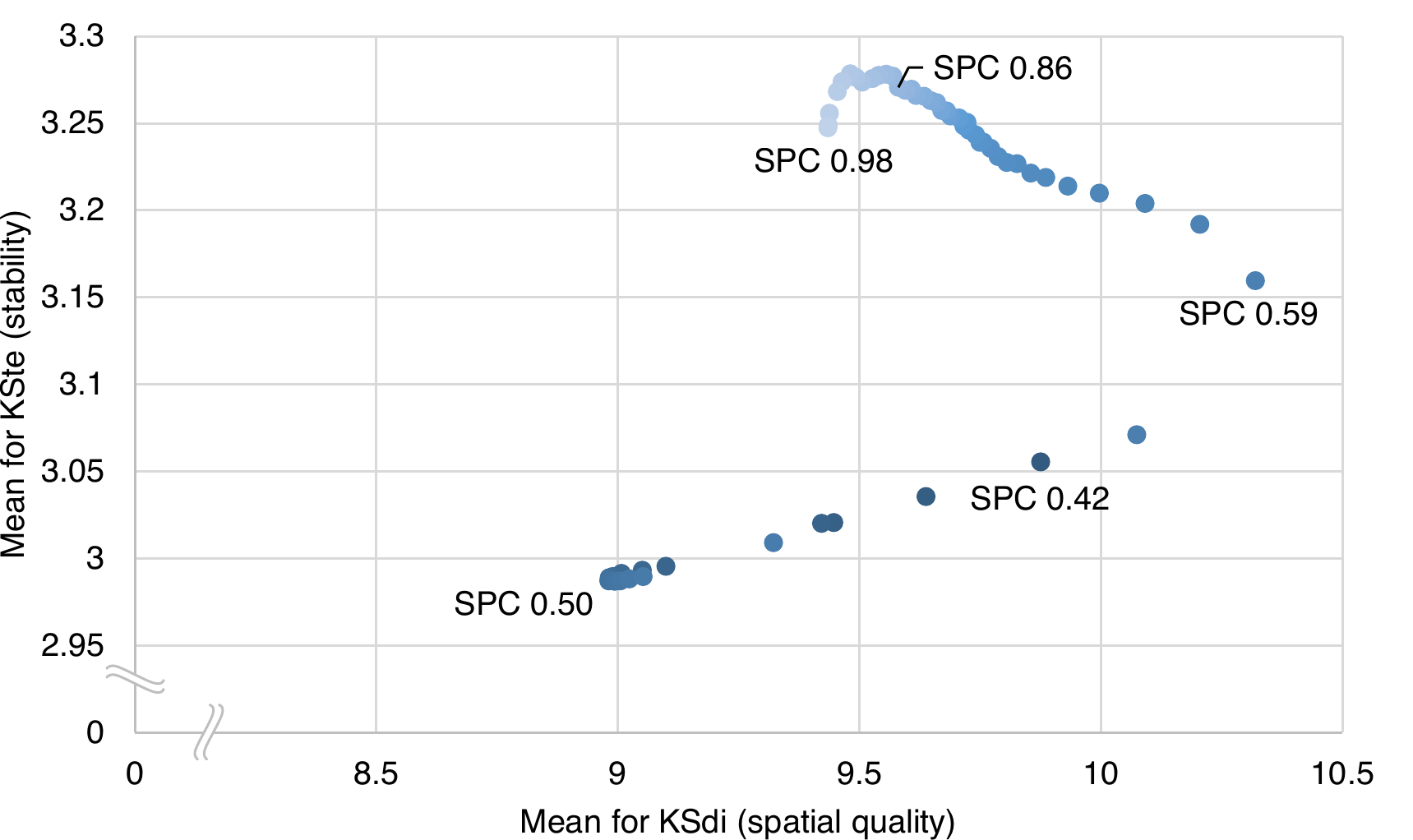}} &
		\subcaptionbox{A comparison between the mean for KSte and for KSdi for all algorithms on Reynolds. \label{fig:clusters-spatialqualityVSstability-chart}}[\linewidth]{\includegraphics[width=\linewidth]{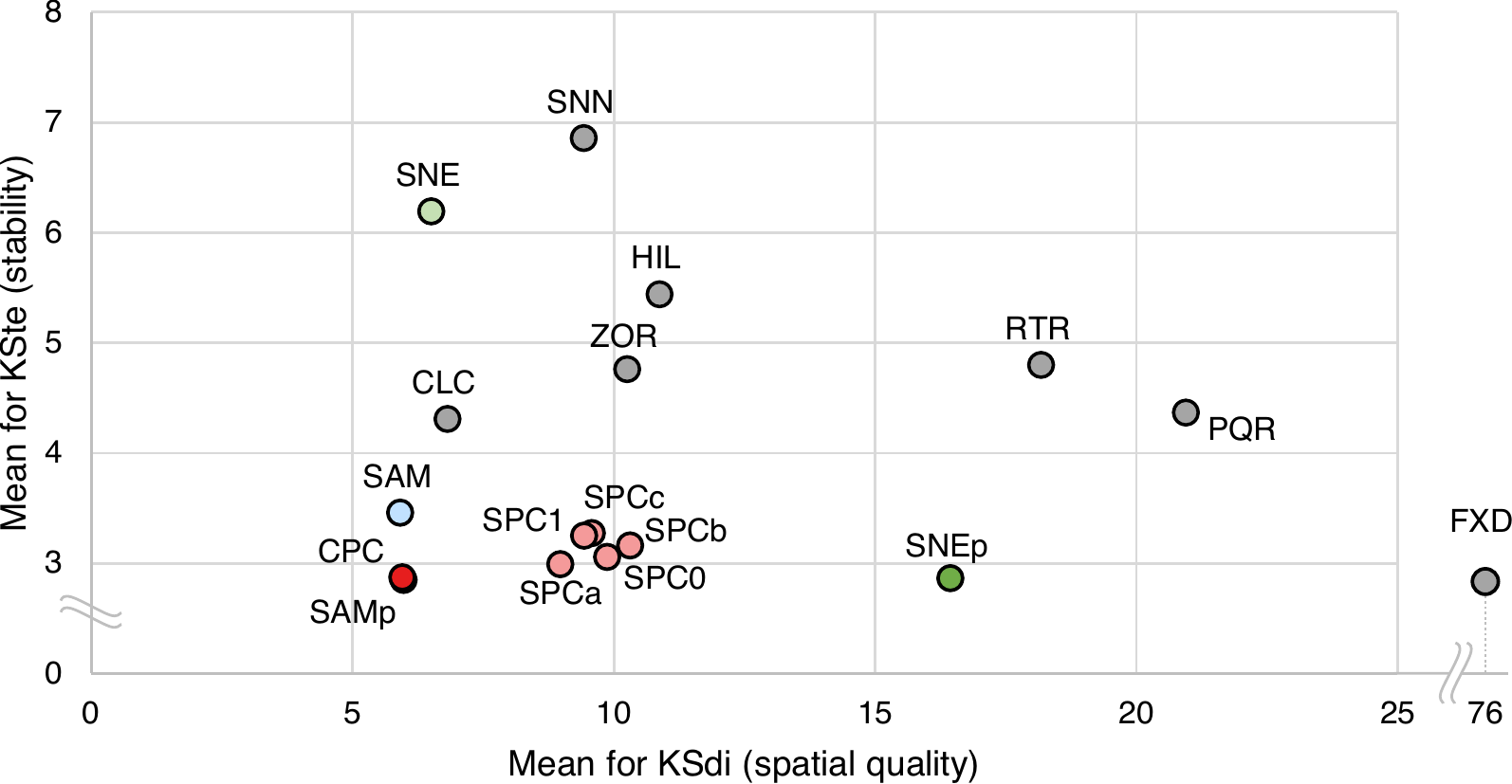}}\\
		\\
		\\
		\subcaptionbox{A comparison between the max for KSte and the mean for KSdi, for uniformly distributed $\sigma$ of SPC$_{\sigma}$ on Reynolds. \label{fig:clusters-parameter-max}}[\linewidth]{\includegraphics[width=\linewidth]{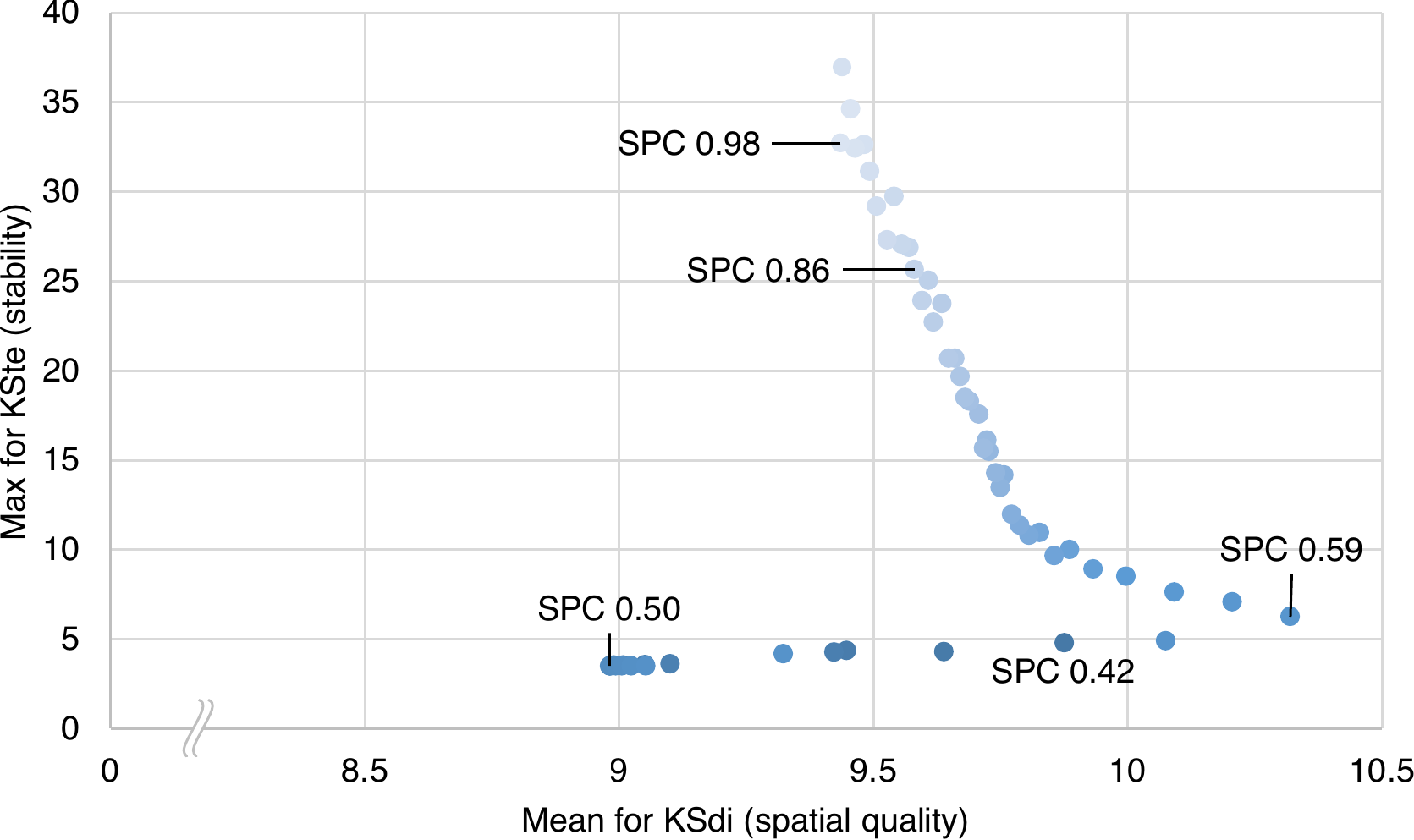}} &
		\subcaptionbox{A comparison between the max for KSte and the mean for KSdi for all algorithms on Reynolds. \label{fig:clusters-spatialqualityVSmaxstability-chart}}[\linewidth]{\includegraphics[width=\linewidth]{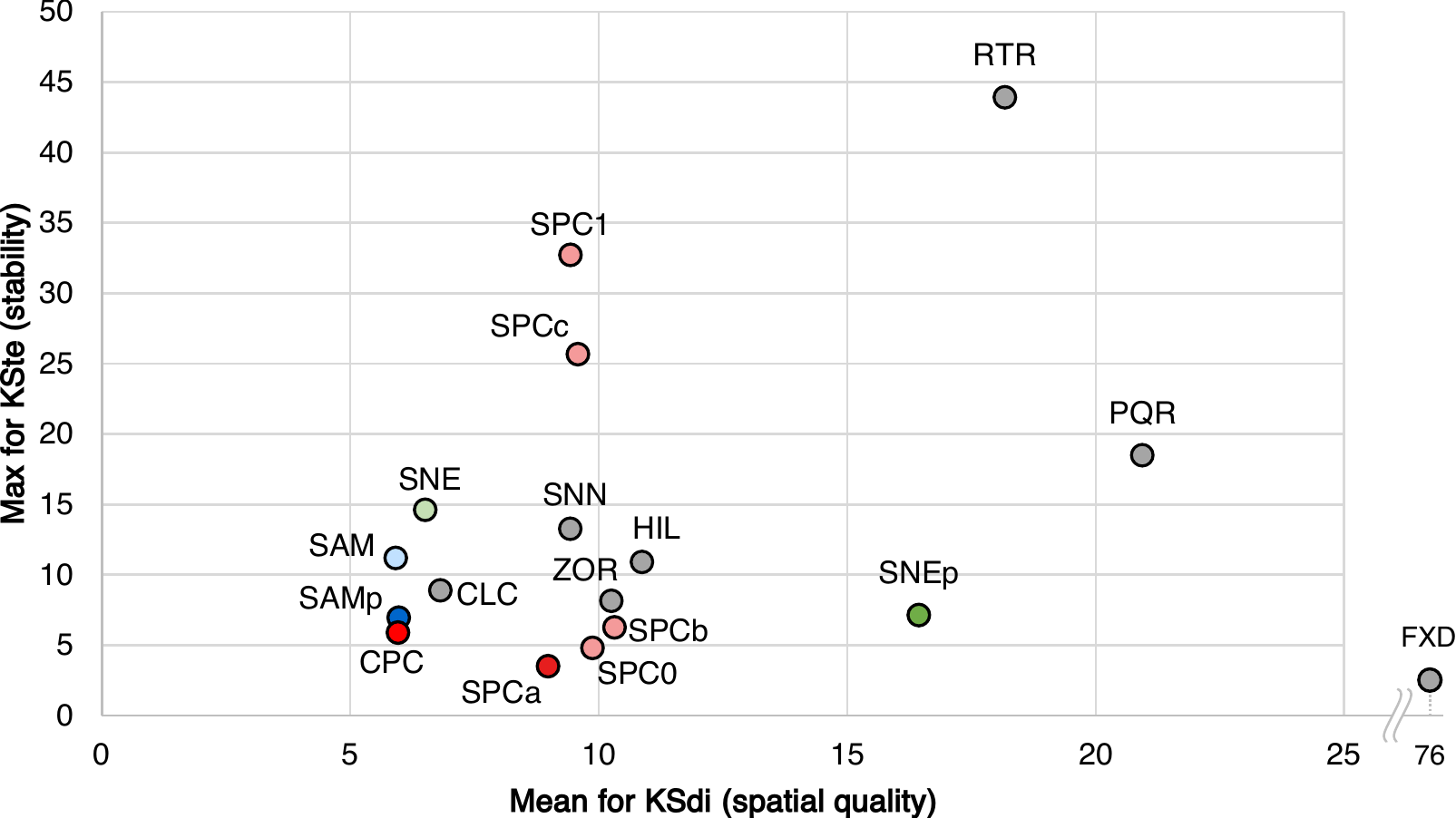}}\\
	\end{tabu}		
	\caption{Quantitative evaluation of Reynolds.}
	\label{fig:clustersevaluation}
\end{figure*}

The results of our experimental evaluation for Reynolds are presented in this section. The structure of this section is similar to the previous two sections, first considering spatial quality and stability in isolation, followed by the parameter experiment. We end the section by considering the trade-off between spatial quality and stability. Charts of all statistics can be found in Figure~\ref{fig:clustersevaluation}.

\mypar{Spatial quality}
The chart in Figure~\ref{fig:clusters-spatialquality-chart} shows the spatial quality for all techniques. Familiar patterns can be found, FXD, RTR, and SNEp are the worst performers, but are this time joined by PQR as another technique that gives relatively worse spatial quality. The other algorithms all score relatively close on spatial quality, in order of increasing spatial quality (and decreasing measure values), the other spatial subdivision techniques are followed by SPC variants, the clustering techniques and finally SNE, SAM, SAMp and CPC. Since there are multiple cluster in this data, which do not interact with each other, it is not surprising that clustering techniques and CPC perform so well.

\mypar{Stability}
The results on the stability of Reynolds are shown in Figure~\ref{fig:clusters-stability-chart}. These results are again very similar to the results on Fish 1 and Fish 2, with the exception that clustering techniques and SNE perform better in comparison to the other techniques. Since the clusters do not interact and contain the same points in every frame, the clustering techniques also perform very well on stability. Especially when considering the KSte we see the clustering techniques and SNE perform better than on the other data sets. The clustering technique CLC even beats some spatial subdivision techniques (HIL and ZOR) on this measure.

\mypar{Parameter experiment}
The parameter experiment also gave some surprising results for Reynolds, as already explained in the main text. Figures~\ref{fig:clusters-parameter-mean}~and~\ref{fig:clusters-parameter-max} show charts containing the results. The cut-off values are $0.98$ and $0.42$ for this data set, meaning that every value above $0.98$ and below $0.42$ uses exactly the projection vectors as the visual summaries using the cut-off values. The parameter values that are indicated by labels in the figures are the values we used in our other experiments. As intermediate values we choose $a=0.50$, $b=0.59$, and $c=0.86$. 

First we consider the parameter values between $0.86$ and $0.59$. These values show the inverse relation between spatial quality and stability: as the parameter value decreases, the stability increases while the spatial quality deteriorate. This is the expected behaviour, which we already saw for Fish 1 and Fish 2.

Between values $0.59$ and $0.50$ we see that lowering $\sigma$ improves both the spatial quality as well as the stability, just as we saw for some parameters in Netlogo. The first principal component does not seem to be the vector that results in the best spatial quality here, hence interpolating more can give better spatial quality, while improving stability. Lowering $\sigma$ further results in worse spatial quality and stability, as we saw in all other data sets.

Finally between $0.98$ and $0.86$ we see better stability when increasing $\sigma$. While this is counter intuitive in general, it can be explained for this data set. As $\sigma$ increases we interpolate less and over configurations where the point set is very rounded. This has a positive effect on the stability in this data set, since it prevents 2 clusters from overlapping a lot: When interpolating, we get a lot of frames where the projected points of two clusters interleave, while the points move in opposite directions. This causes many changes in the neighborhood of all the points in those two clusters. If we interpolate less, this behaviour is prominent and contained in a few frames, leading to less instability according to KSte.

\mypar{Trade-offs}
The trade-offs between spatial quality and stability for Reynolds can be observed in Figures~\ref{fig:clusters-spatialqualityVSstability-chart} and \ref{fig:clusters-spatialqualityVSmaxstability-chart}. As we already observed when considering stability in isolation, clustering techniques and SNE perform really well on this data set, especially when considering maximum values for KSte. These techniques end up in the bottom left corner, making them viable techniques for data sets that are clustered. However, they are still outperformed by SPC variants for low $\sigma$ values, SAMp, SNEp and CLC, when it comes to stability. For $\sigma = 0.50$ SPC performs particularly well, even better than SAMp and CLC on maximum KSte. However SAMp and CLC also have very good spatial quality, making them the best techniques for this data set. 

\section{Summary Statistics}\label{app:summary-stats}
The two tables on the next page provide the statistics over all time steps, for each metric and on every algorithm in our experiment. Each table shows the results for one of the data sets, the fish data set and Netlogo data set respectively. The data in the tables is used throughout the paper in a variety of charts and diagrams.

In Appendix~\ref{subsec:dimreduc} we explained that t-SNE was implemented from scratch, using the simplest form of the algorithm. There are extensions that approximate the gradient during gradient descent to improve run times. These extensions are integrated in most libraries that are currently available. When using the libraries we indeed saw faster run times for both SNE and SNEp, but stability was influenced by the approximations used in the libraries. We therefore chose to implement t-SNE from scratch to show its true capability to produce stable visual summaries. While this lead to slower run times than one would expect from state-of-the-art t-SNE, the run times of the libraries still greatly exceeded the run times observed for the other algorithms in our experiments.

\newpage

\newpage

\newpage
\begin{table*}[b]
	\scriptsize{
		\hfill
		\rotatebox{-90}{
			\begin{minipage}{664.4pt}
				\includegraphics[width=664.4pt]{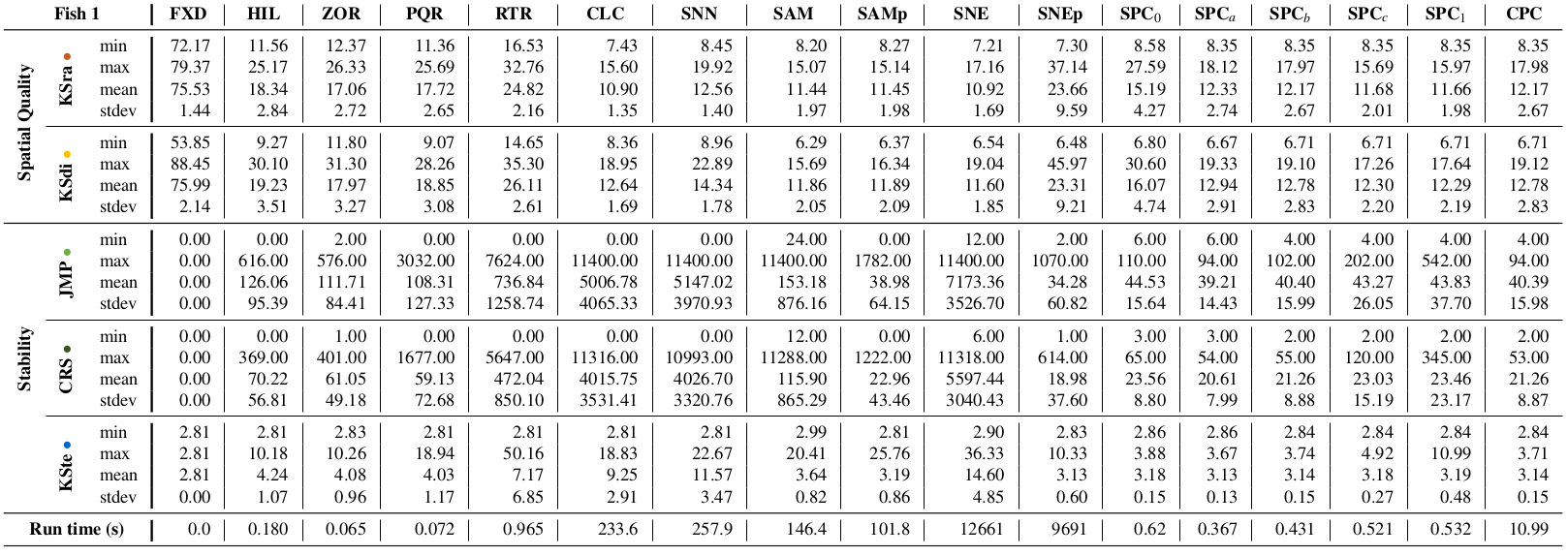}
				\caption{Statistics on Fish 1 for all algorithms and all metrics, including run time in seconds.}
				\label{tab:experiments}
			\end{minipage}
		}
		\hfill
	}
\end{table*}

\newpage
\begin{table*}[b]
	\scriptsize{
		\hfill
		\rotatebox{-90}{
			\begin{minipage}{664.4pt}
				\includegraphics[width=664.4pt]{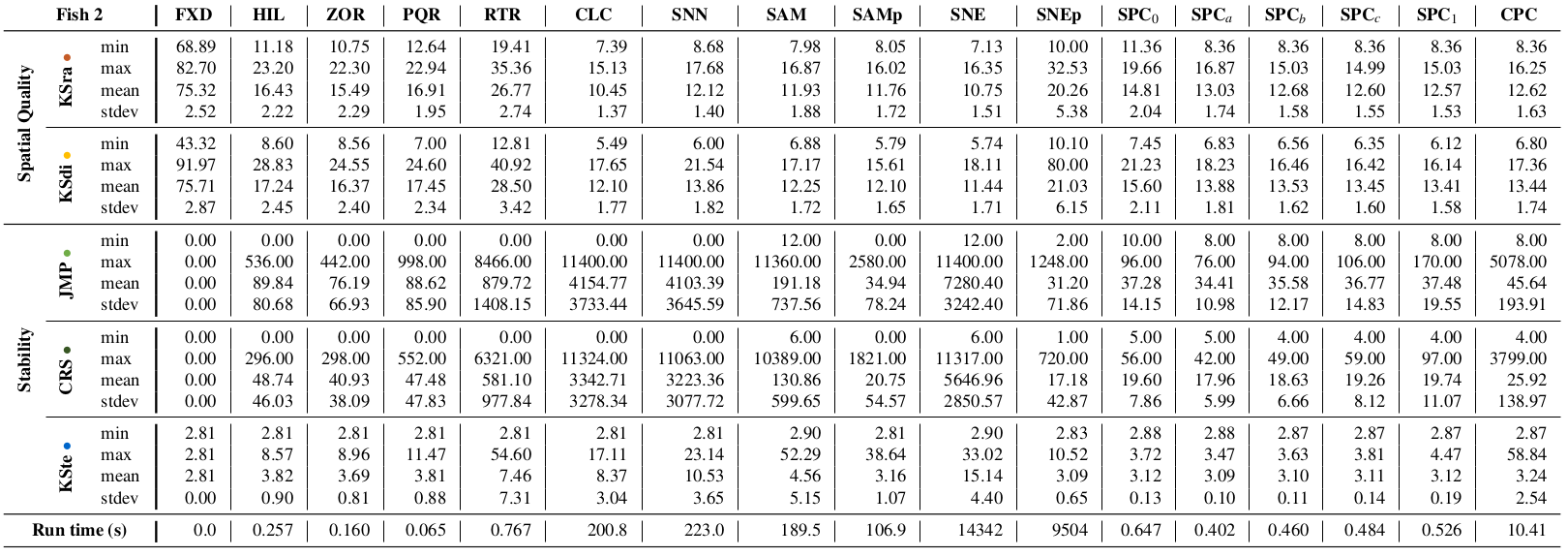}
				\caption{Statistics on Fish 2 for all algorithms and all metrics, including run time in seconds.}
				\label{tab:experiments3}
			\end{minipage}
		}
		\hfill
	}
\end{table*}

\newpage
\begin{table*}[b]
	\scriptsize{
		\hfill
		\rotatebox{-90}{
			\begin{minipage}{664.4pt}
				\includegraphics[width=664.4pt]{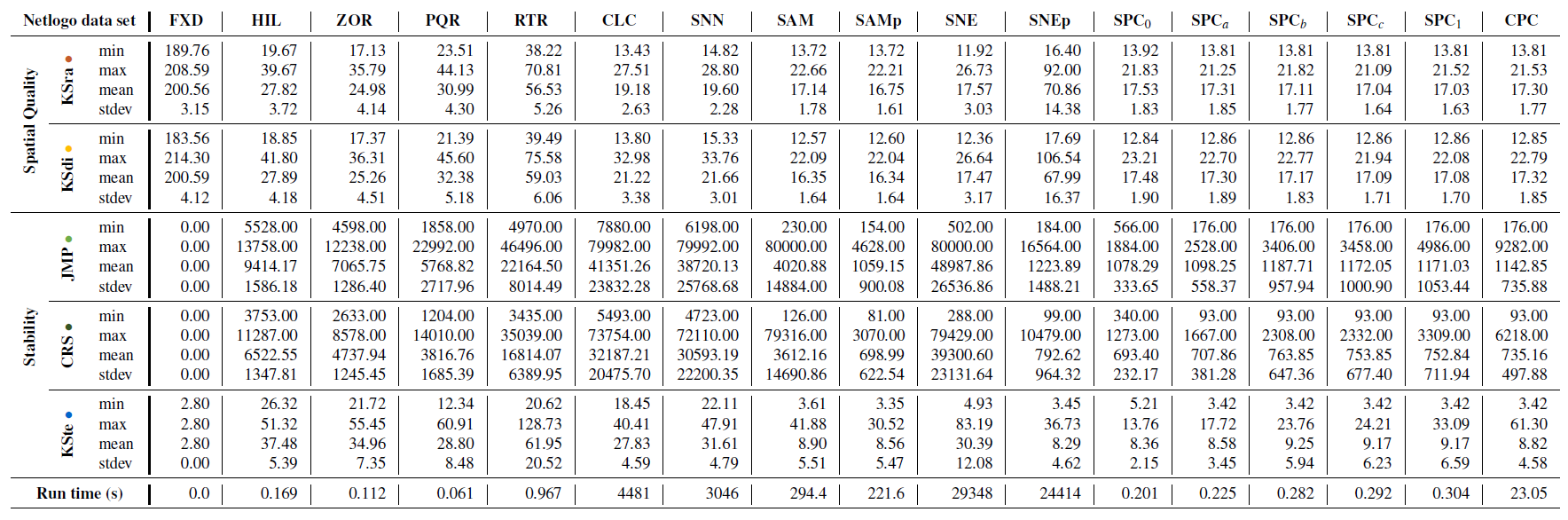}
				\caption{Statistics on Netlogo for all algorithms and all metrics, including run time in seconds.}
				\label{tab:experiments2}
			\end{minipage}
		}
		\hfill
	}
\end{table*}

\newpage
\begin{table*}[b]
	\scriptsize{
		\hfill
		\rotatebox{-90}{
			\begin{minipage}{664.4pt}
				\includegraphics[width=664.4pt]{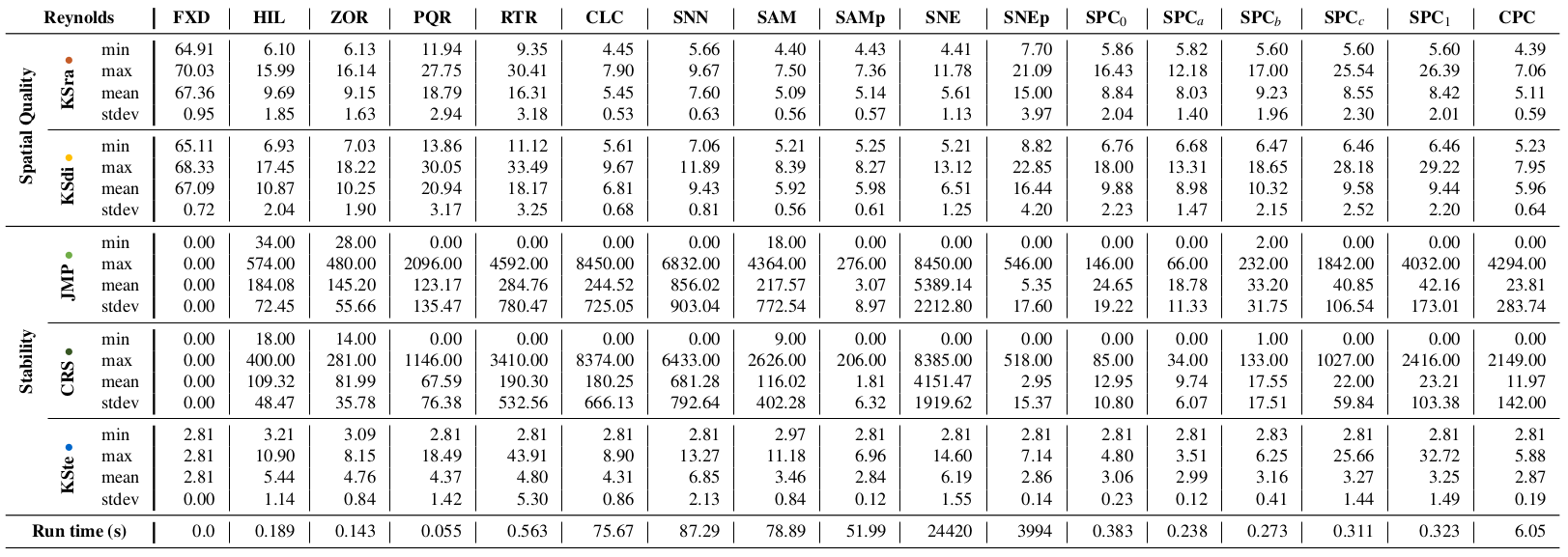}
				\caption{Statistics on Reynolds for all algorithms and all metrics, including run time in seconds.}
				\label{tab:experiments4}
			\end{minipage}
		}
		\hfill
	}
\end{table*}

\end{document}